\documentclass[aps,pra,twocolumn,amsmath,amssymb,nofootinbib,superscriptaddress, showpacs]{revtex4-2}

\usepackage{times}
\usepackage[pdftex]{graphicx}
\usepackage{dcolumn}
\usepackage{bm}
\usepackage{amsmath}
\usepackage{indentfirst}
\usepackage{float}
\usepackage[colorlinks]{hyperref}
\usepackage[dvipsnames]{xcolor}
\usepackage{xcolor}
\usepackage{subfigure}
\usepackage{algorithm}
\usepackage{algorithmicx}
\usepackage{algpseudocode}
\usepackage{amsmath}
\usepackage{verbatim}
\usepackage{makecell}
\usepackage[normalem]{ulem}
\usepackage{url}
\usepackage{times}
\usepackage[pdftex]{graphicx}
\usepackage{dcolumn}
\usepackage{bm}
\usepackage{amsmath}
\usepackage{indentfirst}
\usepackage{float}
\usepackage[colorlinks]{hyperref}
\usepackage[dvipsnames]{xcolor}
\usepackage{xcolor}
\usepackage{subfigure}
\usepackage{algorithm}
\usepackage{algorithmicx}
\usepackage{algpseudocode}
\usepackage{amsmath}
\usepackage{verbatim}
\usepackage{makecell}
\usepackage[normalem]{ulem}
\usepackage{multirow}

\newcommand{\ket}[1]{|{#1}\rangle}

\newcommand{\loss}{\mathcal{L}}
\newcommand{\Hop}{\hat{H}}
\newcommand{\Qop}{\hat{Q}}

\newcommand{\cpx}{\mathcal{S}}
\newcommand{\SVD}{{\rm SVD}}
\newcommand{\circuit}{\hat{\mathcal{C}}}
\newcommand{\rhoop}{\hat{\rho}}
\newcommand{\trace}{{\rm tr}}
\newcommand{\im}{{\rm i}}
\newcommand{\hc}{{\rm H.c.}}
\newcommand{\real}{{\rm Re}}
\usepackage{url}

\usepackage[marginal]{footmisc}\renewcommand{\thefootnote}{}

\begin{document}

\title{Near-Term Quantum Computing Techniques: Variational Quantum Algorithms, Error Mitigation, Circuit Compilation, Benchmarking and Classical Simulation}

\author{He-Liang Huang\textsuperscript{\#}}
\email{quanhhl@ustc.edu.cn}
\affiliation{Henan Key Laboratory of Quantum Information and Cryptography, Zhengzhou, Henan 450000, China}

\author{Xiao-Yue Xu\textsuperscript{\#}}
\affiliation{Henan Key Laboratory of Quantum Information and Cryptography, Zhengzhou, Henan 450000, China}

\author{Chu Guo\textsuperscript{\#}}
\affiliation{Henan Key Laboratory of Quantum Information and Cryptography, Zhengzhou, Henan 450000, China}

\author{Guojing Tian\textsuperscript{\#}}
\affiliation{State Key Lab of Processors, Institute of Computing Technology, Chinese Academy of Sciences, 100190 Beijing, China}
\affiliation{
School of Computer Science and Technology, University of Chinese Academy of Sciences, Beijing 100049, China}

\author{Shi-Jie Wei}
\affiliation{Beijing Academy of Quantum Information Sciences, Beijing 100193, China}

\author{Xiaoming Sun}
\email{sunxiaoming@ict.ac.cn}
\affiliation{State Key Lab of Processors, Institute of Computing Technology, Chinese Academy of Sciences, 100190 Beijing, China}
\affiliation{
School of Computer Science and Technology, University of Chinese Academy of Sciences, Beijing 100049, China}

\author{Wan-Su Bao}
\email{bws@qiclab.cn}
\affiliation{Henan Key Laboratory of Quantum Information and Cryptography, Zhengzhou, Henan 450000, China}

\author{Gui-Lu Long}
\email{gllong@mail.tsinghua.edu.cn}
\affiliation{Department of Physics, Tsinghua University, Beijing 100084, China}
\affiliation{Beijing Academy of Quantum Information Sciences, Beijing 100193, China}



\begin{abstract}
Quantum computing is a game-changing technology for global academia, research centers and industries including computational science, mathematics, finance, pharmaceutical, materials science, chemistry and cryptography. Although it has seen a major boost in the last decade, we are still a long way from reaching the maturity of a full-fledged quantum computer. That said, we will be in the \textit{Noisy-Intermediate Scale Quantum} (NISQ) era for a long time, working on dozens or even thousands of qubits quantum computing systems. An outstanding challenge, then, is to come up with an application that can reliably carry out a nontrivial task of interest on the near-term quantum devices with non-negligible quantum noise. To address this challenge, several near-term quantum computing techniques, including variational quantum algorithms, error mitigation, quantum circuit compilation and benchmarking protocols, have been proposed to characterize and mitigate errors, and to implement algorithms with a certain resistance to noise, so as to enhance the capabilities of near-term quantum devices and explore the boundaries of their ability to realize useful applications. Besides, the development of near-term quantum devices is inseparable from the efficient classical simulation, which plays a vital role in quantum algorithm design and verification, error-tolerant verification and other applications. This review will provide a thorough introduction of these near-term quantum computing techniques, report on their progress, and finally discuss the future prospect of these techniques, which we hope will motivate researchers to undertake additional studies in this field.

\end{abstract}

\maketitle

\def\thefootnote{~~~~~~~\#}\footnotetext{~~~~These four authors contributed equally.}

\section{Introduction}

Quantum computing is a rapidly emerging new-generation computing paradigm that harnesses the laws of quantum mechanics, offering the potential of exponential speedup over classical computation for certain problems~\cite{Nielsen2002QuantumComputation,ladd2010quantum,shor1999polynomial,grover1996fast}. Over the last decade, quantum computing technologies have advanced by leaps and bounds into the NISQ era~\cite{superconducting_2020review,preskill2018quantum}, an important sign of which is the significant achievement of quantum computational advantage on quantum sampling problems has been realized in practice~\cite{arute2019quantum,wu2021strong,zhu2022quantum,zhong2020quantum,zhong2021phase,madsen2022quantum}. The near-term quantum processors, including superconducting qubits~\cite{superconducting_2020review,krantz2019quantum,kjaergaard2020superconducting}, trapped ions~\cite{haffner2008quantum,bruzewicz2019trapped}, and photons~\cite{wang2020integrated,slussarenko2019photonic,flamini2018photonic}, \textit{etc.}, produced during this period contain only a few dozen or even a few thousand qubits, falling well short of the specifications for fault-tolerant quantum computing~\cite{shor1996fault,zhao2022realization,krinner2022realizing,huang2021emulating,liu2019demonstration} (see TABLE.~\ref{tab:quantumsystem} for system parameters of some prototypes). They do, however, serve as fantastic testing grounds for investigating a variety of applications, including machine learning~\cite{biamonte2017quantum}, secure cloud quantum computing~\cite{broadbent2009universal,huang2017experimental,huang2017homomorphic,huang2017universal,barz2012demonstration}, computational science, and complicated quantum chemistry~\cite{cao2019quantum} and many-body quantum systems~\cite{coleman2015introduction} that are not feasible to be simulated with the state-of-the-art supercomputers.

To fully exploit the potential of near-term quantum devices, algorithms/protocols have to be tailored to the constraint of present quantum hardware, especially a modest of qubits with non-negligible errors and limited qubit connectivity. To mitigate quantum errors and achieve valid computational results, several prominent classes of algorithms and tools have been developed specifically for near-term quantum computers, including:

\begin{itemize} 
\item \textit{Variational Quantum Algorithms (VQAs)}~\cite{cerezo2021variational,bharti2022noisy}. VQAs have emerged as one of the leading candidates to achieve application-oriented quantum computational advantage on NISQ devices, owing to their hybrid quantum-classical approach which has potential noise resilience.

\item \textit{Quantum Error Mitigation (QEM)}~\cite{Cai2022quantum}. QEM refers to a series of techniques that allow us to reduce the computational errors and then evaluate accurate results from noisy quantum circuits, although we still lack practical quantum error-correction technique.
\item \textit{Quantum Circuit Compilation (QCC)}~\cite{botea2018complexity,venturelli2019quantum,kusyk2021survey}. QCC is a key technique to transform the nonconforming quantum circuit to an executable circuit on the target quantum platform according to its constraints, including the native gateset, connectivity and so on.

\item \textit{Benchmarking Protocols}~\cite{eisert2020quantum,knill2008randomized}. Quantum benchmarking helps to evaluate the basic performance of a quantum computer and even the capacity to solve real-world problems.

\item \textit{Classical Simulation}~\cite{guo2019general,guo2021verifying,liu2021closing,PanZhang2021,PanZhang2021b}.  Classical simulation of quantum circuits is one of the core tools for designing quantum algorithms and validating quantum devices.

\end{itemize}

\begin{figure}[!htbp]
    \begin{center}
    \includegraphics[width=1\linewidth]{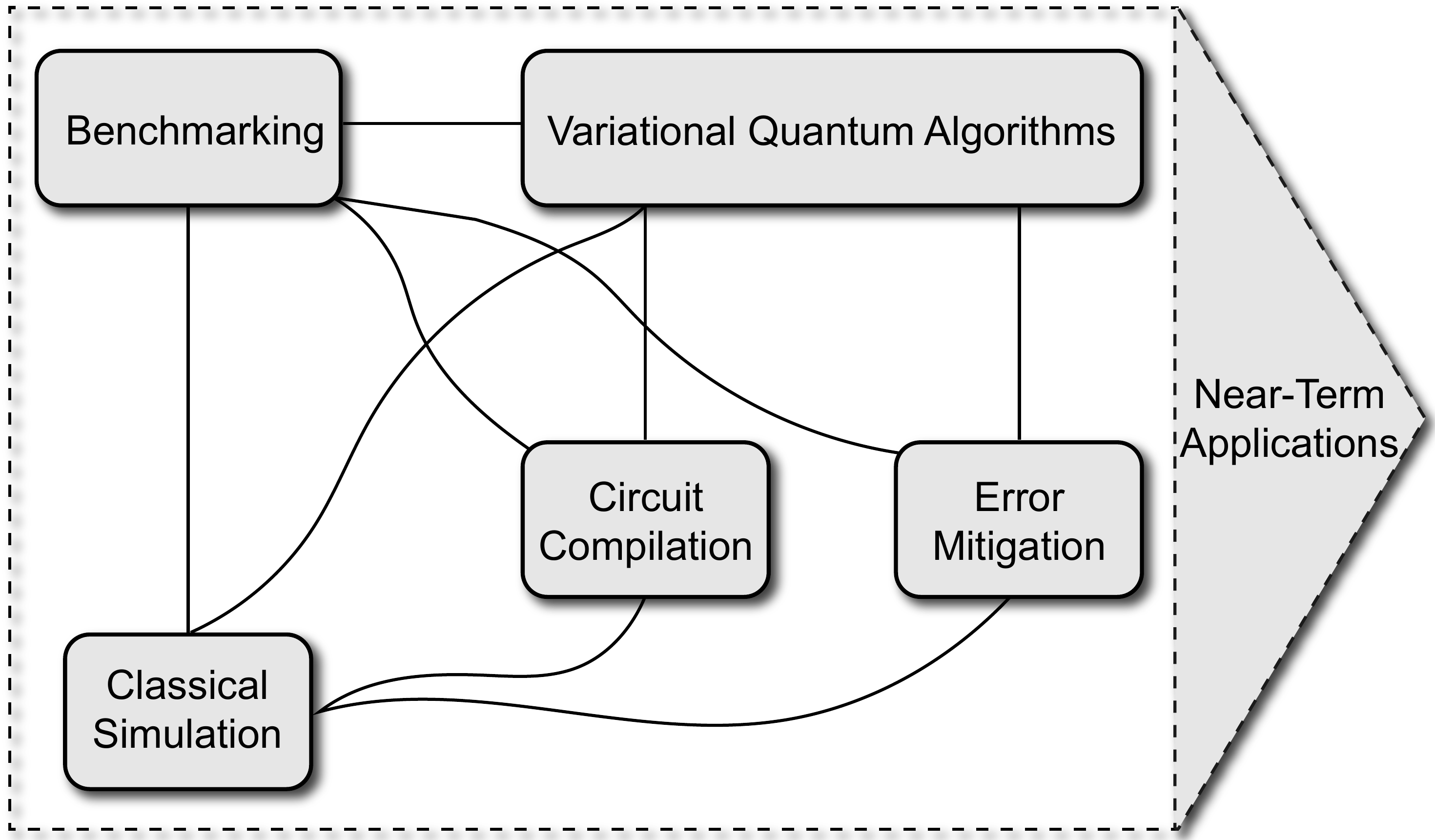}
    \end{center}
    \caption{\textbf{Relationship between near-term quantum techniques.} These techniques collaborate to execute various computing tasks.}
    \label{Relationship}
\end{figure}

These techniques work together to accomplish computational tasks rather than independently of each other (see a rough relationship provided in Fig.~\ref{Relationship}). For instance, variational quantum algorithms, quantum error mitigation, quantum circuit compilation, and quantum benchmarking may all require the help of classical simulation for verification or algorithm design. These near-term quantum computing techniques offer the opportunity to attain practical quantum advantages that can be applied to significant applications such as chemistry~\cite{kandala2017hardware,google2020hartree} and machine learning~\cite{cong2019quantum,havlivcek2019supervised,liu2021hybrid,huang2018demonstration, lloyd2018quantum,huang2021experimental,rudolph2022generation,ebadi2022quantum,Harrigan2021Quantum,zhou2020quantum,gong2022quantum,ding2022active,liu2020variational,xue2022variational}, and they also provide a continuous path for the advancement of quantum computing from today's near-term quantum hardware to tomorrow's fault-tolerant quantum computers.

This review aims to present an overview of these crucial techniques, part of which may serve as a supplement and update to related published review works~\cite{cerezo2021variational,bharti2022noisy,Cai2022quantum,eisert2020quantum}, and some algorithms and protocols are step-by-step taught to help the reader rapidly understand the specifics. We also discuss the challenges and opportunities, and offer some insights on potential future advances. 

\begin{table}[!htb]
\centering
\caption{System parameters of some quantum computing prototypes that have achieved quantum computational advantage.}
\label{tab:quantumsystem}
\begin{tabular}{lcc}
\hline \hline
\textbf{Superconducting system~}& Sycamore~\cite{arute2019quantum}& \textit{Zuchongzhi} 2.1~\cite{zhu2022quantum,wu2021strong}\\
\hline
Number of qubits & 53 &  66 \\
Single-qubit gate error & 0.16\% & 0.16\%\\
Two-qubit gate error & 0.62\% & 0.60\% \\
Readout error & 3.8\% & 2.26\% \\
$T_1$ & 16.04 $\mu$s & 26.5 $\mu$s \\
\hline\hline
\textbf{Photonic system~}& \textit{Jiuzhang} 2.0~\cite{zhong2021phase} & Borealis~\cite{madsen2022quantum}\\
\hline
Detected photons & 113 &  219  \\
Interferometer & 144-mode & 216-mode \\
\hline\hline
\end{tabular}
\end{table}

\section{Variational Quantum Algorithms}

Although some quantum algorithms, such as Shor's algorithm~\cite{shor1999polynomial} and Grover's search algorithm~\cite{grover1996fast}, promise to be much faster than classical algorithms, these algorithms are unattainable in the NISQ era due to the lack of error correction capability. Variational quantum algorithms (VQAs) have been shown to be naturally resistant~\cite{mcclean2016theory} to, and even benefit from, noise, making them particularly suitable for near-term quantum devices, and are therefore considered the most promising route to quantum advantage on practical problems in the NISQ era~\cite{cerezo2021variational}. In this section, we will first introduce the basic concepts, architectures and applications of VQAs, and then discuss the opportunities and challenges for future industrial applications or scientific research.

\subsection{Basic concepts}

\begin{figure*}[!htbp]
\begin{center}
\includegraphics[width=1\linewidth]{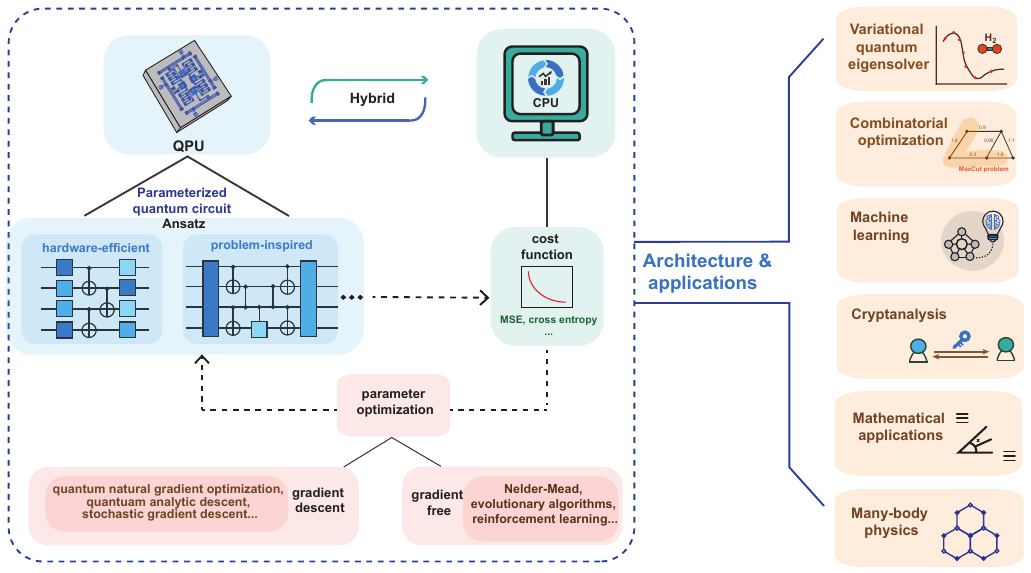}
\end{center}
\caption{\textbf{A high-level overview of variational quantum algorithms (VQAs).} VQAs are a hybrid quantum-classical optimization algorithm. During the training procedure, the quantum processing unit (QPU) estimates the cost function via the parameterized quantum circuit (PQC). Then a classical processing unit (CPU) is employed to implement the parameter optimization. VQAs can be analogous to the quantum counterpart for deep learning, and thus have the potential to be applied to a range of applications, such as find the ground state, combinatorial optimization, machine learning, \textit{etc}.
}
\label{VQAframework}
\end{figure*}

Similar to the classical neural networks, the VQAs is implemented by training a parametrized quantum circuit (PQC) to minimize a problem-specific cost function. Thus, the basic components of VQAs include cost function, PQC and optimization algorithms, as shown in Fig.~\ref{VQAframework}.

\subsubsection{Cost function}

The cost function (sometimes referred to as the loss or objective function) is an integral part of VQAs for encoding problems, and serves essentially the same role as the cost function in classical machine learning. During the training process, the cost function can be considered as a measure of the performance of a VQA with respect to the given training samples and the expected output, which helps to find the global minima. Without loss of generality, the cost function can be expressed as
\begin{equation}
C(\bm{\theta}) = f(O(U(\bm{\theta}),\rho_\text{in}))
\label{equation:costfunction}
\end{equation}
where $O$ are a set of observables on the output state obtained by the input state $\rho_\text{in}$ under the action of the PQC $U(\bm{\theta})$ with tunable parameters $\bm{\theta}$, and $f$ is some function, which is designed according to the specific problem. For example, the cost function of variational quantum eigensolver is often set as
\begin{equation}
\langle H\rangle_{O(U(\bm{\theta}),\rho_\text{in})},
\end{equation}
which is expectation value of qubit Hamiltonian $H$. In addition, for quantum machine learning tasks, one can usually follow the cost functions commonly used in classical community, such as using mean squared error (MSE) or cross entropy as the cost function for the classification task.

The choice of cost function could affect the trainability of VQAs. Cerezo \textit{et al.} proved that barren plateaus are cost-function-dependent in shallow PQCs~\cite{cerezo2021cost}. Specifically, the cost function with global observables leads to an exponentially vanishing gradient (\textit{i.e.}, barren plateau), while the gradient vanishes polynomially when the cost function is defined in terms of local observables. Recently, Liu \textit{et al.} found a similar phenomenon in tensor-network based machine learning~\cite{liu2021presence}.

\subsubsection{Parameterized quantum circuits (PQCs)}

PQCs are the main part that differentiates VQAs from classical neural networks. A PQC consists of a number of fixed quantum gates and trainable quantum gates, and may even includes some measurement and feedback operations. The trainable quantum gate is usually a single-qubit rotation, $R_x(\theta)$, $R_y(\theta)$, and $R_z(\theta)$ rotations, and the trainable parameter is its rotation angle $\theta$. These trainable parameters are trained to minimize the cost function.

The structure of the variational ansatz plays an essential role in the performance of VQA, including the convergence speed and the closeness of the output to the optimal solution of the problem. The design of an effective structure is a challenge for VQAs, which requires optimization in many aspects, such as strong expressibility, shallow circuit depth, and small number of parameters. In the following, we will present some common ansatzes.

\textbf{Problem-inspired ansatzes.} The problem-inspired ansatz is constructed by using knowledge about a specific problem. A typical example is the quantum approximate optimization algorithm (QAOA)~\cite{farhi2014quantum}, whose ansatz is set as
\begin{equation}
U(\bm{\gamma} ,\bm{\beta} ) = \prod\nolimits_{l = 1}^p {{e^{ - i{\beta _l}{H_M}}}{e^{ - i{\gamma _l}{H_C}}}},
\end{equation}
where $H_C$ is the problem Hamiltonian, $H_M=\sum\limits_{i = 1}^n {{\sigma_x^i}}$ is the mixing Hamiltonian, and $\sigma_x^i$ denotes the Pauli $X$ operator acting on qubit $i$. This ansatz is inspired by quantum annealing to map the initial state to the ground state of the problem Hamiltonian $H_C$, by optimizing the parameters $\bm{\gamma}  = ({\gamma _1},{\gamma _2},...,{\gamma _p})$ and $\bm{\beta}  = ({\beta _1},{\beta _2},...,{\beta _p})$.

The problem-inspired ansatz is also widely used in quantum chemistry problems. The unitary coupled cluster (UCC)~\cite{taube2006new} ansatz is a commonly used structure for variational quantum eigensolver (VQE). And historically,  the first VQE experiment by Peruzzo \textit{et al.}~\cite{peruzzo2014variational} utilized the UCC with singles and doubles (UCCSD) ansatz. Subsequently, some improved ansatzes enable shallower depths and higher accuracy. Lee \textit{et al.} proposed the unitary pair coupled cluster with generalized singles and doubles (k-UpCCGSD) method to reduce circuit depth~\cite{lee2018generalized}. The orbital optimized UCC (OO-UCC) ansatz~\cite{mizukami2020orbital} is another variant of UCC to reduce the number of parameters and circuit depth, while maintains a similar level of accuracy to that of UCCSD. Besides reducing circuit depth, Metcalf \textit{et al.}  employed the double unitary coupled-cluster (DUCC) method to effectively reduce the required number of qubits~\cite{metcalf2020resource}. Here we only make a list of these ansatzes, more detailed introduction can be found in the Ref.~\cite{fedorov2022vqe}.

In general, the problem-inspired ansatz is designed according to the characteristics of the problem, so that the output has a high accuracy. However, limited by the connectivity of quantum processors, implementing problem-inspired ansatz on real quantum devices may necessitate a relatively deep PQC, which might be a challenge.

\textbf{Hardware-efficient ansatzes.} Unlike the problem-inspired ansatz, the hardware-efficient ansatz~\cite{kandala2017hardware} is designed around properties of quantum hardware, such as the qubit connectivity and restricted gate sets, to ensure efficient implementation on near-term quantum devices.

The trainable unitary operator $U(\bm{\theta})$, represented by the PQC of this anstz, is composed of $L$ layers and each layer merges trainable single-qubit gates with a entangling block. Mathematically, we have $U(\bm{\theta}) := \prod_{l=1}^L (U_EU_{l}(\bm{\theta}))$, where $U_{l}(\bm{\theta})$ is  the $l$-th  trainable layer  and $U_{E}$ is the entanglement layer. In particular,  we have  $U_{l}(\bm{\theta})=\bigotimes_{i=1}^N (U_S(\bm{\theta}^{(i,l)}))$, where $\bm{\theta}^{(i,l)}$ represents the $(i,l)$-th entry of $\bm{\theta}\in \mathcal{R}^{N\times L}$,  $U_S$ is the trainable unitary with  $U_S\in SU(2)$, \textit{e.g.}, the rotation single qubit gates $R_x$, $R_y$, and $R_z$. The entangle layer $U_E$ is constructed from the entanglement operations that the hardware can provide, such as CNOT, CZ gates, iSWAP gate, and even many-body quantum evolution. These entanglement operations should follow the connectivity of quantum hardware, such as one-dimensional chains, lattices, \textit{etc}.

\textbf{Variable-structure ansatzes.} Unlike the problem-inspired ansatz and hardware-efficient ansatz, which take a fixed structure, the circuit structure in the variable-structure ansatz is no longer fixed and trained as a parameter with the aim of further reducing the circuit depth and the number of gates. A typical example is ADAPT-VQE~\cite{grimsley2019adaptive} and qubit-ADAPT-VQE~\cite{tang2021qubit}, which are proposed to optimize the circuit structure of the ansatz to reduce the circuit depth in quantum chemistry applications. Zhu \textit{et al.} also developed an adaptive QAOA for solving combinatorial problems~\cite{zhu2022adaptive}. Interested readers can refer to the Refs.~\cite{yao2021adaptive, zhang2021mutual,claudino2020benchmarking, liu2021efficient, zhang2022differentiable,zhang2021neural,funcke2021dimensional} for other variants or optimizations, and we will not introduce them one by one.

\subsubsection{Parameter optimization}

The procedure of parameter optimization is implemented after the cost function and ansatz are determined, to minimize the cost function. An appropriate optimizer should be determined by considering the requirements of the application. There are mainly two classes of optimization methods, gradient-based and gradient-free methods, which we will introduce below.

\textbf{Gradient descent methods.} Gradient descent is a general-purpose optimization algorithm for finding a local minimum of the given cost function. For this purpose, the gradient of a cost function $C(\bm{\theta})$ with respect to its parameters $\bm{\theta}$, \textit{i.e.}, ${\nabla _ {\bm{\theta}}}C(\bm{\theta})$, is calculated to obtain the direction of steepest descent. Then a small updates to the parameters is implemented according to 
\begin{equation}
\bm{\theta}  \to \bm{\theta}  - \eta \nabla _ {\bm{\theta}}C(\bm{\theta}),
\end{equation}
where $\eta$ is the step size, to iteratively approach the minimum value of the cost function.

The most commonly used gradient descent method for VQAs is parameter-shift rule~\cite{li2017hybrid,mitarai2018quantum,SchuldKilloran2019,banchi2021measuring}. As shown in Eq.~\ref{equation:costfunction}, the cost functions are usually phrased in terms of expectation values of some observables $O$, evaluated on a output quantum state $\rho_\text{out}$ of the PQC $U(\bm{\theta})$.
In the case that $\rho_\text{out}$ depends on a parameter $\theta_i$ that parametrizes a Pauli-rotation gate $e^{-i\theta_i P/2}$, where $P$ is a Pauli operator, we can compute the derivative with respect to $\theta_i$ as
\begin{equation}
{\nabla _{{\theta _i}}}\langle O\rangle ({\theta _i}) = \frac{1}{2}\left( {\langle O\rangle ({\theta _i} + \frac{\pi }{2}) - \langle O\rangle ({\theta _i} - \frac{\pi }{2})} \right).
\end{equation}
The original parameter-shift rule is design for single-parameter gates. Wierichs \textit{et al.} then extended this method to be applicable to multi-parameter quantum gates~\cite{wierichs2022general}.

The gradient descent methods can also be transformed into the quantum version~\cite{li2021optimizing,stokes2020quantum,yamamoto2019natural,wierichs2020avoiding,koczor2019quantum,jones2020efficient}. The quantum versions of the gradient descent was firstly  proposed by Rebentrost \textit{et al.} for high-dimensional optimization problems~\cite{rebentrost2019quantum}. The gradient  algorithm in Ref.~\cite{rebentrost2019quantum} involves phase estimation, which requires substantial circuit depth and is difficult to implement with current quantum hardwares. Instead of using phase estimation, Li \textit{et al.} proposed an experimental friendly quantum gradient algorithm~\cite{li2021optimizing} using the linear combination of unitaries (LCU). 

In addition to above methods, some other gradient descent methods have been proposed, such as quantum analytic descent~\cite{koczor2022quantum} and stochastic gradient descent~\cite{harrow2021low,sweke2020stochastic,banchi2021measuring}, \textit{etc}.

\textbf{Gradient-free methods.} Similar to the classical machine learning, gradient-free optimization methods can also be used to optimize PQCs. The  Nelder-Mead method is one of commonly used  gradient-free optimization for finding a local minimum of a function of several variables, and thus is naturally suitable for VQAs as well. Besides, evolutionary algorithms and reinforcement learning can be also used to train VQAs~\cite{anand2021natural,zhao2020natural,khairy2019reinforcement,wauters2020reinforcement,yao2020policy}. These optimization methods are standard in the classical machine learning community and we will not go into detail in this review.

\subsection{Architectures and applications}

VQAs can be applied to a wide range of fields, including finding ground states of molecules, combinatorial optimization, machine learning, \textit{etc}. Here, we will introduce some typical applications and their experimental progress.

\subsubsection{Variational quantum eigensolver (VQE)}

The VQE~\cite{peruzzo2014variational,mcclean2016theory} is a flagship algorithm for NISQ applications. This type of algorithms allows one to find the ground state of a given Hamiltonian $H$, which may be used for simulating molecules and chemical reactions. According to the Rayliegh-Ritz variational principle~\cite{rayleigh1870finding,ritz1909neue,gould2012variational}, the ground state energy $E_0$ associated with this Hamiltonian $H$ is bounded by
\begin{equation}
{E_0} \le \frac{{\langle \psi |H|\psi \rangle }}{{\langle \psi |\psi \rangle }},
\end{equation}
where $|\psi \rangle$ is a trial quantum state. The original VQE is to find a quantum state approximating the ground state by training a PQC, such that the expected value of the Hamiltonian is constantly approaching the minimum. In addition to its original purpose, VQE has been widely extended to other objectives, such as finding excited states or the spectrum of Hamiltonian~\cite{McClean2017Hybrid,nakanishi2019subspace,higgott2019variational,parrish2019quantum,jones2019variational,zhang2021adaptive,tilly2020computation}, nonequilibrium steady state~\cite{yoshioka2020variational,endo2020calculation,kreula2016non}, and calculating energy derivatives~\cite{mitarai2020theory,parrish2019hybrid,o2019calculating}. Recently, Wei \textit{et al.} propose a full quantum eigensolver~(FQE)~\cite{wei2020full} which treats the gradient descent part in a full quantum mechanical manner.


VQE has been demonstrated on various quantum architectures such as superconducting qubits~\cite{zhang2021simulating,Kandala2019Errormitigation, kandala2017hardware,Colless2018Computation,montanaro2020compressed,chen2020demonstration,google2020hartree,tazhigulov2022simulating, huggins2022unbiasing}, photonic system~\cite{peruzzo2014variational,lee2022error}, and trapped ions~\cite{hempel2018quantum,nam2020ground,zhang2020probabilistic}. These experiments are still boil down to proof-of-principle demonstrations on small-molecule systems, and thus further experimental efforts are required to scale up this approach to larger molecular systems of chemical interest. Meanwhile, the algorithmic approaches also needs to be further developed to relax the hardware limitation, such as reducing the number of qubits~\cite{bravyi2017tapering,setia2020reducing,McClean2017Hybrid,romero2018strategies,mizukami2020orbital,takeshita2020increasing,tang2021qubit,liu2019variational,fujii2022deep} and circuit depth~\cite{tang2021qubit,zhang2022variational,tkachenko2021correlation,mitarai2019generalization,fan2021circuit,zhang2022variational1}, error-mitigation techniques~\cite{Endo2018Practical,Endo2021Hybrid,rosenberg2022experimental}, accelerated VQE~\cite{wang2019accelerated}, measurement-based VQE~\cite{ferguson2021measurement} and so on.

\subsubsection{Combinatorial optimization}
Combinatorial optimization is a promising path to demonstrating practical quantum advantage on near-term quantum devices, most notably the QAOA~\cite{farhi2014quantum,hadfield2019quantum} for solving the combinatorial optimization problem MaxCut. Max-Cut is the NP-complete problem, which is defined to partition the nodes of a graph into two distinct sets $A$ and $B$ that maximizes the number of edges connecting nodes in opposite sets. It has been shown that to achieve an approximation ratio of $r \ge 16/17 \approx 0.9412$ for Max-Cut
on all graphs is NP-Hard~\cite{haastad2001some}.

Mathematically, consider a graph with 
$m$ edges and $n$ vertices, we can use bitstring $z = {z_1} \ldots {z_n}$ to represent the assignment of vertices to the two sets, where ${z_i}=0$ if $i$-th vertex is belongs to set $A$ and ${z_i}=1$ if it belongs to the other set $B$. The objective is to maximize the number of edges cut, denoted as $C(z)$. When  implementing the QAOA to find approximate solutions to the MaxCut problem, we denote the bitstring using computational basis states $|z\rangle $, and define the objective function to maximize as 
\begin{equation}
C(z) = \sum\limits_{\alpha  = 1}^m {{C_\alpha }(z)}  = \sum\limits_{\text{edge}(j,k)}^m {\frac{1}{2}(1 - \sigma _z^j\sigma _z^k)} |z\rangle,
\end{equation}
where $(j,k)$ vertices connect the $\alpha$ edge. ${C_\alpha }$ has the value 1 only if the $j$-th and $k$-th qubits have different measurement results on the $Z$ basis, representing separate partitions. The $p$-layer QAOA parametered circuit is usually governed by the problem Hamiltonian $H_C=\sum\limits_{\text{edge}(j,k)}^m {\frac{1}{2}(1 - \sigma _z^j\sigma _z^k)}$ and a mixing Hamiltonian $H_M=\sum\limits_{i=1}^n{\sigma_x^i}$ alternatively in each layer, as 
\begin{equation}
U({\bm{\gamma }},{\bm{\beta}}) = \prod\nolimits_{l = 1}^p {{e^{ - i{\beta _l}{H_M}}}{e^{ - i{\gamma _l}{H_C}}}},
\end{equation}
where $\bm{\gamma}=(\gamma_1,\ldots,\gamma_p)$ and $\bm{\beta}=(\beta_1,\ldots,\beta_p)$ are variational parameters. The input state of QAOA circuit is set as an $n$-qubit uniform superposition state $|+\rangle^n$, and the final state $\varphi({\bm{\gamma }},{\bm{\beta }}) =U({\bm{\gamma }},{\bm{\beta }}) |+\rangle^n$ is measured to output the bitstring $z$. We train the QAOA parametered circuit to evolve from the initial state into the bitstring $z$ with the maximum partition.

There has been a lot of theoretical works discussing the performance of QAOA and evaluating the resources needed to achieve quantum computational supremacy~\cite{zhou2020quantum, moussa2020quantum, streif2019comparison, guerreschi2019qaoa, crooks2018performance, dalzell2020many,bravyi2022hybrid,egger2021warm}. In Ref.~\cite{streif2019comparison}, Streif \textit{et al.} presented a comparison between the QAOA
with competing methods, quantum annealing and simulated annealing. Guerreschi \textit{et al.} claimed that at least several hundreds of qubits are required for QAOA to achieve quantum speed-up~\cite{guerreschi2019qaoa}. Dalzell \textit{et al.} concluded
that 420 qubits QAOA would be sufficient for quantum computational supremacy~\cite{dalzell2020many}. Furthermore, it was proven that classical Goemans-Williamson algorithm outperforms the QAOA for certain instances of MaxCut at any constant level~\cite{bravyi1910obstacles}. However, Bravyi \textit{et al.}~\cite{bravyi2022hybrid} and Egger \textit{et al.}~\cite{egger2021warm} suggested that higher-level Recursive-QAOA is competitive (and often better than) to best
known generic classical algorithm based on rounding an SDP relaxation. Overall, whether QAOA has advantages over classical algorithms, and in which issues, is still a controversial issue that needs more research. Besides, to improve the performance of the QAOA,  great efforts have been made, such as finding a good classical optimizer~\cite{crooks2018performance, yang2017optimizing, wecker2016training,khairy2020learning,cain2022qaoa}, ansatz design~\cite{zhu2022adaptive,tan2020optimal,patel2022reinforcement,majumdar2021optimizing}, reducing circuit depth~\cite{herrman2021globally,majumdar2021depth} and number of qubits~\cite{zhou2022qaoa,amaro2022filtering}, and so on.

In terms of experimental progress, QAOA has been experimentally implemented in superconducting~\cite{bengtsson2019quantum,Harrigan2021Quantum}, trapped-ion~\cite{zhu2022multi}, and Rydberg atom~\cite{ebadi2022quantum} platform. In particular, in 2021, Harrigan \textit{et al.} realized the QAOA with up to 23 qubits on the Google's Sycamore superconducting quantum processor~\cite{Harrigan2021Quantum}. Ebadi \textit{et al.} experimentally investigated quantum algorithms for solving the maximum independent set problem using Rydberg atom arrays with up to 289 qubits in two spatial dimensions~\cite{ebadi2022quantum}. 

\subsubsection{Machine learning}
Nowadays, artificial intelligence and machine learning have become an integral part of modern life. As quantum computing promises to enhance our capacity to do some crucial computational tasks, it is natural to look for linkages between these two cutting-edge techniques in the goal of gaining a variety of advantages. In fact, VQAs can be seen as quantum analogs of classical neural networks, and thus can easily be used for various machine learning tasks, such as classification and generative models. 

\textbf{QCNN.} Convolutional neural networks (CNNs) stand out from other neural networks for their outstanding capabilities in computer vision~\cite{krizhevsky2012advances,russakovsky2015imagenet,simonyan2014very,he2016deep}, and natural language processing~\cite{dauphin2017language,gehring2017convolutional,zhang2015sensitivity,kirillov2016joint,song2019abstractive,yu2018qanet}. A CNN usually consists of three main types of layers: 1) convolutional layer, 2) pooling layer, and 3) fully-connected layer, where convolutional layers use filters that perform convolution operations on the input, and pooling layers are a type of downsampling operation.

Inspired by CNNs, Cong \textit{et al.} proposed a quantum convolutional neural network (QCNN), and showed its ability to solve quantum many-body problems~\cite{cong2019quantum}. In the QCNN architecture, the convolutional and pooling layers, as well as the fully-connected layer, are designed as quantum circuits, although their actual functions do not correspond exactly to the classical ones. The analysis by Pesah \textit{et al.} implied that QCNNs do not exhibit barren plateaus~\cite{pesah2021absence}, which are important for trainability as the system scales. The QCNN has been realized on a 7-qubit superconducting quantum processor to identify symmetry-protected topological phases of a spin model~\cite{herrmann2022realizing}. 
QCNN is more suitable for quantum problems, in contrast, the hybrid quantum-classical convolutional neural network (QCCNN)~\cite{liu2021hybrid} is specially proposed for real-world problems. The central idea of QCCNN is to implement the feature map in the convolutional layer with a PQC, while other layers remain the classical operation. Such a design not only introduces a quantum-enhanced feature map, but also preserves important features of classical CNN, such as nonlinearity and scalability. Numerical experiments on the Tetris dataset show that QCCNN can accomplish classification tasks with learning accuracy surpassing that of classical CNN with the same structure. Besides, Wei \textit{et al.} proposed a QCNN framework based on LCU~\cite{gui2006general,wei2016duality} that can transform CNN to QCNN directly~\cite{wei2022quantum}.

\textbf{Quantum GAN.} Generative adversarial networks (GANs) are at the forefront of the generative learning and have been widely
used for image processing, video processing, and molecule development~\cite{goodfellow2020generative}. GANs exploit a two-player minimax game between a generator and a discriminator, where generator aims to output the generated data to fool discriminator, and meanwhile, discriminator tries to distinguish the true example from generator. 

Recently, theoretical works show that quantum generative models may exhibit an exponential advantage over classical counterparts~\cite{lloyd2018quantum,gao2018quantum,romero2021variational}, arousing extensive research interest in the theory and experiment of quantum GANs~\cite{situ2020quantum,dallaire2018quantum,kiani2021quantum,chakrabarti2019quantum,bartkiewicz2021collaborative,zhu2021generative,niu2022entangling,rudolph2022generation,braccia2021enhance,coyle2021quantum}. In particular, Huang \textit{et al.} developed a resource-efficient quantum GAN, and for the first time accomplished the generative task of real-world handwritten digit image on a superconducting quantum processor~\cite{huang2021experimental}. Being able to handle real-world data generation tasks is an exciting thing for an ``infant'' quantum computer. Subsequently, Rudolph \textit{et al.} provided an experimental implementation of generating more high-resolution images of handwritten digits with quantum GAN on an ion-trap quantum computer~\cite{rudolph2022generation}. Niu \textit{et al.} showed that by learning a shallow quantum circuit to generate a superposition of classical data, their proposed entangling quantum GAN can be used to create an approximate quantum random
access memory (QRAM)~\cite{niu2022entangling}.

\textbf{Quantum kernel estimation.} Quantum Kernel estimation is a common method for supervised learning classification task on the NISQ device. In Ref.~\cite{schuld2019quantum}, Schuld \textit{et al.} interpreted the process of encoding classical information into a quantum state as a nonlinear feature map which assigns data to the Hilbert space of the quantum system, and then proposed to use a variational quantum circuit to process the feature vectors. Havl{\'\i}{\v{c}}ek \textit{et al.} proposed and experimentally implemented binary classifiers that use the quantum state space as feature space on a superconducting processor~\cite{havlivcek2019supervised}, providing a possible path to quantum advantage.

\textbf{Quantum auto-encoder.} Quantum auto-encoder (QAE) is an efficient VQA for quantum data compression. QAE starts from a large-scale quantum state $\rho_{AB}$, utilizes a PQC $U({\bm{\theta}})$ to encode $\rho_{AB}$ into the $A$ subsystem (latent space), and then recovers the initial state $\rho_{AB}$ from this compressed state with high fidelity via a decoder constructed by another PQC ${V({\bm{\theta'}}) }$. Usually $V$ is set as ${V({\bm{\theta'}})}={U({\bm{\theta}})^\dag }$. QAE has recently attracted great attentions~\cite{bondarenko2020quantum,bravo2021quantum, cao2021noise,locher2022quantum,srikumar2021clustering,du2021exploring}, and its experimental implementations have been carried out on linear optical systems~\cite{pepper2019experimental,huang2020realization} and superconducting systems~\cite{ding2019experimental}.

\subsubsection{Cryptanalysis} Shor's integer factorization algorithm is considered to be one of the most influential algorithms that shape the research interest in quantum computing today. However, implementing Shor's algorithm requires a fault-tolerant quantum computer, which is far from attainable. Variational Quantum Factoring (VQF) algorithm is heuristic alternative to Shor's algorithm suitable for NISQ devices~\cite{anschuetz2018variational}. It works by reducing the factoring problem to the ground state of the Ising Hamiltonian, which can then be found using VQA. Karamlou \textit{et al.} reported a experimental demonstrations of VQF using a superconducting quantum processor~\cite{karamlou2021analyzing}, the integers 1099551473989, 3127, and 6557 are factored with 3, 4, and 5 qubits, respectively. Besides, the performance and resource analysis of VQF are presented in Refs.~\cite{phan2022quantum,qiu2020resiliency}. 

In addition to the factoring task, VQAs can also be used to attack advanced encryption standard (AES)-like symmetric cryptography~\cite{wang2022variational}. In their simulation results, sometimes the variational quantum attack algorithm is even faster than Grover's algorithm. Moreover, Coyle \textit{et al.} proposed cryptanalysis algorithm based on variational quantum cloning (VarQlone), which allows an adversary to obtain optimal approximate cloning strategies for unknown quantum states with shallow quantum circuits~\cite{coyle2022progress}.

\subsubsection{Mathematical applications}
Numerical solvers such as solving linear systems, non-linear differential equations, $etc$, have a very wide range of engineering application scenarios. Several classical-quantum hybrid algorithms derived from VQAs make it possible for near-term quantum devices to solve large-scale numerical problems without the use of fault-tolerant quantum computing.

\textbf{Variational quantum linear solver.} Solving systems of linear equations is a fundamental computational problem, and the vast majority of numerical problems in science and engineering boil down to solving systems of linear equations. Unlike the Harrow-Hassidim-Lloyd (HHL) algorithm~\cite{harrow2009quantum}, the variational quantum linear solver (VQLS) proposed by Bravo-Prieto \textit{et al.} can run on
near-term quantum computers~\cite{bravo2019variational}. By employing a hybrid quantum-classical algorithm, the goal of VQLS is to variationally prepare $|x\rangle $ such that $A|x\rangle  \propto {\kern 1pt} {\kern 1pt} |b\rangle$. The numerical simulations give the evidence that the run time of VQLS has linear scaling in $\kappa $, logarithmic scaling in $1/\varepsilon$, and polylogarithmic scaling in $N$, where $\kappa $ is the condition number of a $N \times N$ matrix $A$, and $\varepsilon$ is the precision of the solution. In the same period, Huang \textit{et al.}~\cite{huang2019near} and Xu \textit{et al.}~\cite{xu2021variational} independently proposed similar quantum algorithms.

\textbf{Solving nonlinear differential equations.} Differential equations are ubiquitous in various fields of science and engineering. Lubasch \textit{et al.} first extended the concept of VQAs to introduce nonlinearity for solving nonlinear problems~\cite{lubasch2020variational}, such as the nonlinear Schr{\"o}dinger equation. Subsequently, several approaches to solve the nonlinear differential equation using the VQAs were proposed~\cite{kyriienko2021solving,joo2021quantum,kubo2021variational,liu2021variational}, and the application to computational fluid dynamics was studied~\cite{jaksch2022variational}.

\subsubsection{Many-body physics} 
It is more natural to use quantum circuits to process quantum data than to process classical data, because it avoids the input and output problems of classical-quantum data conversion. VQAs enable us to realize the simulation of many-body dynamics~\cite{lee2021neural,benedetti2021hardware}, as well as identifying many-body phase transition~\cite{gong2022quantum,uvarov2020machine,herrmann2022realizing,liu2021probing}. In particular, Gong \textit{et al.} proposed a quantum neuronal sensing based on digital-analog variational quantum circuit, and exprimentally realized this scheme to classify the ergodic and localized phases of matter on a 61 qubit superconducting quantum processor~\cite{gong2022quantum}. This experiment demonstrates the experimental feasibility of large-scale VQAs, and opens new avenues for exploring quantum many-body phenomena in larger-scale systems.

We noticed that the application fields of VQAs are far more than those listed above. It can also be applied to finding quantum error-correcting codes~\cite{cao2022quantum}, entanglement purification~\cite{zhang2022variational2}, amplitude estimation~\cite{plekhanov2022variational}, variational Inference~\cite{benedetti2021variational}, $etc.$, and we will not introduce them one by one.

\subsection{Opportunities and Challenges}


Due to their wide applications and noise resistance to NISQ devices, VQAs have a huge potential to achieve quantum advantages for practically interesting problems. Over the past few years, VQAs have received unprecedented attention, and papers related to VQAs appear on the arXiv website almost every day. The following points deserve our attention in advancing VQAs into future industrial applications or truly meaningful scientific research.
 
\textbf{Where is the main advantage?} Although VQAs have been hotly researched for several years, this core question has never been well answered. Perhaps using VQAs to handle quantum problems is a natural advantage~\cite{gong2022quantum}, since either simulating quantum systems or measuring quantum systems using classical methods would be exponentially resource-intensive. However, for real-world problems, it becomes very tricky to answer why we need VQAs. Does it offer speed or accuracy advantages over classical neural networks? Maybe we can try to answer this question in two ways: 1) Find the provable advantages of VQAs theoretically. This should be difficult, since today's machine learning algorithms are still theoretically difficult to study; 2) Experimentally achieve milestones that beat current state-of-the-art classical machine learning in terms of speed or accuracy on large-scale practical datasets (rather than toy models). Either way, it will take a huge effort.

\textbf{I/O problems for real-world data.} To fully exploit the superposition properties of quantum systems, we actually want  the classical data to be encoded into the amplitude of quantum system. However, this is extremely resource-intensive, whether using QRAM~\cite{giovannetti2008quantum} or pure quantum circuits~\cite{sun2021asymptotically,zhang2022quantum}. Thus, for real-world applications, perhaps we need to find some specfic classical problems suitable for quantum computing. For example, the input to this classical problem has a certain sparsity or symmetry, which is conducive to being encoded as a quantum state. For the output, in general we need to avoid the exponential consumption caused by a large number of measurements, which requires us to extract and analyze only part of the  information of the output quantum state, such as the expected value of observables, principal component analysis~\cite{lloyd2014quantum}, $etc$.

\textbf{Trainability and training efficiency.} Avoiding barren plateaus is now an important area of research, as barren plateaus in training landscapes will appear if care is not taken enough. Some results suggest that the exponential parameter space, the noise and decoherence, and even entanglement can induce barren plateaus~\cite{mcclean2018barren,wang2021noise,marrero2021entanglement,martin2022barren}. At present, some methods have discussed how to mitigate barren plateaus~\cite{sack2022avoiding,haug2021optimal,broers2021optimization,anschuetz2022beyond,cerezo2021cost}, and more research is still needed.

One of the weaknesses of quantum computers compared to classical computers is quantum noise, which also greatly affects the training performance.  Generally, the resillience of VQAs exists for a wide class of noise, as analyzed in Refs.~\cite{mcclean2016theory,1gentini_noise-resilient_2020,3sharma_noise_2020}. Yet such noise resillience is highly limited as the noise level rising. As suggested in Refs.~\cite{5saib_effect_2021,7wright_numerical_2021,8gowrishankar_numerical_2021,wang2021noise,6ito_universal_nodate}, large noise may leads to problems such as performance degradation or barren plateaus. Some special noises, such as leakage error, also have a bad effect on the performance of VQAs~\cite{ding2022evaluating}. Thus, the analysis of iteration complexity and behaviour with noise is necessary~\cite{martin2022barren}, and efficient error mitigation schemes for VQAs need to be developed~\cite{wang2021can,barron2020measurement,botelho2022error,Chen2021Compression}.

Besides, the training process of VQA is very time-consuming, due to the incompatibility with backpropagation and the cost of a large number of measurements, posing a great challenge to the large-scale development of VQAs. Some approaches try to use multiple QPUs for parallel training to alleviate this deficiency, such as data parallelism~\cite{du2022distributed} or parameter parallelism~\cite{niu2022parameter}. However, more fundamental solutions are urgently needed.

\textbf{Hardware development.} For practical NISQ applications, high-quality quantum processors with more qubits, longer coherence times and lower error rates are prerequisites. For VQAs, iterative training requires frequent interaction between classical and quantum computers, which imposes some new demands on the quantum hardware. The cryogenic electronic control architecture is a hardware-level solution for accelerating classical-quantum interactions~\cite{hornibrook2015cryogenic,sebastiano2017cryo,mehrpoo2019benefits,das2018cryogenic,sebastiano2017cryogenic}, 
and also an advanced technique required for the future development of quantum computing.

\section{quantum error mitigation}
\begin{figure*}[!htbp]
    \begin{center}
    \includegraphics[width=1\linewidth]{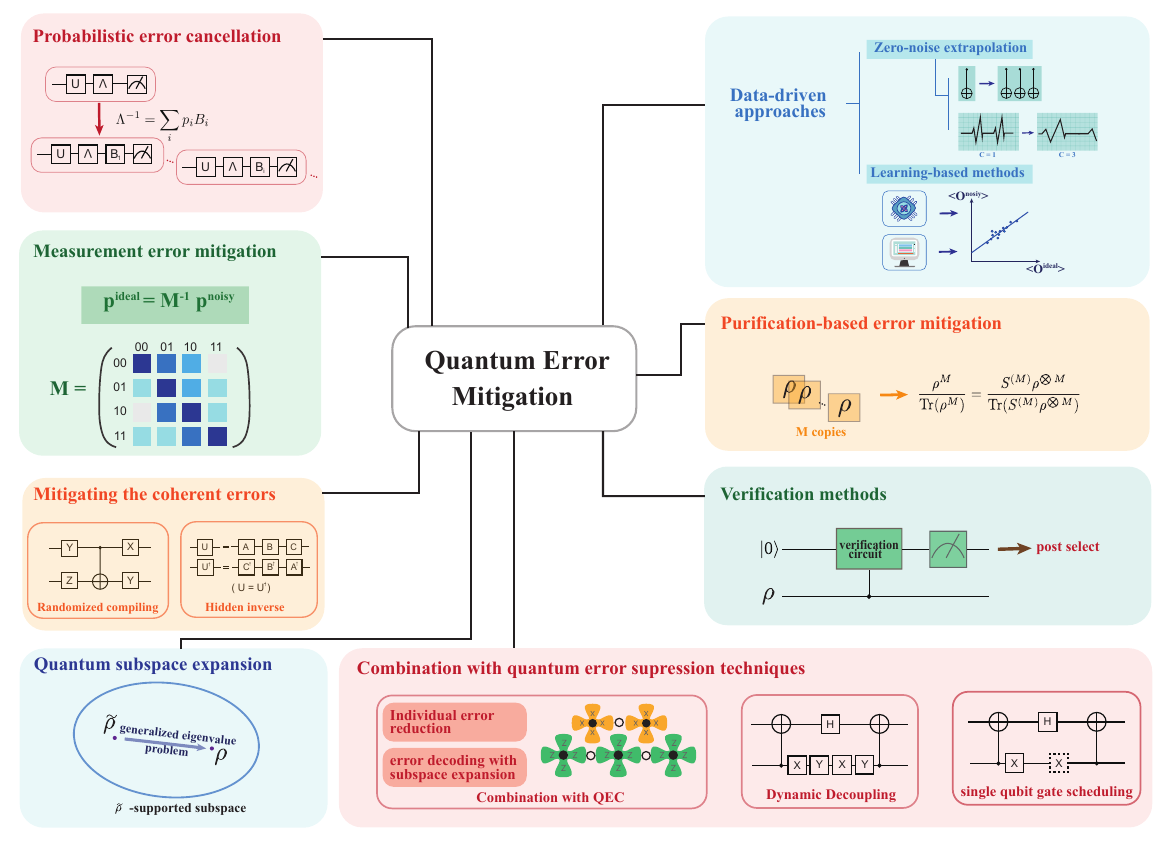}
    \end{center}
    \caption{\textbf{Summary of popular Quantum Error Mitigation(QEM) methods.} Probabilistic error cancellation estimates the noiseless expectation value using a linear combination of expectation values with different noise terms, which is based on the quasi-decomposition. The data-driven approaches including Zero-noise extrapolation and learning-based methods collect expectation values of circuits with different error rates, or circuits with similar structure. Purification-based error mitigation schemes use multiple ($M$) copies of the noisy state and estimate $\frac{\text{Tr}(\rho^MO)}{\text{Tr}(\rho^M)}$ using collective measurements. Verification methods are designed for systems with certain symmetries and combined with classical post-processing. Quantum subspace expansion also uses post-processing to mitigate errors for some VQE algorithms. Other mitigation methods for certain noise types, such as measurement error mitigation (for measurement errors) are also widely applied.}
    \label{fig-qem}
\end{figure*}

In the context of current NISQ devices, the impact of noise remains the greatest challenge for the practical applications ~\cite{preskill2018quantum}. Although quantum error correction (QEC) promises to enable quantum computation with
arbitrary levels of noise, it is out of reach for near-term quantum processors. Quantum error mitigation (QEM) provides us a feasible alternative to mitigate errors of near-term quantum processors, and is also the continuous path that will take us from today's quantum hardware to tomorrow's fault-tolerant quantum computers. Instead of active error-correction, QEM methods usually estimate the error-free expectation value by classical post-processing of the noisy measurement results. While it introduces the additional sampling overhead caused by the increase in the variance of the mitigated observable, QEM requires fewer qubits and gate resources and is therefore more suitable for practical NISQ devices. In recent years, a wide variety of QEM methods have been proposed~\cite{Endo2021Hybrid,Cai2022quantum, Endo2018Practical, Qin2022aoverview, Huggins2021Virtual, Cai2021multi, Li2017Efficient}, and we will discuss some typical QEM techniques in the following subsections. 

\subsection{Probabilistic error cancellation}
We first introduce probabilistic error cancellation (PEC)~\cite{Temme2017Error,Endo2018Practical}. The key idea of PEC is to apply the quasi-probability decomposition of the inverse noise process, leading to a linear combination of a set of noisy circuits. The implementation of PEC requires the full knowledge of the noise model in the target circuit and is usually combined with characterization of the noise process. The PEC method was first introduced by Temme {\it{et al.}}~\cite{Temme2017Error} and further detailed for a practical scheme by Endo {\it{et al.}}~\cite{Endo2018Practical}. It has been shown to mitigate Markovian noise~\cite{Endo2018Practical} and has been further generalized to the non-Markovian noise~\cite{Huo2021selfconsistent}. And it was experimentally demonstrated on superconducting~\cite{Song2019Quantumcomputation} and trapped ion~\cite{Zhang2020Errormitigated} quantum computers. 

\subsubsection{Standard PEC with discrete gate-based circuits}
\textbf{Quasi-probability decomposition of inverse noise model.} 
Suppose we have a noisy quantum gate $\mathcal{E}\circ\mathcal{U}$ contaminated by a noise channel $\mathcal{E}$, where $\mathcal{U}(\rho)=U\rho U^\dagger$ denotes the noiseless gate with the ideal operator $U$ and the initial quantum state $\rho$. With approaches to accurately characterize the noise channel $\mathcal{E}$, we can have the mathematical form of the inverse noise channel $\mathcal{E}^{-1}$, and then apply $\mathcal{E}^{-1}$ after the noisy gate $\mathcal{E}\circ\mathcal{U}$ to recover the ideal gate $\mathcal{U} = \mathcal{E}^{-1}\circ\mathcal{E}\circ\mathcal{U}$. Since the inverse noise channel $\mathcal{E}^{-1}$ may be unphysical, we need a set of basis operations $\{\mathcal{B}_k\}_k$ to decompose it as
\begin{equation}\label{eqn-em-ep}
    \mathcal{E}^{-1}=\sum_kq_k\mathcal{B}_k,
\end{equation}
where $q_k$ is the the combination coefficients, \textit{i.e.}, quasi-probabilities. So we can estimate the noiseless expectation value of the observable $O$ using a linear combination of results obtained from circuits applied by different basis operations
\begin{equation}\label{quasi-de}
    \begin{aligned}
    \langle O \rangle_{\text{PEC}} =&\sum_kq_k\text{Tr}[\mathcal{B}_k\circ\mathcal{E}\circ\mathcal{U}(\rho)]  \\
    =&Q_{\mathcal{E}}\sum_k \text{sgn}(q_k)\frac{|q_k|}{Q_{\mathcal{E}}}\text{Tr}[\mathcal{B}_k\circ\mathcal{E}\circ\mathcal{U}(\rho)],
    \end{aligned}
\end{equation}
where $Q_{\mathcal{E}}=\sum_k|q_k|$. Given above quasi-probability decomposition, we can use the Monte Carlo sampling to randomly apply the basis operation set $\{\mathcal{B}_k\}_k$ with the quasi-probability distribution $\{\frac{|q_k|}{Q_{\mathcal{E}}}\}$. Then we multiply the measurement outcome by the corresponding sgn($q_k$) and weight the Monte Carlo average by $Q_{\mathcal{E}}$. For the general circuit composed of a $N$-gate sequences $\mathcal{U} = \prod_{n=1}^N\mathcal{U}_n$, we can just find the quasi-probability decomposition for noise process $\mathcal{E}_n$ associated with each gate 
\begin{equation}
    \mathcal{E}_n^{-1} = \sum_{k_n}q_{k_n}\mathcal{B}_{k_n}= Q_n\sum_{k_n}\text{sgn}(q_{k_n})\frac{|q_{k_n}|}{Q_n}\mathcal{B}_{k_n},
\end{equation}
in which $Q_n = \sum_k|q_{n_k}|$.

Then we can obtain the error mitigated value in a similar way
\begin{equation}
    \begin{aligned}
        \langle O \rangle_{\text{PEC}} =&\text{Tr}\left[\prod_{n}^{N}\left(\sum_{k_n}q_{k_n}\mathcal{B}_{k_n}\circ\mathcal{E}_n\circ\mathcal{U}_n\right)(\rho)\right]\\
        =&Q_{\mathcal{E}}\sum_{\vec{k}} \text{sgn}(q_{\vec{k}})\frac{|q_{\vec{k}}|}{Q_{\mathcal{E}}}\text{Tr}\left[\prod_{n=1}^{N}\left(\mathcal{B}_{k_n}\circ\mathcal{E}_n\circ\mathcal{U}\right)(\rho)\right],  
    \end{aligned}
\end{equation}
with $\vec{k} = (k_1, k_2,...,k_N)$ for $q_{\vec{k}}=\prod_{n=1}^{N}q_{k_n}$ and $Q_{\mathcal{E}} = \prod_{n=1}^NQ_n$.

\textbf{Sampling overhead.} Note that the total resource overhead for PEC involves both the characterization cost of the noise channel, and the sampling overhead to estimate the linear combinations in Eq~\ref{quasi-de}. Considering  that the gate noise characterization is usually performed during the device calibration stage~\cite{Cai2021Apractical}, we here focus on the sampling overhead for the implementation of PEC.

As for the factor $Q_{\mathcal{E}}$ multiplied with the averaged expectation value, the variance is increased by $Q_{\mathcal{E}}^2 = (\prod_{n=1}^NQ_n)^2$. Therefore, it needs $Q_{\mathcal{E}}^2$ times more samples to achieve the same measurement accuracy as the unmitigated case. The square of the multiplier $Q_{\mathcal{E}}^2$ can be regarded as the sampling overhead $\Upsilon_{\text{PEC}}$ for PEC, which grows exponentially with the overall gate error $\varepsilon N$ if we assume $Q_n = 1+\mathcal{O}(\varepsilon)$ for stochastic noise process~\cite{Endo2018Practical}. This sets limitations on the efficiency of PEC schemes to $\varepsilon N = \mathcal{O}(1)$. 
Moreover, the theoretical analysis of the sample complexity for implementing PEC has been explored~\cite{Takagi2021optimal, takagi2022fundamental} and it has been shown the optimality of PEC among all strategies in mitigating a certain type of noise~\cite{takagi2022fundamental}. 

\textbf{Approaches to extract error parameters.}
To implement the PEC in practice, we need to accurately characterize the noise model $\mathcal{E}$ and obtain the analytical decomposition of the inverse noise map as Eq.~\ref{eqn-em-ep} shown. 

{\it{Gate set tomography.}} Endo {\it{et al.}}~\cite{Endo2018Practical} has introduced gate set tomography (GST)~\cite{Merkel2013selfconsistent, Daniel2015introduction} to PEC, which is free from the SPAM errors. Although GST can reconstruct arbitrary noise process, the number of samples required for its reconstruction grows exponentially with the size of the system $N_q$ and is quite costly. If only specific types of noise are considered, there are some more efficient characterization protocols~\cite{Harper2020efficient, Steven2021averaged, Flammia2020Efficient}. 
         
{\it{Cycle error reconstruction for Pauli noise.}}
Cycle error reconstruction (CER) is an efficient approach to identify the Pauli noise in circuits~\cite{Wallman2016Noisetailoring,Hashim2021Randomizedcompiling}. It utilizes Cycle Benchmarking~\cite{Erhard2019Characterizing} to characterize the Pauli noise channel and extract the Pauli error rates after post-processing. The implementation of PEC with CER has been shown on a 4-qubit superconducting processor~\cite{Samuele2022Efficiently}. Under the Pauli noise assumption, the quasi-probability decomposition can be further simplified using sparse Pauli-Lindblad models~\cite{Ewout2022Probabilistic}. The sparse Pauli noise model considers Pauli operators with only one or two non-trivial terms, but is sufficient to capture the correlated errors. Moreover, the number of noise parameters in the sparse model scales polynomially with the system size, so it remains efficient in large-scale quantum systems. This approach has been experimentally demonstrated to learn the noise model on a superconducting quantum processor of up to 20 qubits~\cite{Ewout2022Probabilistic}.
    
\subsubsection{PEC with continuous time evolutions}
The standard PEC scheme we discussed above is based on the discrete gate-based quantum circuits with the gate-independent Markovian noise. To overcome this limitation, the stochastic QEM method~\cite{Sun2021Mitigating} is proposed to extend the standard PEC to more general scenarios that may have strong gate-dependence and complicated nonlocal effects, and general computing models such as analog quantum simulators. 

Specifically, the time evolution of a noisy quantum system can be described by the Lindblad master equation as 
\begin{equation}
    \begin{aligned}
       \frac{\partial \rho(t)}{\partial t} = -i[H^{\text{noisy}}(t), \rho(t)]+\lambda\mathcal{L}[\rho(t)], \\
       \mathcal{L}(\rho)=\sum_i(2L_i\rho L_i^\dagger-L_i^\dagger L _i\rho - \rho L_i^\dagger L_i) 
    \end{aligned}
\end{equation}
where $H^{\text{noisy}}=H + \delta H_0$  ($H$ is the ideal evolution Hamiltonian, $H_0$ denotes the coherent errors), and $\mathcal{L}(\rho)$ is the noise Lindblad operator corresponding to the decoherent coupling with the environment with noise strength $\lambda$. We denote the ideal and noisy process by $\mathcal{E}_I$ and $\mathcal{E}_{N}$ respectively. Given a small time step $\delta t$ with a time interval $\delta$, both the ideal and noisy case of the evolution can be written as
\begin{equation}
    \rho_i(t+\delta)=\mathcal{E}_i(t)\rho_i(t),
\end{equation}
with $i=I,N$. Similar to the idea of PEC in Eq.~\ref{eqn-em-ep}, we can find a recovery process $\mathcal{E}_Q$ satisfying
\begin{equation}
    \begin{aligned}
        \mathcal{E}_I=\mathcal{E}_Q\mathcal{E}_N + \mathcal{O}(\delta t^2)\approx\mathcal{E}_Q\mathcal{E}_N, \\
        \mathcal{E}_Q=\sum_kq_k\mathcal{B}_k=Q\sum_k\text{sgn}(q_i)\frac{|q_i|}{Q},
    \end{aligned}
\end{equation}
with $Q = \sum_k|q_k|$. Given the full evolution time $T$, we can apply the Monte Carlo sampling to realize the continuous recovery $\mathcal{E}_Q$ with a time interval $\delta$ at the mitigation cost of $\Upsilon_{\text{SEM}}=Q^{\frac{T}{\delta t}}$~\cite{Sun2021Mitigating,Endo2021Hybrid}.

\subsection{Data-driven approaches}
We now introduce the data-driven approaches for error mitigation, including zero noise extrapolation (ZNE) and error mitigation with learning process such as Clifford data regression (CDR). These methods utilize different types of data from circuits with various error rates (\textit{e.g.}, ZNE), or from near-Clifford circuits under similar circuit structure (\textit{e.g.}, CDR), and they can be naturally combined with each other to better utilize data resources~\cite{Lowe2021Unified,Daniel2021Unifying}.

\subsubsection{Zero-Noise Extrapolation}
ZNE is a typical error mitigation scheme using  classical post-processing. It uses data collected at different error rates to fit the function of expectation values with respect to the error rates, and then extrapolate to the zero noise limit~\cite{Li2017Efficient, Endo2018Practical, Temme2017Error, Cai2021multi, Lowe2021Unified, Giurgica2020Digital,Cao2022algorithmic}.

\textbf{Fitting methods of ZNE.} The chosen fitting method is essential to the performance of ZNE. A basic fitting method is the Richardson extrapolation, which utilizes the polynomial relationship between noisy and ideal expectation values yielded by Tailor expansion when the error rate is low. This estimation is slightly coarse. More refined models, such as exponential and poly-exponential fitting functions, can be introduced in the extrapolation method to improve  the actual mitigation performance in specific cases. We discuss these fitting methods in detail.

{\it{Richardson extrapolation}}.
Richardson extrapolation was the first introduced extrapolation method~\cite{Li2017Efficient, Temme2017Error} and later experimentally demonstrated using superconducting qubits~\cite{Kandala2019Errormitigation}.

Suppose the quantum circuit outputs a noisy $N_q$-qubit quantum state $\rho_\varepsilon$, where $\varepsilon$ denotes the noise parameter which characterizes the error rate in the circuit. Then the expectation value of the target observable $O$ under error rate $\varepsilon$ can be regarded as a function towards $\varepsilon$: 
$\langle O \rangle(\varepsilon) = \text{Tr}(\rho_\varepsilon O)$. In order to evaluate the error-free expectation value $\langle O \rangle(0)$, we choose a set of error rates $\{\lambda_0\varepsilon, ..., \lambda_n\varepsilon\}$ with $n + 1$ different coefficients $\{\lambda_i\}_{i=0,...,n}$($\lambda_0 = 1$) , and then run circuits under these error rates to obtain the set of $n+1$ expectation values $\{\langle O \rangle(\lambda_i\varepsilon)\}_i$. It is known to be difficult to reduce the error rate in the circuit, we thus usually boost it with amplified coefficients $1=\lambda_0 < \lambda_1 < ... < \lambda_n$ using some noise scaling methods. Linear spacing of the amplified coefficients (\textit{e.g.} $\{\lambda_i = i+1\}_{i=0,...,n}$) is commonly used~\cite{Temme2017Error, Cai2021Apractical} and some specific spacing methods have been explored for better performance~
\cite{Michael2022Optimization}.

The expectation value for $\lambda = 0$ (the error-free case) can be estimated using
\begin{equation}\label{eqn-em-re}
    \langle O\rangle_{\text{Rid}} = \sum_{i=0}^{n}\gamma_i\langle O \rangle(\lambda_i\varepsilon),
\end{equation}
where the fitting coefficients $\{\gamma_i\}$ are chosen to satisfy both $\sum_{i=0}^n\gamma_i = 1$ and $\sum_{i=0}^n\gamma_i\lambda_i^j = 0 \ \text{for} \ j = 1,...,n$, and the solution gives
\begin{equation}\label{eqn-em-re3}
    \gamma_i = \prod_{i\neq j}\frac{\lambda_j}{\lambda_j - \lambda_i}.
\end{equation}

If the error rate $\varepsilon$ is sufficiently low, we can expand the function $\langle O \rangle(\varepsilon)$ according to the Taylor expansion
\begin{equation}\label{eqn-em-re2}
    \langle O \rangle(\varepsilon) = \langle O \rangle(0)+\sum_{j=1}^{n} a_j\varepsilon^j + \mathcal{O}(\varepsilon^{n+1})
\end{equation}
with some constant coefficients $a_j$. Then the Richardson mitigated value $\langle O\rangle_{\text{Rid}}$ can be rewritten using Eq.~\ref{eqn-em-re} and ~\ref{eqn-em-re2}
\begin{equation*}
    \begin{aligned}
        \langle O\rangle_{\text{Rid}} &= \sum_{i=0}^n \gamma_i\left(\langle O\rangle (0)+\sum_{j=1}^{n} a_j\lambda^j\varepsilon^j + \mathcal{O}(\varepsilon^{n+1})\right) \\
        &= \langle O\rangle (0) + \sum_{j=1}^n\left(\sum_{i=0}^n\gamma_i\lambda_i^j\right)\varepsilon^j + \mathcal{O}(\varepsilon^{n+1}) \\
        &= \langle O\rangle (0) + \mathcal{O}(\varepsilon^{n+1}),
    \end{aligned}
\end{equation*}
Here we use $\sum_{i=0}^n\gamma_i = 1$ and $\sum_{i=0}^n\gamma_i\lambda_i^j = 0 \ \text{for} \ j = 1,...,n$. We can see that the error rate is suppressed from the original $\varepsilon$ to the order $\mathcal{O}(\varepsilon^{n+1})$, under the weak noise assumption. However, the assumption of a valid Taylor expansion for low error rate may be inaccurate in the large circuit limit, and the lack of specific information on the noise channel sets limitation to the efficacy of Richardson extrapolation.

As for the sampling overhead of Richardson extrapolation, we need to take into account the variance increase introduced in Eq.~\ref{eqn-em-re}, which reads~\cite{Endo2021Hybrid}
\begin{equation}
    \text{Var}(\langle O\rangle_{\text{Rid}}) = \sum_{i=0}^n\gamma_i^2\text{Var}(\langle O\rangle_{\lambda_i\varepsilon}).
\end{equation}
The variance of estimating $\langle O\rangle_{\text{Rid}}$ is about $\sum_{i=0}^n\gamma_i^2$ larger than evaluating $\langle O\rangle$ without error mitigation, so it requires around $\Upsilon_{\text{Rid}} = \sum_{i=0}^n\gamma_i^2$ more samples as the sampling overhead of Richardson extrapolation.

Besides, an optimized protocol~\cite{Michael2022Optimization} for the implementation of Richardson extrapolation has been proposed to further explore the relevant parameters of Richardson extrapolation.

{\it{Exponential Extrapolation}}.  
The error rate $\varepsilon$ used in Richardson extrapolation quantifies the local noise strength, but in the context of NISQ error mitigation, it is more natural to consider the mean circuit error count~\cite{Cai2021multi} $\mu$, where $\mu \approx N\varepsilon$ and $N$ denotes the number of gates in the circuit. In the large circuit limit when $N \gg 1$, the number of errors happened in the circuit (denoted by $k$) follows the Poisson distribution with probability~\cite{Cai2021multi} 
\begin{equation}
    p_k = e^{-\mu}\frac{\mu^k}{k!}.
\end{equation}
Denoting the expectation value with $k$ errors occurred as $\langle O_k \rangle$, the expectation value with mean circuit error count $\mu$ is
\begin{equation}\label{em-ee}
    \langle O_\mu \rangle = \sum_{k=0}^\infty p_k\langle O_k\rangle = e^{-\mu}\sum_{k=0}^\infty \frac{\mu^k}{k!}\langle O_k\rangle,
\end{equation}
where the factor $e^{-\mu}$ implies an exponential decay with $\mu$ of the expectation value.
By simply assuming an exponential function 
\begin{equation}\label{em-ee2}
    \langle O \rangle(\mu) = e^{-f\mu}\langle O \rangle(0)
\end{equation}
with a parameter $f$ which denotes the observable decay rate, we can then obtain an two-points exponential extrapolation result with mean circuit error count $\mu$ and $\lambda\mu$ ($\lambda > 1$)
\begin{equation}
    \langle O \rangle_{\text{exp}} = {\left(\frac{\langle O \rangle^\lambda (\mu)}{\langle O \rangle (\lambda\mu)}\right)}^\frac{1}{\lambda -1},
\end{equation}

Endo {\it{et al.}}~\cite{Endo2018Practical} first considered the exponential decay curve for error extrapolation, where the advantage of exponential extrapolation over Richardson extrapolation has been numerically demonstrated~\cite{Endo2018Practical, Giurgica2020Digital}. 
And Cai~\cite{Cai2021multi} provided a general multi-exponential extrapolation framework 
\begin{equation}
    \langle O \rangle(\mu) = \sum_{i=1} A_i(e^{-f_i})^\mu, 
    \text{and} \ \sum_{i=1} A_i = \langle O \rangle(0),
\end{equation}
which achieves a much lower estimation bias in numerical simulation.

{\it{Poly-exponential extrapolation}}. The poly-exponential extrapolation~\cite{Giurgica2020Digital, Cristina2022Volumetric} assumes a more general model that the exponential decay with  the mean circuit error count $\mu$ has a polynomial expansion which can be fitted to the function
\begin{equation}
    \langle O \rangle(\mu) = e^{\sum_{i=0}f_i \mu^i}\langle O \rangle(0).
\end{equation}
Then the single-exponential extrapolation modeled by Eq.~\ref{em-ee2} is a particular case of the more general poly-exponential extrapolation.

\textbf{Noise scaling methods}.
The implementation of ZNE, especially Richardson extrapolation, depends non-trivially on the ability to boost the error rate $\varepsilon$ in the circuit by different amplified coefficients$\{\lambda_i\}_i$~\cite{Michael2022Optimization}. Next we will introduce a variety of methods to amplify the error rate in a controlled manner. 

{\it{Identity insertion}}~\cite{Endo2018Practical, Dumitrescu2018Cloud} is a hardware-agnostic approach which replaces a particular gate or layer $\mathcal{C}$ by $\mathcal{C}\{\mathcal{C}^\dagger \mathcal{C}\}^n$ for non-negative integer $n$. The circuit after insertion is logically equivalent to the original circuit but the error rate is amplified by a factor $2n+1$ as the circuit depth increases.  There are many variant identity insertion methods proposed~\cite{He2020Zeronoise,Pascuzzi2022Computationally} for ZNE, which explore the trade-off between the inserted gate number and the required measurement to achieve the same (or higher) accuracy.
And more general approaches using unitary folding can obtain an arbitrary real scaling factor~\cite{Giurgica2020Digital}. The limitation of identify insertion method is that the amplified error rate may arbitrarily deviate from the desired one~\cite{Youngseok2021Scalable}. Given a self-adjoint noisy gate $\tilde{\mathcal{U}} = \mathcal{E}\circ\mathcal{U}$, after identity insertion of $\tilde{\mathcal{U}}^\dagger \tilde{\mathcal{U}} = \tilde{\mathcal{U}}\tilde{\mathcal{U}}$ we have $\tilde{\mathcal{U}}\tilde{\mathcal{U}}^\dagger \tilde{\mathcal{U}} = \mathcal{E}(\mathcal{U}\mathcal{E}\mathcal{U})\mathcal{E}\circ\mathcal{U} = \tilde{\mathcal{E}}\circ\mathcal{U}$. The actually amplified noise $\tilde{\mathcal{E}} = \mathcal{E}(\mathcal{U}\mathcal{E}\mathcal{U})\mathcal{E}$ is gate-dependent for general noise channel $\mathcal{E}$, leading to an unpredictable amplified error rate in the circuit.

 {\it{Stretching the control pulses in time}}~\cite{Temme2017Error,Kandala2019Errormitigation,Youngseok2021Scalable} is also used to implement circuits under different error rate. Given an open quantum system described by the time-dependent multi-qubit Hamiltonian $H(t)$, the time evolution is
\begin{equation}
    \frac{\partial}{\partial t}\rho_\varepsilon(t) = -i[H(t), \rho_\varepsilon(t)] + \varepsilon \mathcal{L}(\rho_\varepsilon(t)),
\end{equation}
where $\varepsilon$ refers to the error rate in the system, and the noise term $\mathcal{L}$ is a Lindblad operator. Then we rescale the Hamiltonian $H(t)$ to $\frac{1}{\lambda} H(\frac{t}{\lambda})$ with a stretched factor $\lambda > 1$ and obtain a new state $\rho_\varepsilon'(t)$.
Under the assumption that the generator $\mathcal{L}$ is invariant after time rescaling, and also independent from the way the Hamiltonian $H(t)$ is encoded, we can increase the evolution time by a factor $\lambda$. Then the error rate is boosted from $\varepsilon$ to $\lambda \varepsilon$ according to $\rho_{\varepsilon}'(ct) = \rho_{\lambda\varepsilon}(t)$. This method has been experimentally demonstrated on superconducting processors~\cite{Kandala2019Errormitigation, Youngseok2021Scalable} using up to 26 qubits~\cite{Youngseok2021Scalable}.

Note that two methods above do not require the exact form of the noise channel, and there are some approaches that utilize the specific structure of certain noise models. {\it{Parameter noise scaling}}~\cite{Giurgica2020Digital} is designed for the pulse error caused by imperfect control or finite precision of the physical parameters in the parametric quantum gates. Considering a quantum gate $G(\vec{\theta}) = \text{exp}(-i\sum_{j=1}^k\theta_jH_j)$ which is parameterized by $k$ classical control parameters $\vec{\theta}=(\theta_1, \theta_2,...,\theta_k)$ with Hamiltonian operators $H_1, H_2,...,H_k$, the actually implemented gate suffered from pulse error is modeled by $G(\vec{\theta^{'}})$ with
\begin{equation}
    \theta^{'}_j = \theta_j + \hat{\varepsilon}_j, \ \text{for} \ j=1,2,...,k,
\end{equation}
where $\hat{\varepsilon}$ is a random variable following Gaussian distribution with zero mean and variance $\sigma_j^2$ which denotes the effect of the stochastic calibration noise. Given the value of variance $\sigma_j^2$, we can rescale the noise by a factor $\lambda >1$ using 
\begin{equation}
    \theta^{re}_j = \theta_j + \hat{\delta}_j,
\end{equation}
here the $\hat{\delta}_j$ is sampled from a zero-mean Gaussian distribution with variance $(\lambda -1)\sigma_j^2$. Therefore, the effective parameter after rescaling is
\begin{equation}
    \theta^{'}_j = \theta_j + \sqrt{\lambda}\hat{\varepsilon}_j, \ \text{for} \ j=1,2,...,k,
\end{equation}
which achieves a noise rescaling by a factor $\lambda$ without knowledge of the Hamiltonian operators $H_j$.

{\it{Pauli twirling techniques}}~\cite{Geller2013Efficient, Cai2019Constructing}
focus on the cases that error rates of single-qubit gates are much lower than those of two-qubit gates and measurement. By applying randomly chosen Pauli gates before and after the Clifford two-qubit gates, arbitrary noise channel can be converted into an effective stochastic Pauli channel~\cite{Wallman2016Noisetailoring,Hashim2021Randomizedcompiling}. Then we can use additional Pauli gates to tune the practical error rate to any target value~\cite{Li2017Efficient}, but it requires the full knowledge of the exact Pauli noise channel. We can also combine efficient Pauli error reconstruction methods such as Cycle Error Reconstruction~\cite{Hashim2021Randomizedcompiling, Erhard2019Characterizing} to amplify the error rate for ZNE~\cite{Samuele2022Efficiently}.

{\it{Gate Trotterization}}~\cite{Kevin2022Reducing} is a local noise scaling technique acting at the level of individual gates. We can replace each gate $U$ of the circuit with the product of $\lambda$ equal gates using the gate Trotterization technique
\begin{equation}
    U\to(U^{\frac{1}{\lambda}})^\lambda, \ \lambda=0,1,2...\ .
\end{equation}
However, the way the $U^{\frac{1}{\lambda}}$ is compiled by the hardware depends on $\lambda$, so the circuit depth may not increase by the expected factor.

{\it{Leveraging other error mitigation schemes}} as a noise scaling tool has also been explored for ZNE~\cite{Cai2021multi, Andrea2021Extending}. Different from above tuning approaches which produces boosted error rate, we can also utilize other error mitigation methods such as PEC to reduce the error rate~\cite{Cai2021multi}.

\subsubsection{Multi-dimensional variant of ZNE}
Least square fitting~\cite{Otten2019Recovering} may be regarded as a multi-dimensional version but not reliant on the Richardson extrapolation. It utilizes a hypersurface fit where one axis refers to the measurement results and the other axes describe the effect of difference noise parameters. The least square fitting method has been experimentally demonstrated on Rigetti's 8-qubit quantum processor~\cite{Otten2019Recovering}.

Considering only one noise parameter $\varepsilon$ with $k$ different noise level $\varepsilon_1, \varepsilon_2, ..., \varepsilon_k$, the polynomial model up to $n$ order in Eq.~\ref{eqn-em-re2} can be generalized to a linear equation
\begin{equation}\label{eqn-em-lsf}
    \left (\begin{matrix}
        1 & \varepsilon_1 & \cdots  & \varepsilon_1^n \\
        1 & \varepsilon_2 & \cdots  & \varepsilon_2^n \\
        \vdots & \vdots   & \ddots  & \vdots          \\
        1 & \varepsilon_k & \cdots  & \varepsilon_k^n \\
        \end{matrix}\right)
    \left(\begin{matrix}
        \langle O \rangle(0) \\
        \alpha_1    \\
        \vdots      \\
        \alpha_n    
    \end{matrix}
    \right)=
    \left(\begin{matrix}
        \langle O \rangle(\varepsilon_1) \\
        \langle O \rangle(\varepsilon_2) \\
        \vdots      \\
        \langle O \rangle(\varepsilon_k)  
    \end{matrix}\right).
\end{equation} 
We can solve this linear equation using standard least-squares methods and obtain the expansion parameters $\left(\begin{array}{cccc}
    \langle O \rangle(0), &
    \alpha_1,  &
    \cdots,    &
    \alpha_n
    \end{array}
    \right)^T$,  
where $\langle O \rangle(0)$ is the error-free expectation value of the observable. 

Then  we  can  extend  the  consideration  of  one  noise  parameter  to  the  multiple  noise  parameters.  Take  the  spontaneous  emission  rate  $\gamma$  and  the  pure  dephasing  rate  $\zeta$  for  example,  suppose  we  have  k  different  noise  level  for  both  $\gamma$  and  $\zeta$  which  are  denoted  by  $\{\gamma_1,  \gamma_2,...,\gamma_k\}$  and  $\{\zeta_1,  \zeta_2,...,\zeta_k\}$,  respectively.
The  linear  equation  in  Eq.~\ref{eqn-em-lsf}  turns  to
\begin{equation}\label{eqn-em-lsf2}
        \left  (\begin{matrix}
                1  &  \gamma_1    &\zeta_1  &\gamma_1\zeta_1  &(\gamma_1)^2    &  (\zeta_1)^2  \\
                1  &  \gamma_2    &\zeta_2  &\gamma_2\zeta_2  &(\gamma_2)^2    &  (\zeta_2)^2  \\
                \vdots  &  \vdots      &  \ddots    &  \vdots                    \\
                1  &  \gamma_k    &\zeta_k  &\gamma_k\zeta_k  &(\gamma_k)^2    &  (\zeta_k)^2  \\
                \end{matrix}\right)
        \left(\begin{matrix}
                \langle  O  \rangle(0)  \\
                \alpha_1        \\
                \vdots            \\
                \alpha_5        
        \end{matrix}
        \right)=
        \left(\begin{matrix}
                \langle  O  \rangle(\gamma_1,  \zeta_1)  \\
                \langle  O  \rangle(\gamma_2,  \zeta_2)  \\
                \vdots            \\
                \langle  O  \rangle(\gamma_k,  \zeta_k)    
        \end{matrix}\right).
\end{equation}  
Here  we  truncate  the  series  to  the  second  order  of  $\gamma$  and  $\zeta$.

\subsubsection{Error mitigation with learning process}
Instead of using the exact knowledge of the noise channel (such as PEC) or the expectation values at different error rates (like ZNE), the learning-based error mitigation optimizes the estimator of the observable's expectation value using automatical learning process~\cite{Czarnik2021errormitigation, Strikis2021Learningbased, Alexander2020adeeplearning, Zhukov2022quantumerror, Urbanek2021noiseestimation, Vovrosh2021Simple, Ashley2021fermionic, Zhang2021Variational}. Specifically, the learning-based error mitigation methods estimate the observable's expectation value of the targeted circuit using an ansatz that typically describes the relationship between the noisy and noiseless case of the expectation value. The training set fed into the ansatz comes from classically simulable quantum circuits, or measurement results from relevant quantum circuits. 

\textbf{Clifford data regression}~\cite{Czarnik2021errormitigation} generates training data set $\{O_i^{\text{noisy}}, O_i^{\text{ideal}}\}_i$ composed of the noisy expectation value $O_i^{\text{noisy}}$ from real quantum devices and the noiseless expectation value $O_i^{\text{ideal}}$ simulated on the classical computers. To enable efficient classical simulation, the quantum circuits for generating training set are replaced with near-Clifford circuits generated by Markov Chain Monte Carlo sampling method. We can fit the training data set to a linear ansatz
\begin{equation}
    O^{\text{ideal}} = f(O^{\text{noisy}}, \vec{\theta})=\theta_1O^{\text{noisy}}+\theta_2,
\end{equation}
where the parameters $\vec{\theta}=(\theta_1, \theta_2)$ are optimized by minimizing the cost function
\begin{equation}
    C = \sum_{i}\left(O_i^{\text{ideal}}-(\theta_1O_i^{\text{noisy}}+\theta_2)\right)^2.
\end{equation}

CDR has been experimentally demonstrated on a 16-qubit IBMQ quantum computer and can achieve an order-of-magnitude improvement for a ground state energy problem~\cite{Czarnik2021errormitigation}, and it also been shown to has the potential to outperform other state-of-the-art approaches~\cite{Daniel2021Unifying}. CDR can be further improved by more efficient training set construction methods and by applying symmetries in certain quantum systems~\cite{Piotr2022Improving}. 

\textbf{Learning-based probabilistic error cancellation}~\cite{Strikis2021Learningbased} does not depend on the full tomography of the noise channel as the original PEC does. For a $L$-layer quantum circuit consisting of single-qubit unitary gates $\mathbf R = (R_1,R_2 ...,R_{N_q(L+1)})$ and multi-qubits Clifford gate layers $\mathbf U=(U_1,U_2,...,U_L)$, the learning-based PEC applies the Pauli gates $\mathbf P=(P_1,P_2,...,P_{2N_q(L+1)})$ as the error mitigating gates before and after each single-qubit unitary gates. It can then establish a circuit configuration where the multi-qubit Clifford gates are fixed, the other gates ($\mathbf{R}$ and $\mathbf P$) are served as variables, and $\mathbf P=\mathbf I$ indicates that all error mitigating gates are identity gates.
 Here we denote the noisy and error-free expectation value by $f(\mathbf R, \mathbf P)$ and $f^{\text{ef}}(\mathbf R, \mathbf P)$. According to the PEC, the error-mitigated expectation value $f^{\text{em}}(\mathbf R, \mathbf I)$ is 
\begin{equation}
    f^{\text{em}}(\mathbf R, \mathbf I) = \sum_{\mathbf R}q(\mathbf P)f(\mathbf R, \mathbf P),
\end{equation}
where $q(\mathbf P)$ is the the combination coefficients, \textit{i.e.}, quasiprobabilities. By minimizing the cost function,
\begin{equation}
    C(\mathbf R) = \frac{1}{|\mathbb{S}|}\sum_{\mathbf R \in \mathbb{S}}|f^{\text{em}}(\mathbf R,\mathbf I)-f^{\text{ef}}(\mathbf R, \mathbf I)|^2,
\end{equation}
where a subset of single-qubit Clifford gates $\mathbb{S}$ is chosen as the training set so that the error-free expectation value $f^{\text{ef}}(\mathbf R, \mathbf I)$ can be efficiently simulated, we can obtain a optimal distribution $q(\mathbf P)$ for the training set $\mathbb{S}$ which can also be applied for the non-Clifford single-qubit gates. This scheme requires that all the single-qubit gates in the configuration are error-free and do not rely on the exact error model of the multi-qubit Clifford gates. For practical reasons, since the spaces of $\mathbf R$ and $\mathbf P$ grow exponentially with the system size $N_q$, we need to truncate the spaces of the training set and error mitigating gates, or resort to variational optimization approaches~\cite{Strikis2021Learningbased}. 

\textbf{Deep learning method}~\cite{Alexander2020adeeplearning,Kim2020Artificial,Zhukov2022quantumerror} trains a deep neural network to model a noise channel utilizing the "black box" nature of the neural network. The training of the neural network requires the measurement results with both noise and ideal cases. However, since classical simulations may not be possible for large-scale non-Clifford quantum circuits, the training set can only consist of measurement results of specific quantum circuits whose ideal measurement outcomes are known~\cite{Alexander2020adeeplearning, Kim2020Artificial}.

\subsection{Measurement error mitigation}

Measurement (or readout) error mitigation (MEM) schemes are designed to improve the accuracy of the measurement results obtained from noisy quantum devices.

\subsubsection{MEM under classical noise assumption}
Under the assumption of a classical noise model, MEM is usually applied via classical post-processing, where measurement error is modeled by a stochastic and invertible response matrix $\Lambda$~\cite{Chen2019Detector, Bravyi2021Mitigate, Maciejewski2021modelingmitigation}. 

\textbf{Classical noise models}.
The ideal quantum measurement in the computational basis can be written in terms of positive operator valued measurement (POVM). For a $n$-qubit system $\rho$, suppose the POVM operator is $\Pi_{\boldsymbol x} = |\boldsymbol x\rangle \langle \boldsymbol x|$ satisfying $\Pi_{\boldsymbol x} \geq 0$ for $\forall {\boldsymbol x}$ and $\sum_{\boldsymbol x}\Pi_{\boldsymbol x}=\mathbb{I}$, where $\boldsymbol x \in \mathbb{Z}_2^{N_q} $ refers to the POVM outcome. The probability distribution of the measurement outcome is represented by a vector $\vec{p}$, where the $\boldsymbol x$ term of $\vec{p}$ is the probability of obtaining the outcome $\boldsymbol x$, given by $p(\boldsymbol x|\rho) = \text{Tr}(\rho\Pi_{\boldsymbol x})$ according to the Born's rule. 

If the POVM elements have no non-trivial off-diagonal terms, we can treat the measurement noise channel as a classical noise channel, where the transformation between ideal and noisy measurement probability distributions can be written using a response matrix $\Lambda$:  
\begin{equation}
    \vec{p}_{ideal} = \Lambda^{-1}\vec{p}_{noisy}.
\end{equation}
Here we use $\vec{p}_{ideal}$ and $\vec{p}_{noisy}$ to represent the ideal and noisy probability distributions. And the element $\Lambda_{\boldsymbol x, \boldsymbol y}$ of the response matrix $\Lambda$ is defined as 
\begin{equation}
    \Lambda_{\boldsymbol x, \boldsymbol y} = \langle\boldsymbol x|\rho|\boldsymbol y\rangle, \quad\boldsymbol x, \boldsymbol y\in \mathbb{Z}_2^{N_q}, 
\end{equation}
which can be estimated directly by preparing the computational state and measuring it in the computational bases, or by using some tomographic means, such as quantum detector tomography (QDT)~\cite{Maciejewski2020mitigationofreadout, Chen2019Detector, Lundeen2009Tomography}. The main idea of QDT is to estimate an unknown set of fixed noisy POVM operators $\{\hat{\Pi}_{\boldsymbol x}\}_{\boldsymbol x}$ with a set of well-known states $\{\rho_i\}_i$. Once the noisy POVM operators are reconstructed, we can extract the response matrix elements using
\begin{equation}
    \hat{\Pi}_{\boldsymbol x} = \sum_{\boldsymbol y}\Lambda_{\boldsymbol x, \boldsymbol y}\Pi_{\boldsymbol y}, \quad \forall \boldsymbol x, 
\end{equation}
where $\Pi_{\boldsymbol y}$ denotes the error-free POVM operator. 

In practice, however, for non-classical noise, the estimated measurement probability distribution obtained from $\vec{p}_{est}=\Lambda^{-1}\vec{p}_{noisy}$ may be unphysical due to the negative terms of the vector $\vec{p}_{est}$. 
Moreover, the statistical uncertainties in the response matirx can be amplified through simply inversion, similar to the challenges faced in high-energy physics. Unfolding methods are thus introduced to readout error mitigation~\cite{Nachman2020unfolding,Ouadah2021Dealing}, which show robustness to some failure cases of matrix inversion and least-squares.

\textbf{Simplified classical models.}
Dealing with the full response matrix of size $2^{N_q} \times 2^{N_q}$ is 
not scalable beyond hundreds of qubits. Therefore, in order to avoid the exponential consumption of measurements, many scalable approaches~\cite{Bravyi2021Mitigate, Nation2021Scalable, Bo2022AnEfficient, Funcke2022Measurement} have been proposed to simplify the response matrix model.

{\it{Tensor Product Noise (TPN) model}}.
The simplest model assumes that the noise acts independently on each qubit and is called the Tensor Product Noise (TPN) model~\cite{Bravyi2021Mitigate,Geller2020Rigorous, Michael2020Efficient}. After such a simplification, the response matrix can be described in terms of the tensor product of $N_q$ single-qubit response matrix as 
\begin{equation}
    \Lambda^{\text{TPN}} = \prod_{i=1}^{N_q} \left (\begin{matrix}
        1-p_i& q_i   \\
        p_i&  1-q_i  \\
        \end{matrix}\right),
\end{equation}
where $p_i$ and $q_i$ denote the single-qubit readout error probability of $1\to0$ and $0\to1$, respectively. We can see that the number of measurements required to construct the TPN response matrix is reduced from $\mathcal{O}(2^{N_q})$ to $\mathcal{O}(N_q)$. However, in the TPN model, the multi-qubit readout error correlations are neglected. 

{\it{Continuous Time Markov Processes (CPTP) noise model}}.
Bravyi {\it{et al.}}~\cite{Bravyi2021Mitigate} have extended the TPN model to the CTMP noise model, which takes the correlated errors into account. It defines a matrix exponential $\Lambda = e^{G}$ to describe the response matrix, and the generator $G = \sum_{i=1}^{2{N_q^2}}r_iG_i$ models the readout error. Here $r_i\geq 0$ refers to error rate and $G_i$ represents the single-qubit or two-qubit readout error. For example, the generator describing falsely measuring $|00\rangle$ to $|11\rangle$ on two qubits is $|11\rangle\langle00|-|00\rangle\langle00|$. The MEM with the CPTP model has been experimentally demonstrated on the superconducting device~\cite{Bravyi2021Mitigate}.     

{\it{Subspace Reduction}}.
When the measured probability distribution only contains a few principal bit strings with high probability, we can mitigate readout errors in a renormalized subspace defined by the observed probability distribution~\cite{Nation2021Scalable, Michael2020Efficient}. The subspace reduction of the full $2^{N_q}$ dimensional response matrix efficiently circumvents the original exponential overhead and avoids matrix inversion by using the matrix-free iterative methods~\cite{Nation2021Scalable}.

\textbf{Readout symmetrizing}.
Due to the observation that mismeasuring $|1\rangle$ to $|0\rangle$ is more frequent than $|0\rangle$ to $|1\rangle$, Several studies symmetrizing the readout with targeted Pauli $\hat{X}$ gates are proposed~\cite{Harrigan2021Quantum,Hicks2021Readoutrebalancing,Alistair2021Qubit,Berg2022Modelfree}.

{\it{Readout balancing}}.  
 We can exploit the readout asymmetry in the practical cases, which aims to rebalance the measurement outcomes by minimizing the expected number of qubits in the $|1\rangle$ state~\cite{Hicks2021Readoutrebalancing}.

{\it{Bit-flip averaging}}.
Bit-flip averaging~\cite{Alistair2021Qubit} reduces the calibration overhead by an exponential factor without making any specific assumptions about the classical noise model for MEM. It applies Pauli $\hat{X}$ gates on the randomly chosen qubits before measurements and offsets the effect of $\hat{X}$ gates using classical post-processing. This leads to a symmetrized effective response matrix $\hat{\Lambda}$ which contains only $\mathcal{O}(2^{N_q})$ free parameters compared to $\mathcal{O}(2^{2N_q})$ in the full matrix. Another independent work~\cite{Berg2022Modelfree} utilizes the same bit-flipping protocol. It transforms the bias between the ideal and noisy expectation value to a multiplicative factor, and eliminates it by deviding the noisy expectation by an estimated factor obtained in the calibration procedure.

\subsubsection{Quantum noise model for MEM}
Almost all of the readout mitigation approaches discussed above are applied under the classical noise model assumption. However, quantum measurements inevitably suffer from the quantum coherent noise~\cite{Chen2019Detector, Maciejewski2020mitigationofreadout, Tang2022Detecting}. With the presence of quantum noise, the noisy POVM operators have non-trivial off-diagonal values, so the classical assumption no longer holds. Fitting the difference between two specific measurement statistics to the Fourier series can effectively detect the presence of quantum noise~\cite{Tang2022Detecting}. Meanwhile, many approaches, such as $IZ$ dephasing~\cite{Chen2019Detector, Tang2022Detecting}, can be applied to eliminate the quantum noise, so that the effective measurement device has only classical noise. Then we can utilize those MEM methods mentioned above to deal with the classical noise.

\subsubsection{MEM combined with other techniques}
\textbf{MEM using QEC.}
On devices whose readout errors dominate over the entangling gate errors, one can combine readout error mitigation with quantum error correction to actively reduce readout errors on a shot-by-shot basis, called active readout error mitigation~\cite{Hicks2022Active}.

\textbf{Combination with readout compression}. The compression readout compresses the large-scale quantum state into one qubit and recovers the state amplitude populations from the one-qubit
measurement results~\cite{Chen2021Compression}. Thus, after appling the compression readout technique, only single-qubit measurements are performed, so this method is free of correlated measurement errors and easy to combine with other MEM methods.

\textbf{Neural network model for MEM.}
We can also utilize a trained artificial neural network to characterize an non-linear model~\cite{Kim2022quantumreadout} for measurement noise, which captures both classical and quantum noise. The neural network model for MEM is experimentally performed on the IBM's 5-qubit quantum device~\cite{Kim2022quantumreadout}.

\subsection{Purification-based QEM scheme}
Recently, a range of purification-based QEM schemes~\cite{Huggins2021Virtual, Koczor2021Exponential, Cotler2019Cooling, Cai2021Resource, Xiong2022Permutation, Hong2022Logical, Alireza2022Shadow, Huo2022Dual, Piotr2021Qubitefficient} have been proposed to extract the ideal state component from the noisy state, without any requirements on the priori knowledge of the quantum noise channel. These  schemes assume that the state of interest is usually a pure state, while the noise tends to corrupt it into a mixed state.

The two typical methods are Virtual Distillation (VD)~\cite{Huggins2021Virtual} and Error Suppression by Derangement (ESD)~\cite{Koczor2021Exponential}, which use multiple copies of the noisy state to reduce the state preparation error. Given an error-free state $|\psi_0\rangle\langle\psi_0|$, the  noisy output state $\rho$ can be expressed via the spectral decomposition 
\begin{equation}
    \rho = (1 - \lambda)|\psi\rangle\langle\psi| + \lambda\rho_{\text{err}},
\end{equation} 
where $\lambda \geq 0$ denotes the error rate. The pure state $|\psi\rangle\langle\psi|$ denotes the dominant eigenvector which may not match the ideal state $|\psi_0\rangle\langle\psi_0|$ due to the coherent error~\cite{Koczor2021Thedominant}. And the mixed state $\rho_{\text{err}}$ refers to the noisy component consisting of the weighted sum of other eigenvectors in the spectral decomposition. 
Using $M$ copies of noisy state, the $M$-degreed mitigated expectation value is
\begin{equation}
\begin{aligned}
    \langle\mathcal{O}\rangle_{\text{em}} = \frac{\text{Tr}(\rho^MO)}{\text{Tr}(\rho^M)} &= \frac{(1-\lambda)^M\langle\psi|O\psi\rangle + \lambda^M\text{Tr}(O\rho_{\text{err}}^M)}{(1-\lambda)^M + \lambda^M\text{Tr}(\rho_{\text{err}}^M)} \\
    &\approx   \langle\psi|O\psi\rangle + \mathcal{O}(\lambda^M)
\end{aligned}
\end{equation}
where $\mathcal{O}(\lambda^M)$ is the approximation error which decays exponentially with the number of copies. 
\begin{figure}[!htbp]
    \begin{center}
    \includegraphics[width=1\linewidth]{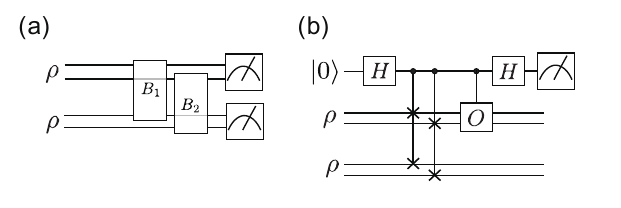}
    \end{center}
    \caption{\textbf{The illustration of two implementations of virtual distillation for 2 qubit states.} (a) We can apply a layer of diagonalization gates $B_i$ to compute Eq.~\ref{QEM-vd}, with a joint measurement on $M$(here $M$=2) copies. (b) We can also add an ancilla qubit and use controlled-$S^{M}$ gates to estimate the expected value $\text{Tr}(O\rho^M)$ using the probability of measuring the ancilla qubit in the $|0\rangle$ state.}
    \label{QEM-vd}
    \end{figure}

We usually don't need to prepare the exact $\rho^M$ state to compute the $\langle\mathcal{O}\rangle_{\text{em}}$. An illustration of the implementation of VD and ESD is given in Fig.~\ref{QEM-vd}. They both utilize the cyclic shift operator (also called derangement operator) $S^{(M)}$ on the system with size $MN_q$. Using the property of $S^{(M)}$
\begin{equation}
    S^{(M)}|\psi_1\rangle|\psi_2\rangle\dots|\psi_M\rangle = |\psi_M\rangle|\psi_1\rangle\dots|\psi_{M-1}\rangle, 
\end{equation}
we have
\begin{equation}\label{eqn-vd}
    \frac{\text{Tr}(O\rho^M)}{\text{Tr}(\rho^M)} = \frac{\text{Tr}(O^iS^{(M)}\rho^{\otimes M})}{\text{Tr}(S^{(M)}\rho^{\otimes M})},
\end{equation}
where $O^i$ refers to the observable $O$ acting on an arbitrary subsystem $i$.
The VD shown in Fig.~\ref{QEM-vd}(a) applies a layer of diagonalization gates $B_i$ to compute Eq.~\ref{QEM-vd}, with a joint measurement on $M$ copies. While the ESD given in Fig.~\ref{QEM-vd}(b) adds an ancilla qubit and uses controlled-$S^{M}$ gates to estimate the expected value $\text{Tr}(O\rho^M)$ (and for $\text{Tr}(\rho^M)$ setting $O =I$) by
\begin{equation}
   \text{Tr}(O\rho^M) = 2P_0 -1.
\end{equation}
Here we denote the probability of measuring the ancilla qubit in the $|0\rangle$ state by $P_0$.

The VD (or ESD) scheme is subject to some limitations. First, the coherent dismatch between the target pure state $|\psi_0\rangle\langle\psi_0|$ and the dominant eigenvector $|\psi\rangle\langle\psi|$ of the noisy state contributes to a noise floor regardless of the increasing of $M$. Second, the increased overhead of qubits and controlled gates with large $M$ is hard to afford for the NISQ devices. Taking these limitations into account, many alternative protocols have been explored. 

\subsubsection{Resource-efficient variants for purification-based schemes}
\textbf{Realization with classical shadows}.
Classical shadows~\cite{Huang2020Predicting, Huang2021Efficient, Elben2020Mixed} are protocols used to efficiently predict many different properties, in particular linear functions. Considering a trade-off between qubits (and controlled gates) and measurements overhead, many methods implement the VD(or ESD) scheme with classical shadows~\cite{Hong2022Logical, Alireza2022Shadow}.

\textbf{Dual-state purification.}
This protocol~\cite{Huo2022Dual} purifies states using $\frac{\rho\bar{\rho}+\bar{\rho\rho}}{2}$, and the $\bar{\rho}$ is the dual state of $\rho$. The ideal expectation value of observable $O$ is estimated as 
\begin{equation}
    \langle\mathcal{O}\rangle_{\text{em}} = \frac{\text{Tr}(\frac{\rho\bar{\rho}+\bar{\rho\rho}}{2}O)}{\text{Tr}(\frac{\rho\bar{\rho}+\bar{\rho\rho}}{2})}.
\end{equation}
It works when the noisy state $\rho$ and its dual state $\bar{\rho}$ share the same dominant eigenvector in their spectral decompositions. Dual-state purification is efficient in terms of qubit overhead since its implementation requires at most one ancilla qubit.

\textbf{Combination with active qubit resets}.
The active qubit reset technique is enabled by major quantum computing architectures including superconducting qubit and trapped-ion devices. With the use of active qubit resets, we can achieve a similar error suppression using 2$N$ + 1 qubits compared to $MN$ + 1 qubits with the original VD~\cite{Piotr2021Qubitefficient}, which shows a space-time trade-off in the computational resources.

\subsubsection{Generalized purification methods}
In order to deal with the coherent dismatch, we can consider a general polynomial function $f(\rho, M) = c_0+c_1\rho+c_2\rho^2+...+c_{M}\rho^M$~\cite{Yoshioka2022Generalized, Hong2022Logical}.
Recently Xiong {\it{et al.}}~\cite{Xiong2022Permutation} provides a more general framework called permutation filters for those schemes using permutations. It uses a filter 
\begin{equation}
    y_{\text{filter}} = \frac{\text{Tr}(O\mathcal{F}_{\vec{\alpha}}(\rho, M))}{\text{Tr}(\mathcal{F}_{\vec{\alpha}}(\rho, M))}
\end{equation}
to substitute the polynomial $\frac{\text{Tr}(\rho^MO)}{\text{Tr}(\rho^M)}$, where $\mathcal{F}$ denotes some complex function and the parameter optimization for $\vec{\alpha}$ is proven to be an invex problem, which can always converge to the global optimum.

\subsection{Quantum subspace expansion}\label{em-section-QSE}
Quantum subspace expansion (QSE) is first proposed to explore the excited states, and it can also contribute to error mitigation in VQE with additional classical resources~\cite{McClean2017Hybrid, McArdle2019errormitigated, Colless2018Computation, Urbanek2020Chemistry}. 
The QSE effectively mitigates coherent errors due to the imperfect variational optimization.
Given a noisy ground state $\rho$ prepared on the quantum device and the target Hamiltonian $H$, we consider a variational subspace spanned by $\rho$, as 
\begin{equation}
    \begin{aligned}
        \{\rho_{\text{sub}} = \sum_{i,j=1}^Mc_ic_j^*P_i\rho P_j|c_i\in \mathcal{C},P_i\in \{I,X,Y,Z\}^{\otimes N_q} \\
        \text{and}\ \text{Tr}(\rho_{\text{sub}}) = 1\},
    \end{aligned}
\end{equation}
where the parameters $\vec{c} = (c_1,...,c_M)$ denote the expansion coefficients and $M$ refers to the number of expansion terms. 
Then, to obtain the error-mitigated value, we need to solve the following minimization problem
\begin{equation}
    \begin{aligned}
        &\text{min}_{\vec{c}}\{\text{Tr}(\rho_{\text{sub}}H)\} \\
        &\text{such that Tr}(\rho_{\text{sub}}) = 1. 
    \end{aligned}
\end{equation}
The spectrum of the Hamiltonian within the classically expanded subspace can be calculated as the solution to a generalised eigenvalue problem
\begin{equation}
    \hat{H}\vec{c} = E\hat{B}\vec{c}.
\end{equation}
Here, $E$ is a diagonal matrix with elements representing the eigenenergies. $\hat{H}_{ij} = \text{Tr}(P_i\rho P_j H)$ and $\hat{B_{ij}} = \text{Tr}(P_i\rho P_j)$ both can be efficiently estimated via quantum computers. With the optimized parameter obtained from minimization, we can compute the error-mitigated expectation value using
\begin{equation}
    \langle O \rangle_{\text{em}} = \sum_{i,j=1}^Mc_ic_j^*\text{Tr}(P_i\rho P_j H).
\end{equation}

Note that the efficiency of the QSE method depends on the number of expansion terms $M$, thus it may not effective in suppressing stochastic errors due to the requirement on exponential expansion terms to project the noisy state to the error-free subspace~\cite{Endo2021Hybrid, McClean2017Hybrid, Yoshioka2022Generalized}. While when combined with certain properties of the target state such as symmetry~\cite{Bonet2018lowcost}, the QSE can handle stochastic errors more efficiently because the complexity of constructing the projection subspace is reduced. Beside, a generalised framework of QSE~\cite{Yoshioka2022Generalized} has extended the expansion operators from Pauli operators to more general operators relevant to the noisy state.

\subsection{Verification methods}
Verification methods~\cite{McArdle2019errormitigated, Bonet2018lowcost, Cai2021quantumerror} exploit the knowledge of inherent symmetry within the quantum system, such as the spin symmetry in quantum many-body physics. This methods focus on the errors that place state outside the symmetry-preserving subspace.

The symmetry commonly used is Pauli symmetry $\hat{S}$, which can be effectively estimated. Suppose the Pauli symmetry of the system with size $n$ is $\hat{S}=\prod \limits_{i=1}^{M}\hat{S}_i$, $\hat{S}_i \in \mathbb{P}^{\otimes N_q}$, where $M$ is the number of non-trivial terms in $\hat{S}$, and $\mathbb{P}^n$ refers to the $n$-qubit Pauli group. The verification of the symmetry can be achieved by dropping the circuit runs which fail the verification or by using the post-processing appraoches, as we will discuss below.

\subsubsection{Symmetry verification}
\begin{figure}[!htbp]
    \begin{center}
    \includegraphics[width=1\linewidth]{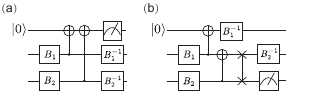}
    \end{center}
    \caption{\textbf{Symmetry Verification.} (a) Symmetry verification circuit. (b) The low-cost version of symmetry verification circuit.}
    \label{QEM-sv}
    \end{figure}

Symmetry verification~\cite{McArdle2019errormitigated, Bonet2018lowcost} discards the circuit runs failed to pass the verification, which costs additional measurement overhead. It utilizes an ancilla qubit interacting with each qubit in the system register, which is of the form shown in Fig.~\ref{QEM-sv}(a). The gates $\{B_i\}_{i=1,2}$ are the basis transformation gates that map the eigenstate of symmetry $\hat{S}$ with eigenvalue $s$ to $|0\rangle$ state. If the ancilla qubit reads 1, we discard this circuit run. Note that the circuit can only detect odd number of error(s). 
In the low-cost version~\cite{Bonet2018lowcost} of symmetry verification shown in Fig.~\ref{QEM-sv}(b), it shuffles the ancilla qubit along the system register, which needs only local CNOT and SWAP two-qubits gates. 

Both cases require $\mathcal{O}(M)$ circuit depth to ensure the entanglement between the ancilla qubit and each register qubit individually, which is general intractable in quantum circuit. Moreover, the verification circuit applied on the noisy state may introduce extra errors, which reduces the reliability of verification. An alternative way to perform symmetry verification will be detailed below.

\subsubsection{Symmetry expansion}
 We can also implement verifications via post-processing approach~\cite{Bonet2018lowcost}. Usually if the target quantum state is the eigenstate of $\hat{S}$ (or in the eigen-subspace) with eigenvalue $s$, we can construct projector valued measurement to project the output state $\rho$ to the $\hat{S}=s$ subspace. Suppose the corresponding projector is ${\hat{\Pi}_s}$ and we have $\hat{\Pi}_s\hat{\Pi}_s = \hat{\Pi}_s$. The effective density matrix after projection is 
 \begin{equation}\label{eqn:QEM-se1}
    \rho_s = \frac{\hat{\Pi}_s\rho\hat{\Pi}_s}{\text{Tr}(\hat{\Pi}_s\rho\hat{\Pi}_s)} = \frac{\hat{\Pi}_s\rho\hat{\Pi}_s}{\text{Tr}(\hat{\Pi}_s\rho)}.
    \end{equation}
If we will measure an observable $O$ which commutes with the symmetry $\hat{S}$ and $[O, \hat{S}] = 0$ and $[O, \hat{\Pi}_s] = 0$. Then the expectation value under the projected state is 
\begin{equation}\label{eqn:QEM-se2}
    \langle O \rangle_{\text{em}} = \text{Tr}(O\rho_s) = \frac{\text{Tr}(O\hat{\Pi}_s\rho)}{\text{Tr}(\hat{\Pi}_s\rho)}.
    \end{equation}
Note that the probability of passing the verification is just the probability of projecting the noisy state to the $\hat{S}=s$ subspace, which is $\text{Tr}(\hat{\Pi}_s\rho)$. Hence we need $\Upsilon_{\text{SE}} = \frac{1}{1-\text{Tr}(\hat{\Pi}_s\rho)}$ more measurement shots as the sampling overhead for mitigation.  

We can combine the post-processing approaches with QSE to further improve the efficiency of mitigation, which called s-QSE~\cite{Bonet2018lowcost}. If we choose $\hat{\Pi}_s = \frac{I + s\hat{S}}{2}$, we will have
\begin{equation}\label{eqn:QEM-se3}
    \langle O \rangle_{\text{em}} = \frac{\text{Tr}(O\rho) + s\text{Tr}(O\hat{\Pi}_s\rho)}{1+s\text{Tr}(\hat{\Pi}_s\rho)}.
    \end{equation}
Then the original minimization problem (as we mentioned in the Section~\ref{em-section-QSE}) is reformulated to a generalized eigenvalue problem, and the terms $\text{Tr}(O\rho)$, $\text{Tr}(O\hat{\Pi}_s\rho)$ and $\text{Tr}(\hat{\Pi}_s\rho)$ can be efficiently estimated via quantum devices. The s-QSE method has been experimentally demonstrated to mitigate errors in the VQE of $\text{H}_2$ with two transmon qubits~\cite{Sagastizabal2019Experimental}.

 Cai has extended the post-processing approaches to a general framework called symmetry expansion~\cite{Cai2021quantumerror}, which encompasses a broader range of symmetry-based error mitigation methods. For example, VD can be seen as a special case of symmetry expansion using permutation symmetry. Moreover, the particle number is also applied as symmetry in the post-processing approach\cite{Huggins2021Efficient}. 

\subsubsection{Other verification methods}
\textbf{Verified phase estimation.}
It applies phase estimation to estimate expectation values while effectively post-selects for the system register to be in the starting state~\cite{Brien2021verifiedpe}. It can also be adapted to the case without the use of control qubits, which simplifies the control circuits. 

\textbf{Pauli check sandwiching.}
The Pauli check sandwiching scheme~\cite{Alvin2022Quantum} applies multiple pairs of Pauli checks to detect the occurrence of errors, and obtains the mitigated results using post-selecting. Each pair of Pauli checks uses one ancilla qubit to detect a component of the error operator.

\subsection{Mitigating the coherent errors}
Coherent errors refer to the imperfect or unwanted unitary rotations acting in the circuits, which can be modeled as 
\begin{equation}
    U(\vec{\theta}) = e^{-\frac{i}{2}\vec{\theta} \cdot \vec{\sigma}},
\end{equation}
where $\vec{\theta} = (\theta_1,...,\theta_{4^{N_q}})$ quantifies the strength of the coherent error on the $4^{N_q}$ Pauli bases. Coherent errors map the noiseless pure states to another pure states, since unitary operators maintain the quantum coherence of the states. While they are purity-preserving, it can pose a threat to the reliable multi-qubit quantum computation. Now we will introduce several typical methods for mitigating coherent errors~\cite{Nielsen2002QuantumComputation, hashim2022noise}.

\subsubsection{Randomized compiling}
Randomized compiling (RC)~\cite{Wallman2016Noisetailoring,Hashim2021Randomizedcompiling} is designed to tailor coherent errors into stochastic Pauli errors, and we can combine RC techniques with other QEM schemes~\cite{Urbanek2021noiseestimation, Ewout2022Probabilistic,Youngseok2021Scalable}. 

When the two-qubit Clifford gate errors dominant over other types of error, we can mitigate the coherent error in the two-qubit gates using RC.
After sandwiching each two-qubit gate between randomly sampled Pauli gates and compiling those twirling Pauli gates into the original single-qubit gates (which can be implemented on the classical computers in advance), the newly generated circuits are logically equivalent to the bare circuits with the same circuit depth. And the averaged results over many logically-equivalent circuits is exactly the desired result with tailored noise.

\subsubsection{Hidden inverses}
Hidden inverses (HI)~\cite{Zhang2022Hidden, Swarnadeep2022Characterizing, Vicente2022Quantumerror} is first introduced to mitigate certain coherent errors (such as over-rotations and phase misalignment) in the trapped-ion quantum computer~\cite{Zhang2022Hidden}, and then it has been extended to implement on the superconducting hardwares~\cite{Vicente2022Quantumerror}. The HI method relies on the self-adjoint unitary operators satisfying $U=U^\dagger$ or self-inverse unitary operators with $U=U^{-1}$, such as $H$ and $CX$. Though these gates represent the same operation in the error-free case, they may suffer from noise with different strength in practice, due to the different compiling ways for $U$ and $U^\dagger(U^{-1})$. Then we can mitigate coherent errors by a local optimization which determines to construct the same gate from the original elementary gate sequence or the sequence of the inverted (conjugate) gate. 

\subsection{Combination with quantum error suppression techniques} 

\subsubsection{Combination with QEC}
\textbf{Individual error reduction.}
The individual error reduction method~\cite{Otten2019Accounting} make sufficient use of the limited QEC techniques which have been well demonstrated on several qubits. It reduces error on each qubit respectively and obtains the mitigated expectation value of the observable by post-processing.
Consider the Lindblad master equation describing the noise process
\begin{equation}
    \frac{\partial \rho}{\partial t} = \mathcal{L}(\rho)=\sum_{i}\mathcal{L}_i(\rho),   
\end{equation}
where the $\mathcal{L}_i$ denotes the Lindblad operator acting on the single qubit. We can give a solution for this equation with the duration $\tau$ of the noise process to the first order approximation
\begin{equation}
    \begin{aligned}
        &\rho(t+\tau)=\mathcal{E}_\tau(\rho(t)), \\
        &\mathcal{E}_\tau(\rho) = e^{[\tau\mathcal{L}(\rho)]} \approx 1 + \tau\sum{i}\mathcal{L}_i(\rho),
    \end{aligned}
\end{equation}
where $\mathcal{E}_\tau$ is the Lindblad evolution operator. Suppose the error on the $i$-th qubit is reduced by a known factor $g_i$ via QEC, then the corresponding Lindblad evolution operator $\mathcal{E}_\tau^i$ turns to 
\begin{equation}
    \mathcal{E}_\tau^i(\rho)\approx 1 + \tau\sum_{j\neq i}\mathcal{L}_j(\rho) + \tau(1-g_i)\mathcal{L}_i(\rho).
\end{equation}
Hence, after QEC applied on the $i$-th qubit, the expectation value of observable $O$ is $\langle O\rangle_i = \text{Tr}(\rho^iO)$ with the output density matrix $\rho^i$. We can estimate the ideal case of the expectation value of observable $O$ by a linear combination of different $\langle O\rangle_i$:
\begin{equation}
    \langle O \rangle_{\text{Ind}} = \text{Tr}(\rho^{\text{Ind}}O) = \langle O \rangle - \sum_{i}\frac{1}{g_i}(\langle O \rangle - \langle O\rangle_i).
\end{equation}
 Note that the individual error reduction has suppressed errors to the first order and can be combined with other error mitigation methods for higher-order error suppression. However, it relies on the accurate estimation of the factor $g_i$ which results in extra cost for characterizations. And additional quantum resources (qubits or gates) are needed to implement QEC on a single qubit.

\textbf{Code space projection.}
The code space projection method~\cite{McClean2020decoding, Hong2022Logical} decodes errors on logical qubits via post-processing, without additional qubits and operations for syndrome measurements. To encode $k$ logical qubits with $N$ physical qubits, we use a stabilizer code $[[N, k]]$ defined by a stabilizer group $S=\langle S_1,S_2,...,S_{N-k}\rangle$ with generators $\{S_i\}_i$. And the code subspace is determined by the projection operator
\begin{equation}
    \Pi = \prod_{i=1}^{N-k}\frac{\mathbb{I}+S_i}{2}=\frac{1}{2^{N-k}}\sum_{j}M_j,
\end{equation}
where $M_j \in S$ refers to the element in the group. Given a logical observable $O$ which can be decomposed by Pauli operators $O=\sum_m\gamma_mP_m$, this method can reduce errors which take the physical state outside the code space. Then we estimate the ideal expectation value using
\begin{equation}\label{eqn-em-code}
    \langle O \rangle_{\text{CS}}=\frac{1}{c2^{N-k}}\sum_{j,m}\gamma_m\text{Tr}(\rho M_jP_m).
\end{equation}
Stochastic and deterministic subspace expansion schemes have been proposed~\cite{McClean2020decoding} for estimation of Eq.~\ref{eqn-em-code}. We can also realize the code space projection using classical shadow techniques~\cite{Hong2022Logical}.  
  
\subsubsection{Dynamic Decoupling}
Dynamical decoupling (DD)~\cite{ Hahn1950Spin, Viola1999Dynamical, Biercuk2009Optimized,Quiroz2013Optimized, Qi2017Optimal, Zeng2018General} methods are designed to suppress decoherence caused by system-environment interaction. The main idea of DD is to apply specific pulse sequence called DD sequence to the idle qubits, which keeps the overall logic of the circuit unchanged. Introducing the DD techniques into the QEM schemes can improve the circuit performance while it dosen't result in additional sampling overhead~\cite{Youngseok2021Scalable, Ravi2022VAQEM}.
\subsubsection{Single-qubit gate scheduling}
Rather than adding additional gates to the circuit as DD does, single-qubit gate scheduling~\cite{Smith2021Error} optimizes circuits by scheduling within idle windows which are periods of idle qubit waiting for the next operation. {\it{As Late As Possible}} (ALAP) scheduling is a typical scheduling technique used as a default approach in IBM Qiskit~\cite{Aleksandrowicz2019qiskit}  to execute the single-qubit gates at the end of the idle windows. We can also tune the positions of single-qubit gates within idle windows by circuit slicing and inverting~\cite{Smith2021Error}. A variational QEM method designed to perform DD and single-gate scheduling within the VQA framework has been proposed recently~\cite{Ravi2022VAQEM}, which avoids specific configuration selections for DD sequences and gate positions.  

\subsection{Framework for QEM}
Recently, much effort has been devoted to exploring the unified framework for QEM schemes and the general bounds for sampling overhead~\cite{Cai2021Apractical,takagi2022fundamental, Qin2021Errorstatistics, Takagi2021optimal, Quek2022Exponentially, Takagi2022Universal}.  
Takagi {\it{et al.}}~\cite{takagi2022fundamental} described the QEM process as the concatenation of
a quantum-classical channel composed of quantum operations and independent POVM measurements, and a classical-classical channel for the classical post-processing. They also introduced the maximum bias to benchmark the worst-case performance of the QEM strategies for an arbitrary state and observable, and then derive a general bound on the sampling overhead based on the maximum bias. And their related work~\cite{Takagi2022Universal} has showed the explicit universal lower bounds on the sampling overhead, which grow exponentially with the circuit depth, 
indicating the fundamental limitations for QEM schemes.

Other metric such as extraction rate was proposed to characterize the cost-effectiveness of QEM schemes within the {\it{linear quantum error mitigation}}~\cite{Cai2021Apractical} framework, where the process of error mitigation is regarded as extracting the effective state out of the unmitigated state. 
Statistics principles are also applied into a general QEM formalism ~\cite{Qin2021Errorstatistics} where an approximate quadratic reduction with gate number $N$ has drawn in error increase.

\section{Quantum Circuit Compilation}

Compilation in the classical computing community is the process of converting a high-level programming language into a machine language that the computer can understand and execute fluently. Similarly, in the field of quantum computation, quantum circuit compilation (QCC) is used to transform a quantum algorithm always in a mathematical form into its corresponding quantum circuit which can be executed on a real quantum device. 

\begin{figure*}[!htbp]
    \begin{center}
    \includegraphics[width=1\linewidth]{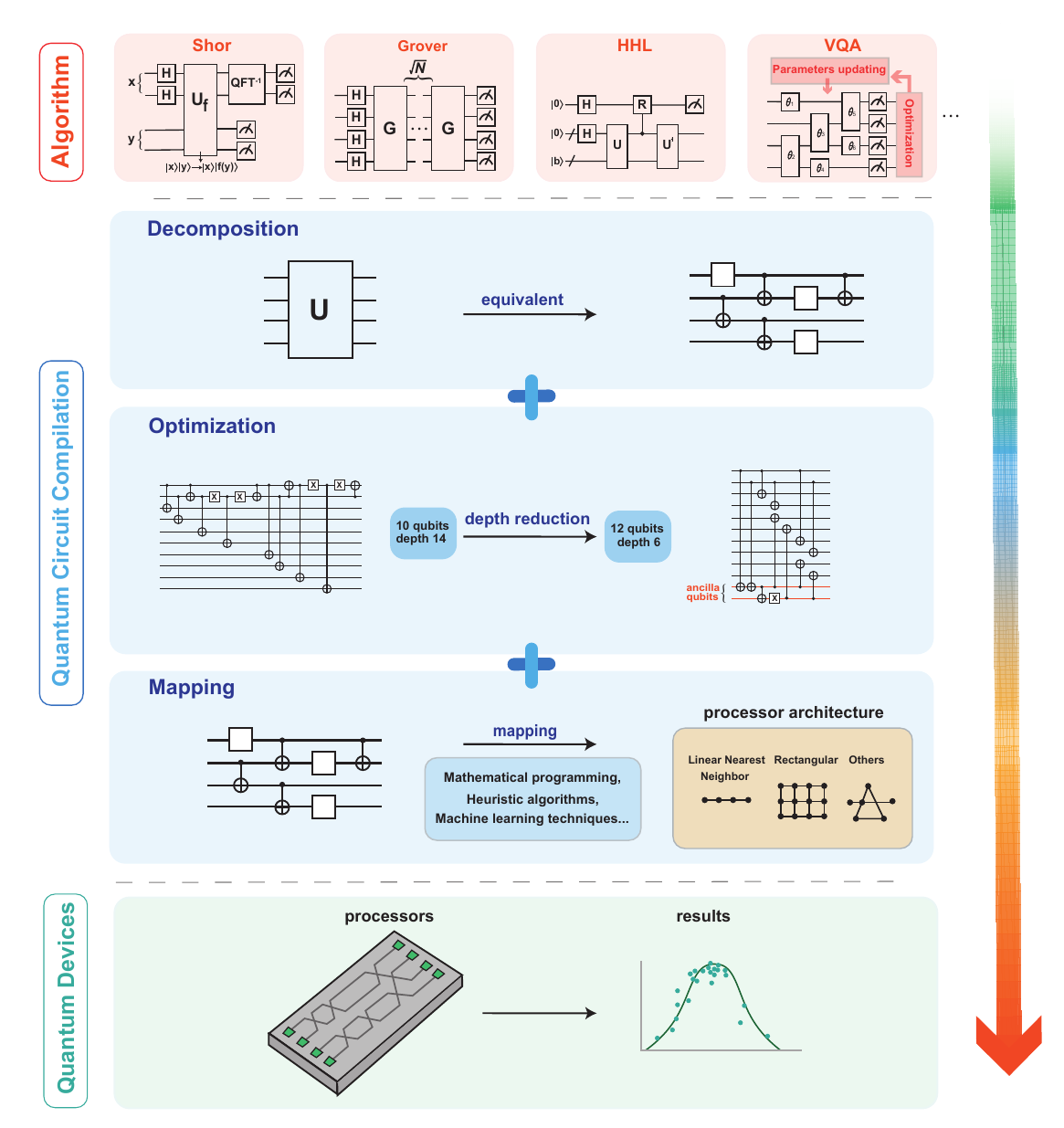}
    \end{center}
    \caption{\textbf{Quantum circuit compilation.}     Quantum circuit compilation translates the abstract quantum algorithm into a compiled circuit optimized for a specific quantum hardware platform, enabling efficient and highly reliable execution through the use of decomposition, optimization, and mapping methods.}
    \label{qcc}
    \end{figure*}

Usually, QCC includes three stages as follows (see Fig.~\ref{qcc}).
\begin{enumerate}
    \item \textit{Decomposion} (also referred to as the synthesis). Usually, a mathematically designed quantum algorithm can be represented by several $n$-qubit unitary operators, such as the Grover iteration in Grover's algorithm, the controlled exponentiation in Shor's algorithm, the Hamiltonian simulation in HHL algorithm and so on. However, real quantum devices, especially NISQ devices, usually provide only some elementary quantum gates, such as single-qubit gates and CNOT gates. Thus, it is one of the most fundamental problems that how to decompose a specific quantum algorithm or unitary operator into as few the provided elementary quantum gates as possible for implementation.
    
    \item \textit{Optimization}. Despite the rapid development of quantum hardware, the qubit lifetime of some quantum systems is still not satisfactory and the quantum noise is also inevitable~\cite{8382253,arute2019quantum}. The optimization aims to further reduce the depth and size of quantum circuits with the help of auxiliary qubits after the decomposition process to alleviate the problem of short qubit lifetime. Thus, the optimization seems like a kind of space-time trade-off, \textit{i.e.}, using ancillary qubits (space) to efficiently reduce circuit size or depth (time).

    \item \textit{Mapping} (also referred to as the qubit routing, qubit allocation, quantum circuit transformation). 
    The final stage is to transform the logical quantum circuit to the physical quantum circuit which can be operated on a real quantum device. Actually after decomposition and optimization, there are only elementary quantum gates in the logical circuit, and the size and depth of which are sufficiently optimized. For the quantum hardware with fully-connected architecture, any logical circuit can be directly implemented. However, for the quantum hardware with limited connectivity architecture~\cite{8382253,arute2019quantum}, we should find the correspondence between logical qubits and physical qubits to complete the transformation.
\end{enumerate}

Although QCC has been separated into the above three stages, sometimes there is no exact order or boundaries. Alternatively, one can perform all the three stages simultaneously to complete the QCC, such as the works in Refs.~\cite{kissinger,nash-qst,ours-topological}. QCC is extremely important as a bridge between quantum algorithms and quantum hardware devices, especially for NISQ devices. There have been a series of results on QCC \cite{PhysRevA.52.3457, https://doi.org/10.48550/arxiv.quant-ph/9508006, RN4646,
RN4645, RN4647, 10.1145/1120725.1120847, 10.5555/1209464,2005quant.ph..4100M, PhysRevA.52.3457, johri2021nearest, ours-state, RN4672, RN4666, 10.5555/2011679.2011685, qft-2000, moore-2001, t-depth-1,  alg-for-T-depth, cnot-ours, kissinger,nash-qst,ours-topological,depth-condition, H-free, RN4684, RN4680, 10.1145/1120725.1120847, childs_et_al:LIPIcs:2019:10395,np-implication, 2015-slover,RN4682,RN4680,foribmq,RN4689,10.1145/3445814.3446706, RN4684, soa-solver, foribmq, 8382253,10.1145/3297858.3304023,RN4692, RN4642, 2020arXiv200715957P,RN4694,10.1145/3400302.3415621, RN4677, RN4679,RN4688, RN4694,10.1145/3400302.3415621,RN4677,2020arXiv200715957P, distributed1,distributed2,distributed3} and we will present them according to the different approaches of the three stages mentioned above.

\subsection{Decomposition}
The first result in the aspect of \textit{Decomposition} was published in 1995, where all $n$-qubit unitary operators can be expressed as compositions of a finite number $O(n^3 4^n)$ of single-qubit gates and CNOT gates exactly~\cite{PhysRevA.52.3457}. 
Soon, this upper bound was lowered to $O(n 4^n)$ by Knill~\cite{https://doi.org/10.48550/arxiv.quant-ph/9508006}. Nine years later, the upper bound for the exact decomposition of $n$-qubit unitary operator has been improved to $O(4^n)$~\cite{RN4646}, which coincides with the theoretical lower bound~\cite{RN4645}. In 2004, M$\ddot{o}$tt$\ddot{o}$nen \textit{et al.} came up with Cosine-Sine Decomposition (CSD), \textit{i.e.},
\begin{equation}
    F_{n}^l(R_a) = 
    \left(
    \begin{array}{ccc}
       R_a(\alpha_1) &         & \\
                     & \ddots  & \\
                     &   & R_a(\alpha_{2^{n-1}})\\
    \end{array}
    \right).
\end{equation}
to express an arbitrary $n$-qubit unitary operator in terms of $n-1$-qubit unitary operators \cite{RN4647}, which further decreased the upper bound to $4^n-2^{n+1}$ CNOT gates.
Next in 2005, Shende \textit{et al.} applied the NQ matrix decomposition to implement an arbitrary $n$-qubit operator only using $\frac{1}{2} \times 4^n -3 \times 2^{n-1} + 1$ CNOT gates \cite{10.1145/1120725.1120847}, where the $n$-qubit operator can be implemented by a circuit containing three uniformly controlled rotations and four ($n-1$)-qubit operators. And then in 2006, M$\ddot{o}$tt$\ddot{o}$nen \textit{et al.} applied CSD recursively to yield a synthesis algorithm, in which the number of CNOT gates is $\frac{23}{48} 4^n -\frac{3}{2} 2^n+\frac{1}{3}$, which firstly decreased the upper bound less than $\frac{1}{2}4^n$ CNOT gates \cite{10.5555/1209464,2005quant.ph..4100M}.

The analysis of lower bound was first came up with by Barenco \textit{et al.} in~\cite{PhysRevA.52.3457} as a conjecture, \textit{i.e.}, $\frac{1}{9} 4^n-\frac{1}{3}n-\frac{1}{9} $, based on dimension counting. In 2004 Shende \textit{et al.} raised the lower bound to $\frac{1}{4}(4^n-3n-1)$ by the technique of parameter counting \cite{RN4645}, which also implies a depth lower bound of $\Omega(4^n/n)$.

We can see that there is a gap on the optimal depth within the range of $[{\Omega}(4^n/n), O(4^n)]$ for general $n$-qubit circuit optimization without ancillary qubits. Most recently, based on Gray code and unary encoding~\cite{johri2021nearest}, Sun~\textit{et al.} have proved a fact that any $n$-qubit unitary operator can be implemented by a quantum circuit of size 
\begin{equation}O(4^n)\end{equation}
and depth
\begin{equation}O\big(n2^n+\frac{4^n}{m+n}\big)\end{equation} 
with $m \le 2^n$ ancillary qubits \cite{ours-state}. They used CSD repeatedly and factored an arbitrary unitary operator into a sequence of uniformly controlled gates. Namely, they presented that any $n$-qubit unitary operator $U$ can be decomposed as $U=V^n_{n}(0) \cdot \prod_{i=1}^{2^{n-1}-1} V^n_{n-\zeta(i)}(i)  \cdot V^n_{n}(2^{n-1}),$
where $\zeta(n)$ is the Ruler function defined as $\zeta(n)=\max\{k:2^{k-1}|n\}$, $V^n_k$ denotes an $n$-qubit uniformly controlled gate whose index of target qubit is $k$, and different $i$ in $V_k^n(i)$ denote different forms of $n$-qubit uniformly controlled gates despite the same target qubit $k$. Combined with their implementation of any $n$-qubit uniformly controlled gate, \textit{i.e.}, any $n$-qubit $V_k^n(i)$ can be implemented by a circuit of size $O(2^n)$ and depth $O\big(n+\frac{2^n}{n+m}\big)$, they derived the above result. What we should note is that their size and depth are simultaneously optimized to asymptotically optimal based on the lower bound in~\cite{RN4645}. In specific without any ancillary qubits, the depth can be optimized to $O(4^n/n)$, which is exactly the lower bound.

Besides the most general decomposition on $n$-qubit unitary, there have already been some results about the special cases. When the elements of unitary belong to the ring $\mathbb{Z}[\frac{1}{\sqrt{2}}, i]$, the single-qubit unitary can be constructed using $H$ and $T$ gates only ~\cite{RN4672}, in which the similar $n$-qubit case was conjectured to be implemented by the Clifford and $T$ gates. This conjecture has been proved soon afterwards. 
An $n$-qubit unitary operator has an exact representation over the Clifford+$T$ gate set assisted by only one ancilla if and only if its entries are in the ring $\mathbb{Z}[\frac{1}{\sqrt{2}}, i]$, and the total gate count is $O(3^{2n}nk)$, where $k$ is the denominator exponent of the unitary~\cite{RN4666}. 
Besides the above results on exact decomposition of unitary, there have already been some works on approximate decomposition. Based on Solovay-Kitaev theorem, Dawson and Nielsen designed an approximate algorithm, which runs in $O(\log ^{2.71} \frac{1}{\epsilon})$ time and produces as output a sequence of $O(\log ^{3.97} \frac{1}{\epsilon})$ quantum gates including only Hadamard, controlled-not, and $\pi/8$ gates \cite{10.5555/2011679.2011685}.

\subsection{Optimization}

As seen in the rapid development of quantum hardware, the number of physical qubits may increase much faster than the qubit lifetime. Thus, when optimizing quantum circuits, we can consider how to reduce the circuit depth using auxiliary qubits.

\begin{figure}[!htbp]
    \begin{center}
    \includegraphics[width=1\linewidth]{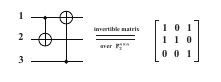}
    \end{center}
    \caption{\textbf{The matirx representation of a CNOT circuit.}}
    \label{cnot-matrix}
\end{figure}

In 2000, Cleve showed the quantum Fourier transform can be approximated to the depth of $O(\log n + \log\log (1/\epsilon))$ with sufficient ancillary qubits~\cite{qft-2000}. Next in 2001, Moore \textit{et al.} proved that a circuit of any size on $n$ qubits composed entirely of CNOT gates (CNOT circuit) can be parallelized to $O(\log n)$ depth with $O(n^2)$ ancilla qubits~\cite{moore-2001}. Similarly for the circuit of controlled-Pauli and $H$ gate, they derived the same result. And for the Clifford$+T$ circuit, Selinger in 2013 made use of four ancillary qubits to represent Toffoli gate with a T-depth of 1 and a total depth of 7~\cite{t-depth-1}. Amy \textit{et al.} described an algorithm to reduce the $T$-depth for Clifford$+T$ circuit~\cite{alg-for-T-depth}, and it has worst case runtime of cubic in the number of $T$ gates, qubits and Hadamard gates. Next let us turn our eyes back to the optimization of CNOT circuit because of its significance in quantum computation~\cite{PhysRevA.52.3457,stab}. In 2020, Jiang \textit{et al.} established an asymptotically optimal space-depth tradeoff for the design of CNOT circuits, that is, any $n$-qubit CNOT circuit can be parallelized to 
$$O ( \max \{ \log n, \frac{n^2}{(n+m)\log (n+m)} \})$$ 
depth with $m$ ancillary qubits, where $m$ being any natural number \cite{cnot-ours}. Firstly, any $n$-qubit CNOT circuit can be regarded as an invertible matrix $\textbf{M}$ over the finite field $\mathbb{F}_2^{n\times n}$. Take a three-qubit CNOT circuit shown in Fig.~\ref{cnot-matrix} as an example, the left circuit can be represented as the right matrix.
Based on the above representation, the CNOT circuit optimization problem will be transformed into a corresponding parallel Gaussian elimination problem mathematically. Combined with a standard technique in reversible computation, they firstly constructed two CNOT circuits $\mathcal{C}_1$, $\mathcal{C}_2$ with $2n$ and $3sn$ qubits respectively, where $1\leq s\leq O(n/\log^2 n)$. Specifically, for $\textbf{x},\textbf{j} \in \mathbb{F}_2^n$,
$$
    \mathcal{C}_1 \ket{\textbf{x}} \ket{\textbf{j}} \ket{0}^{\otimes 3sn} = \ket{\textbf{x}} \ket{\textbf{j} \oplus \textbf{Mx}} \ket{0}^{\otimes 3sn},
$$
$$
    \mathcal{C}_2 \ket{\textbf{x}} \ket{\textbf{j}} \ket{0}^{\otimes 3sn} = \ket{\textbf{x}} \ket{\textbf{j} \oplus \textbf{M}^{-1}\textbf{x}} \ket{0}^{\otimes 3sn}.
$$
Then they applied $\mathcal{C}_1, \mathcal{C}_2$ to transform the initial state $\ket{\textbf{x}} \ket{0}^{\otimes n} \ket{0}^{\otimes 3sn}$ to the final state $\ket{\textbf{Mx}}$ after permuting the first and second $n$ qubits. This boun
d is tight and this algorithm can be directly extended to stabilizer circuits with the reason that any stabilizer circuit has a canonical form ``H-C-P-C-P-C-H-P-C-P-C'', where H and P are one layer of Hadamard gates and Phase gates respectively \cite{stabilizer-cano}. 

Near-term quantum devices are limited not only by the qubit lifetime, but also by the connectivity between qubits. All the above optimization results focus on the optimal depth, without considering the constraints on connectivity. Thus, it is necessary to consider the quantum circuit optimization problem for the quantum processors with sparsely connected structure. We should note that the optimization here usually refers to reducing the circuit depth with the help of ancillary qubits, which is quite different from the mapping introduced in the next section, which focuses on reducing  the number of SWAP gates added to realize the logical circuit physically and keeping the number of qubits constant.

Actually there have already existed some results on reducing the circuit size under the limited connectivity architecture~\cite{kissinger,nash-qst,ours-topological,depth-condition}. In 2019, based on Gaussian elimination, Kissinger and Griend presented an algorithm which can extract any $n$-qubit CNOT circuit to $2n^2$-size equivalent CNOT circuit if the architecture contains a Hamiltonian path~\cite{kissinger}. But if there is no Hamiltonian in the architecture, their optimized size is $O(n^3)$. Next in Ref.~\cite{nash-qst}, Nash \textit{et al.} studied the CNOT circuit optimization on any connected graph, and they proposed an algorithm which gives a $4n^2$-size equivalent CNOT circuit for any $n$-qubit CNOT circuit. This bound has been reduced to $2n^2$~\cite{ours-topological}. Furthermore for size optimization, Wu \textit{et al.} came up with an algorithm, inspired by the result in~\cite{patel-alg-inspi}, which can reduce the size of any $n$-qubit CNOT circuit to 
$$O(n^2/\log \delta)$$ 
on any connected graph with $\delta $ being the minimum degree, and this bound is optimal when the limited connectivity architecture is a regular graph~\cite{ours-topological}. In addition, they also optimized the depth of CNOT circuit with the assistance of ancillary qubits. Specifically, their algorithm can reduce the depth of any given $n$-qubit CNOT circuit to 
$$O(n^2/\min \{m_1, m_2\}),$$ 
with $3n\leq m_1m_2 \leq n^2$ ancillary qubits and the limited connectivity architecture is $m_1 \times m_2$ grid graph \cite{ours-topological}. And this bound is asymptotically optimal when $m_1 \times m_2 =n$. Also this result can be similarly generalized to any constant dimensional grid graph. Most recently, Maslov and Yang considered the optimization of Hadamard-free Clifford circuit on Linear Nearest Neighbor (LNN) \cite{H-free} qubit connectivity architectures, and derived this circuit can be implemented over LNN in depth $5n$.

\subsection{Mapping}
The study of mapping problem firstly focused on its complexity. In 2009, it has been conjectured that mapping a non-LNN (Linear Nearest Neighbor) circuit into an LNN circuit is NP-complete \cite{RN4684}, because the number of possible combinations of all permutations for all two-qubit gates in circuit is $((n-1)!)^k$, where $n$ is the number of qubits and $k$ is the number of two-qubit gates. Obviously, the number $((n-1)!)^k$ will become huge with $n$ and $k$ increasing. In 2014, Shafaei \textit{et al.} modeled the logical quantum circuit as an interaction graph and the structure of quantum device as a connectivity graph. Thus the mapping problem is a standard graph embedding problem with connectivity and interaction graphs as the host and guest graphs, respectively. And the objective is then to minimize the total distance between adjacent nodes of the interaction graph. For a 2D grid connectivity graph, the mapping problem is NP-complete \cite{RN4680}. In 2018, Siraichi \textit{et al.} proved the mapping problem, similar to the classical register allocation, is NP-complete in general \cite{10.1145/1120725.1120847}, which is also derived in \cite{childs_et_al:LIPIcs:2019:10395,np-implication}. Due to the NP-completeness of mapping problem, researchers have taken more attention on how to reduce the SWAP gates needed approximately. We will introduce and analyze the relevant results in three aspects based on different approaches to solve the mapping problem.

\textbf{Mathematical programming.} It is a most natural idea to reformulate the mapping problem on some specific physical architectures into a corresponding optimisation problem and to solve it with state-of-the-art tools and solvers~\cite{2015-slover,RN4682,RN4680,foribmq,RN4689,10.1145/3445814.3446706}.
In 2009, Hirata \textit{et al.} considered a sorting of the initial qubit order according to a function they defined at first, then sorted this qubit order to the final one. Thus the objective is to minimize the sum of SWAP gates of the first and second step \cite{RN4684}. 
Next in 2013, Shafaei \textit{et al.} regarded the mapping problem as improving locality of a given quantum circuit, \textit{i.e.}, the minimum linear arrangement (MinLA) problem in graph theory, and they showed the effectiveness of this approach for quantum Fourier transformation and reversible benchmarks through experimental results \cite{RN4682}. Also they focused on 2D grid architecture \cite{RN4680} and refactored the mapping problem to Mixed Integer Programming problem. In 2015, Lye \textit{et al.} formulated the mapping problem on multi-dimensional quantum architectures as Pseudo-Boolean Optimization (PBO) problem, and used a state-of-the-art PBO solver \cite{soa-solver} to achieve a minimal number of SWAP gates \cite{2015-slover}. In 2019, Wille \textit{et al.} formulated the mapping problem as a symbolic optimization problem that is solved using reasoning engines like Boolean satisfiability solvers, and they provided a method that maps small-size logical circuits to IBM's QX architectures with a minimal number of SWAP and $H$ gates \cite{foribmq}. 


\textbf{Heuristic algorithms.} To adapt real quantum devices with complex connectivity architectures, a series of heuristic algorithms~\cite{8382253,10.1145/3297858.3304023,RN4692, RN4642, 2020arXiv200715957P,RN4694,10.1145/3400302.3415621, RN4677} on the mapping problem were developed. Actually for mapping problem, the most important thing is to determine how to insert the SWAP gate effectively and efficiently. Here ``effectively'' means mapping the logical quantum circuit to physical quantum circuit and running that circuit on a quantum hardware device successfully, while ``efficently'' means adding as few SWAP gates as possible during the mapping procedure. These two aspects correspond to the two steps in the heuristic algorithm design. The first step is to determine the candidate SWAP-gate sets at present, and the second step is to design the evaluation function that can evaluate the effectiveness of the present SWAP gate.

In 2011, Saeedi \textit{et al.} proposed two heuristic methods to determine which qubits should be reordered, a template matching optimization and an exact synthesis approach~\cite{RN4679}, which, from experimental results, improved the quantum cost by more than $50\%$ on average compared to the naive method. In 2016, Wille~\textit{et al.} considered the effect of inserting SWAP gate on the following two-qubit gates and came up with a look-ahead scheme, which reduced the number of SWAP gates of $56\%$ in the best experimental evaluation~\cite{RN4688}. This look-ahead scheme also leads to the improvement for 2D architectures. In 2019, Zulehner, the first-place winner of IBM Q Award, devised a much better algorithm in which quantum gates firstly would be partitioned into layers such that each layer contains only gates acted on distinct sets of qubit. Then they employed the $A^*$ algorithm~\cite{4082128} to search and determine how to insert the SWAP gate for the current layer, which can help to decrease the circuit depth simultaneously. But this algorithm is not suitable to large-scale logical circuit because of the space and time cost of $A*$ algorithm. In order to improve the efficiency of the searching algorithm, in 2019, Li \textit{et al.} proposed a SWAP-based BidiREctional heuristic search algorithm~\cite{10.1145/3297858.3304023}. Instead of searching for a mapping in the full space, they only searched for SWAPs associated with the qubits and designed a heuristic cost function to help find the SWAP that can reduce the sum of distances between each qubit pairs in the front layer. Based on this idea, in 2020, Zhou \textit{et al.} calculated the costs of SWAP gates on all front layers heuristically and chose the one with the minimum cost~\cite{RN4692}, which would further reduce the complexity in time polynomial in all parameters. Most recently, they proposed a Monte Carlo Tree Search framework to tackle the mapping problem~\cite{RN4642}, which enabled the search process to go much deeper. 

\textbf{Machine learning techniques.} Machine learning techniques have also been exploited to provide a more precise evaluation tool for mapping problem~\cite{RN4694,10.1145/3400302.3415621,RN4677}, as they can provide a more precise assessment of the effects of the present SWAP gate. In 2020, Pozzi \textit{et al.} framed the mapping problem as a reinforcement learning problem, then applied the combinatorial optimization techniques to search for the action leading to the highest-quality under their defined quality function, and finally scheduled SWAP gates~\cite{2020arXiv200715957P}. In Ref.~\cite{RN4694}, graph neural networks were used as an improved architecture to help guide the tree search used to find an optimal set of SWAP gates, then they used an array of mutex locks to represent the current parallelization state to help to reduce the depth of the output circuits. In Ref.~\cite{10.1145/3400302.3415621}, Zhou \textit{et al.} took both short- and long-term rewards into consideration to design a scoring mechanism, and proposed a search algorithm which is polynomial in all relevant parameters.

\section{benchmarking Protocols}

Benchmarking is a technique to assess the performance and capabilities of quantum devices, and is therefore critical to accelerating progress in quantum computing. Design a ``good'' benchmarking protocol for quantum devices is complicated for the following reasons: 1) There are different physical ways to implement quantum computing, such as ion traps, photons, superconducting, \textit{etc}, and each has its own strengths and weaknesses. It is difficult for us to design a comprehensive protocol that can evaluate the performance of each physical system well. 2) Different types of applications may have different forms of assessment and core metrics. 3) When the benchmarking protocol is applied to large-scale systems, significant resource consumption may be required.

To date, a number of benchmarking protocols have been proposed and widely used in experiments. These benchmarking protocols can be broadly divided into two major categories, gate-level benchmarking and circuit-level benchmarking, which will be introduced in this section.

\begin{figure*}[!htbp]
    \begin{center}
    \includegraphics[width=1\linewidth]{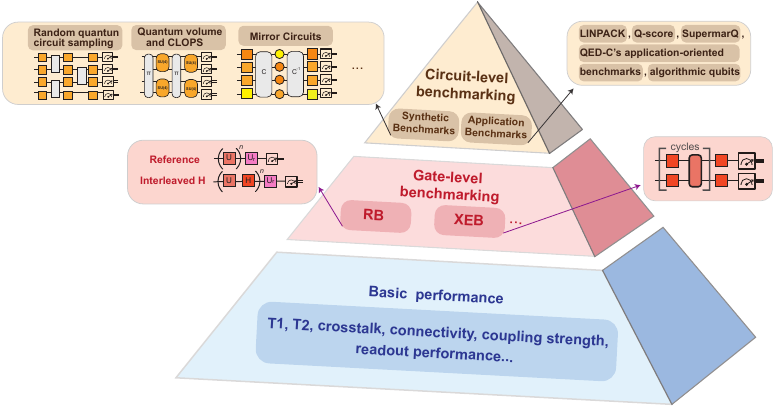}
    \end{center}
    \caption{\textbf{Benchmarking methods for different levels.} These methods can be divided into three levels, basic performance benckmarking, gate-level benckmarking and circuit-level benchmarking. At the lowest level, the basic performance of quantum devices, such as $T_1$, $T_2$, and chip connectivity, is calibrated, while in the gate-level benckmarking, each individual quantum operations, such as single-qubit gate and two-qubit gate, are optimized to enable the implementation of quantum circuits. In the circuit-level benchmarking, the performance of the quantum system is tested by executing complete quantum circuits or particular applications. Higher-level benchmarks reflect a integrated performance, while lower-level benchmarks are the basic steps performed prior to the utilization of the processor.
    }
    \label{benchmark}
    \end{figure*}

\subsection{Gate-level benchmarking}
\subsubsection{Randomized benchmarking (RB)}
Quantum process tomography (QPT) is a standard approach to fully characterizing quantum process, but it is sensitive to the state preparation and measurement (SPAM) errors  and infeasible for large systems. RB is a technique for estimating the average fidelity of a set of quantum gates without relying on accurate SPAM~\cite{emerson2005scalable,magesan2011scalable}. 

RB is originally proposed for random unitary gates~\cite{emerson2005scalable,dankert2009exact,levi2007efficient}, and now is developed as a widely used experimental technique for characterizing the average error of quantum operations. Typically, RB is executed with gates from the Clifford group, which is referred to Clifford randomized benchmarking (CRB)~\cite{magesan2011scalable,magesan2012efficient,knill2008randomized,emerson2007symmetrized}. A CRB protocol consists of the following steps:

Step 1: Generate RB sequences. First randomly sample a sequence of gates of a fixed length $m$ from the Clifford group $G$ on $n$-qubits, and then a global inversion gate is added to return the qubits to the initial state in the ideal case without noise. For each length $m$, we choose $K_m$ RB sequences.

Step 2: Execute the RB sequences on the noisy quantum devices. Measure the output state to estimate its overlap with the initial state, which is named as survival probability $\alpha_m$. Repeat $K_m$ sequences to obtain the a single average survival probability $\overline{\alpha_m}$.

Step 3: Fit the results. Repeat Steps 2 for various sequence lengths $m$ to produce a list of average survival probabilities $\{\overline{\alpha_m}\}_m$, and then fit $\{\overline{\alpha_m}\}_m$ to the a single exponential decay model:

\begin{equation} \label{eq:zerothorder}
\overline{\alpha_m}\ = A_0 p^{m} + B_0
\end{equation}
where $A_0$ and $B_0$ absorb state preparation and measurement errors as well as an \emph{edge effect} from the error on the final gate. The quality parameter $p$ determines the average error-rate $r$ according to the relation
\begin{equation}\label{eq:average error-rate}
r = \frac{{{2^n} - 1}}{{{2^n}}}(1 - p ).
\end{equation}

The RB protocol has been extensively developed in recent years. It has been extended for other finite groups~\cite{barends2014rolling,onorati2019randomized,carignan2015characterizing, cross2016scalable,helsen2019new,Erhard2019Characterizing,francca2018approximate,proctor2019direct,zhang2022scalable}, and it has been also extensively experimentally studied~\cite{knill2008randomized,gaebler2012randomized,gambetta2012characterization,mckay2019three,garion2021experimental}. More recently, a general framework for randomized benchmarking has been discussed~\cite{helsen2022general}. 

\subsubsection{Cross-entropy benchmarking (XEB)}
Linear XEB is a sampling-based approach to characterize an arbitrary quantum gate~\cite{boixo2018characterizing,neill2018blueprint,barends2019diabatic}, and is experimentally friendly for large-scale quantum devices with dozens of qubits. In particular, XEB provides a measure to approximate the fidelity of random quantum circuit (RQC) in the quantum computational advantage experiments with limited bitstring samples~\cite{arute2019quantum,wu2021strong,zhu2022quantum}.

For single-qubit gates XEB, the steps are:

Step 1: Generate XEB sequences. First randomly sample a sequence of gates of a fixed length $m$ from the single-qubit gateset formed by the eight $\pi /2$ rotations along the following axes in the Bloch sphere representation: $\pm X$, $\pm Y$, and $\pm (X \pm Y)$ on a single-qubit, and then end with
a final random single-qubit gate and measurement. For each length $m$, we choose $K_m$ XEB sequences. 

Step 2: Execute the XEB sequences on the noisy quantum devices. By comparing the measured state probabilities with the ideal ones, we can calculate the relative cross-entropy differences between probability distributions~\cite{neill2018blueprint},
\begin{equation} 
\alpha {\rm{ = }}\frac{{{H_{\text{uniform},\text{ideal} }} - {H_{\text{experiment},\text{ideal} }}}}{{{H_{\text{uniform},\text{ideal}}} - {H_{\text{ideal}}}}}
\end{equation}
where $H= - \sum\nolimits_i {{p_i}\log ({q_i})}$ is the cross entropy between two probability distributions $\{p_i\}$ and $\{q_i\}$, and $H_{\text{uniform},\text{ideal}}$ is the cross-entropy between the uniform and ideal distributions, $H_{\text{experiment},\text{ideal}}$ is the cross-entropy between the experimentally measured and ideal distributions, and $H_{\text{ideal}}$ is the self-entropy of the ideal distribution. Repeat $K_m$ sequences to obtain the a single average survival probability $\overline{\alpha_m}$.

Step 3: Fit the results. Repeat Steps 2 for various sequence lengths $m$ produces a list of average survival probabilities $\{\overline{\alpha_m}\}_m$, and then fit $\{\overline{\alpha_m}\}_m$ to the a single exponential decay model:
\begin{equation}
\overline{\alpha_m}\ = A_0 p^{m} + B_0
\end{equation}
where $A_0$ and $B_0$ absorb state preparation and measurement errors. The decay parameter $p$ can be converted to the average error $r$ and Pauli error $r_P$,
\begin{equation}
r = \frac{{N - 1}}{N}(1 - p)
\end{equation}

\begin{equation}
{r_P} = \frac{{N - 1}}{N}(r)
\end{equation}
with $N=2^n$ the dimensionality of the system having $n$ qubits. 

The procedure of characterizing two-qubit gates using XEB is similar to that of single-qubit gates, except that the circuit sequence generated in Step 1 is different. Specifically, each cycle of the XEB circuit consisting of a layer of random single-qubit gates and the two-qubit gate, with a final round of random single-qubit gates appended at the end, where single-qubit gate set is still $\pm X$, $\pm Y$, and $\pm (X \pm Y)$. Finally, divide the Pauli fidelity of the cycle by the Pauli fidelity of the two single-qubit gates to obtain the Pauli fidelity of the two two-qubit gates.

XEB is a powerful calibration tool for near-term quantum devices. Some works have attempted to analyze how to spoof linear XEB, an important topic related to the quantum computational advantage experiments~\cite{aaronson2019classical,barak2020spoofing}. In addition, Chen \textit{et al.} proposed the \textit{Clifford} XEB~\cite{chen2022linear}, which replaces the random circuits with Clifford circuits, to enable the characterization of large-scale quantum systems, since Clifford circuits can be simulated in polynomial time.


\subsection{Circuit-level benchmarking}

\subsubsection{Random Quantum Circuit (RQC) Sampling}

RQC sampling is outstanding candidate to demonstrate quantum computational advantages, and also a tool to characterize the overall performance of the quantum processor. The steps of RQC sampling are:

Step 1: Generate RQC. Each RQC is composed of $m$ cycles, and each cycle is composed of a single-qubit gate layer and a two-qubit gate layer. In the single-qubit gate layer, single-qubit gates are applied on all qubits and chosen randomly from the set of $\{ \sqrt X ,\sqrt Y ,\sqrt W \}$, where $\sqrt X {\rm{ = }}{R_X}(\pi /2)$, $\sqrt Y {\rm{ = }}{R_Y}(\pi /2)$, and $\sqrt W {\rm{ = }}{R_{(X+Y)}}(\pi /2)$ are $\pi /2$-rotation around specific axis. Each single-qubit gate on a qubit in subsequent cycle is independently and randomly chosen from the subset of $\{ \sqrt X ,\sqrt Y ,\sqrt W \}$, which does not include the single-qubit gate to this qubit in the preceding cycle. In the two-qubit gate layer, two-qubit gates are applied according to a specified pattern, labeled by A, B, C, and D, in sequence of ABCDCDAB. By executing the four-cycle sequence ABCD once, the two-qubit gate between all qubits can be executed exactly once. Finally, an additional single-qubit gate layer is applied after $m$ cycles and before measurement.

Step 2: Execute the RQC on the noisy quantum devices and collect the measured bitstrings $\{x_i\}$.

Step 3: Compute the linear cross-entropy benchmarking fidelity,

\begin{equation}
{F_{{\rm{XEB}}}} = {2^n}{\langle p({x_i})\rangle _i} - 1
\end{equation}
where $n$ is the number of qubits, $p({x_i})$ is the probability of bitstring $x_i$ calculated for the ideal quantum circuit.

\subsubsection{Quantum volume (QV) and circuit layer operations per second (CLOPS)}

QV is a single-number metric for quantifying the computational power of a near-term quantum computers of modest size~\cite{cross2019validating}. In general, to achieve higher QVs, quantum computing systems not only require more qubits, but also to maintain high-fidelity operations, high qubit connectivity, high gate parallelism, as well as high-performance circuit compilation techniques.

A QV benchmark procedure is as follows:

Step 1: Generate QV circuits. The circuits $U$ consist of $d$ sequential layers that act on $m$ qubits. Each layer consists of a random permutation of the qubits, followed by Haar-random two-qubit gates (from SU(4)) performed on neighbouring pairs of qubits. 

Step 2: Execute the QV protocol on the noisy quantum devices. If we cannot execute circuit $U$ perfectly or efficiently with the provided gate set provided by the target system, then we need to resort to a circuit-to-circuit transpiler to find an approximate $U'$ with infidelity $1 - {F_{{\rm{avg}}}}(U,U') \le 1$, where
\begin{equation}
    {F_{{\rm{avg}}}}(U,U') = \frac{{|Tr({U^\dag }U'){|^2}/{2^m} + 1}}{{{2^m} + 1}}
\end{equation}
is the average gate fidelity~\cite{horodecki1999general} between $m$-qubit unitaries $U$ and $U'$. Arbitrary circuit transpiler techniques can be used to facilitate efficient execution on hardware. Finally, observe the distribution $q_U(x)$ for the implementation $U'$.

Step 3: Determine whether the result satisfies the heavy output. Calculate the ideal output distribution ${p_U}(x) = |\langle x|U|0\rangle {|^2}$ of the circuit $U$, and sort it in ascending order
${p_0} \le {p_1} \le  \cdots  \le {p_{{2^m} - 1}}$, where $x \in {\{ 0,1\} ^m}$ is an observable bit-string. The heavy ${H_U}$ outputs are 
\begin{equation}
{H_U} = {\{ x \in \{ 0,1\} \} ^m}~~\text{such that}~~{p_U}(x) > {p_{\text{med}}}\},
\end{equation}
where $p_{\text{med}}=(p_{2^{(m-1)}}+p_{2^{(m-1)}-1})/2$ is the median of the set of probability. The probability that the experimental sampling satisfies the heavy output is:
\begin{equation}
{h_U} = \sum\limits_{x \in {H_U}} {{q_U}(x)}
\end{equation}

If ${h_U} > 2/3$, the test is passed.

Step 4: Calculate the QV. For $m$-qubit quantum system, repeat steps 1-3 to get the largest achievable depth $d(m)$ such that 
\begin{equation}
{h_1},{h_2}, \ldots ,{h_{d(m)}} > 2/3~~\text{and}~~{h_{d(m) + 1}} \le 2/3.
\end{equation}

By increasing the system size until $d(m)<m$, we then get the QV $V_Q$, which is defined as 
\begin{equation}
{\log _2}{V_Q} = \arg {\mathop{\rm \mathop{max}\limits_{m}~min}\nolimits} (m,d(m))
\end{equation}

QV has become a widely used benchmarking tool as it reflects the comprehensive performance of quantum devices. Recently, Quantinuum claimed a 8,192 QV is measured on their trapped-ion quantum computing~\cite{Quantinuum2022}. IBM Quantum has has achieved a 256 QV on their superconducting quantum processor Falcon r10~\cite{IBMQuantum2022}.

CLOPS~\cite{wack2021scale} is a measure correlated with how many QV circuits a QPU can execute per unit of time, and it is a speed benchmark that serves as a complement to QV. Thus, IBM Quantum team take the number of qubits, QV, and CLOPS as three key attributes to measure the performance of near-term quantum computers, in terms of scale, quality, and speed.

\subsubsection{Mirror Circuits}
Both QV and RQC sampling require predicting the ideal outcomes, whose classical complexity grows exponentially. Mirror circuits -- quantum circuits with a reflection structure -- is an efficiently verifiable benchmarking method~\cite{proctor2022measuring}, which does not require huge classical computing resources. Specifically, a mirror circuit consists of: 1) A layer to prepare each qubit in a randomized Pauli eigenstate; 2) The layer of quantum circuit $C$; 3) A randomly chosen Pauli layer $Q$; 3) The `quasi-inverse' circuit $C^{-1}$ so that $C^{-1}QC$ is a Pauli operation. By construction, every mirror circuit has a unique and easy-to-calculate outcome bitstring in the ideal case without noise. 

Run the mirror circuit on the noisy quantum devices, and observe the probability $S$ of the bitstring that should be output theoretically. Calculate the polarization 
\begin{equation}
p=(S-1/2^w)/(1-1/2^w),
\end{equation}
which is a metric represents performance on $C$, where $w$ is the number of qubits.

\subsubsection{Application-Oriented Benchmarks}

Application-oriented benchmarks aim to measure the usefulness of quantum system at handling actual problems, rather than simply measuring its theoretical performance. Recently, various application-oriented benchmarks have been introduced~\cite{dong2021random,tomesh2022supermarq, van2022evaluating, lubinski2021application, AlgorithmicQubits2022}. These benchmarks deal with different practical problems. Inspired by the classical LINPACK benchmark, the quantum LINPACK benchmark~\cite{dong2021random} is to measure the performance of quantum computers for scientific computing applications by considering the problem of solving quantum linear system problem. The Q-score metric~\cite{van2022evaluating} is defined as the size of the largest graph for which the quantum device can approximately solve the Max-Cut problem with a solution that significantly outperforms a random guessing algorithm. While, the SupermarQ~\cite{tomesh2022supermarq}, QED-C's application-oriented benchmarks~\cite{lubinski2021application}, and algorithmic qubits~\cite{AlgorithmicQubits2022} are more integrated by executing a series of common quantum algorithms and programs.

\section{classical simulation}

\begin{figure*}[!htbp]
    \begin{center}
    \includegraphics[width=1\linewidth]{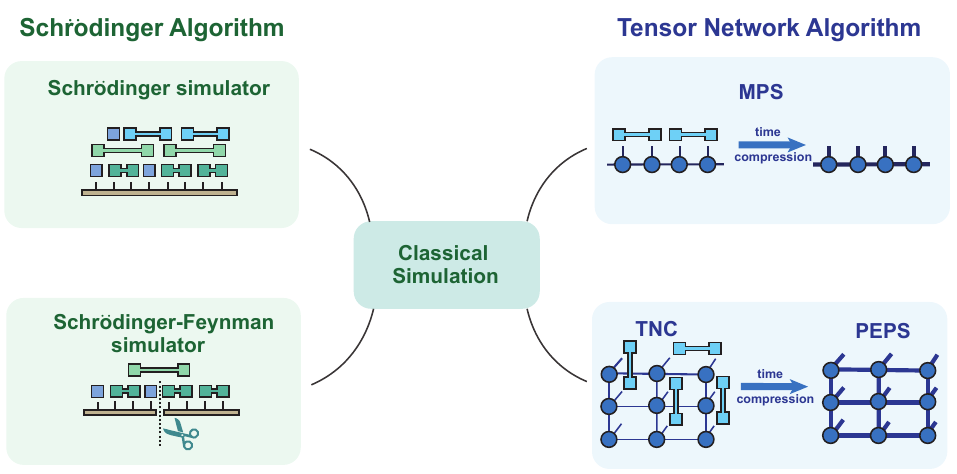}
    \end{center}
    \caption{\textbf{Classical algorithms for simulating quantum circuits.} The Schr$\ddot{\text{o}}$dinger simulator directly stores the quantum state as a state vector and its memory requirement grows exponentially with the system size. The Schr$\ddot{\text{o}}$dinger-Feynman simulator divides the system into two subsystems and store the two smaller quantum states accordingly, the complexity of which grows exponentially with the number of ``cross gates'' acting on both subsystems simultaneously. In tensor network based algorithms the quantum state (and the quantum circuit) are indirectly stored as a tensor network. Concretely, the matrix product states based simulator represents the quantum state as an MPS, which is a special instance of the one-dimensional tensor network. The tensor network contraction algorithm represents the quantum state together with the quantum circuit as a whole tensor network. Pre-contracting the tensor network formed in the TNC algorithm in the time direction will result in a projected-entangled pair state (assuming that the underlying quantum circuit has a clear two-dimensional structure, which are physically motivated by the geometry of the current quantum processors).    }
    \label{fig-classicalsimulate}
\end{figure*}

In the NISQ era, classical simulators for quantum algorithms are particularly important. On one hand, classical simulators are perfect benchmarking baselines for noisy quantum computers. As an example, classical simulators have been extensively used in RQC sampling experiments, ranging from calibrating the gate-level fidelities and the circuit-level fidelities, to estimating the fidelities of the hardest RQC sampling experiments~\cite{arute2019quantum,wu2021strong,zhu2022quantum}. Moreover, the claim of \textit{quantum supremacy} for RQC sampling also requires an efficient classical simulator to set the performance baseline for a fair comparison. On the other hand, the classical simulators also provide an alternative platform for running quantum algorithms without resorting to an actual quantum computer, which could be very helpful, for example, for exploring near-term quantum algorithms such as VQE, or for testing quantum noise models or quantum error mitigation schemes. 


In this section we will first review several important classical algorithms used to implement classical simulators for both noiseless and noisy quantum circuits, and then we will briefly review different ways to compute the gradients of parametric quantum circuits on classical computers.

\subsection{Simulating noiseless quantum circuits}

Here we introduce some definitions and notations which will be used throughout this section. 
Mathematically, an $n$-qubit pure quantum state $\vert \phi\rangle$ can be written as
\begin{align}\label{eq:phi}
\vert \phi\rangle = \sum_{\sigma_1, \sigma_2, \dots, \sigma_n} c_{\sigma_1, \sigma_2, \dots, \sigma_n} \vert \sigma_1, \sigma_2, \dots, \sigma_n\rangle,
\end{align}
where $\vert \sigma_1, \sigma_2, \dots, \sigma_n\rangle$ represents a specific computational basis and $c_{\sigma_1, \sigma_2, \dots, \sigma_n}$ is the coefficient (amplitude) for it. All the coefficients constitute a rank-$n$ tensor, which contains $2^n$ complex numbers since each $\sigma_l$ can be $0$ or $1$. The coefficient tensor is usually equivalently viewed as a long \textit{state vector} of length $2^n$. 

Generally, a quantum algorithm starts by initializing the $n$-qubit quantum register in a particular state $\vert 0^{\otimes n}\rangle$ where all the qubits are set to state $\vert 0\rangle$. Then one applies a series of quantum gate operations onto the initial state to obtain the final quantum state $\vert \psi\rangle$:
\begin{align}\label{eq:circuitevolve}
\vert \psi \rangle = \circuit \vert 0^{\otimes n}\rangle = \Qop^m \cdots \Qop^1 \vert 0^{\otimes n}\rangle,
\end{align}
where $\Qop^j$ denotes the $j$-th quantum gate operation, $m$ is the total number of gates and $\circuit$ is an abbreviation for all the gate operations. We will also use $Q$ to denote the tensor which corresponds to the quantum operator $\Qop$. Finally, a quantum algorithm usually ends by measuring the final quantum state $\vert \psi\rangle$ to obtain the measurement outcomes in the form of bitstrings, which are either used as the final outputs or used to further compute the expectation values of some quantum observables depending on the specific quantum algorithms. We note that a general quantum algorithm could also allow quantum measurements to be performed in between the quantum gate operations, which however adds no difficulty for classical simulators as long as each elementary quantum operation could be simulated.

For classical simulators, one could usually perform more flexible ``measurements'' compared to quantum computers since the quantum state is often stored in a certain data structure on classical computers which can be easily copied and reused. We will primarily focus on the following three types of measurements that can be performed using classical simulators: 1) computing the amplitude of a given bitstring $\vec{b} = \{b_1, b_2, \dots, b_n\}$, which is useful to verify the fidelity of the measured bitstrings by noisy quantum computers (the XEB test is one such example); 2) simulating the sampling process, namely generating a number of samples (bitstrings), this could be done as long as one can compute amplitudes for any given bitstrings, for example one could first generate a number of random bitstrings and then perform a rejective sampling to select a subset of them as samples according to their amplitudes~\cite{MarkovBoixo2018}; and 3) computing the expectation values of some quantum observables (Pauli strings for example). We note that although expectation values could be computed in a similar way to quantum computers as long as one can generate samples, for classical simulators there often exist shortcuts in which they can be directly computed without sampling errors, and these exact expectation values could be used as perfect references for noisy quantum computers.

\subsubsection{Schr$\ddot{\text{o}}$dinger simulator}

The most straightforward approach to simulate quantum circuits is the so-called Schr$\ddot{\text{o}}$dinger algorithm, which stores a quantum state $\vert \phi\rangle$ on a classical computer by directly storing its coefficient tensor $c_{\sigma_1, \sigma_2, \dots, \sigma_n}$ as a state vector~\cite{RaedtIto2007,SmelyanskiyGuzik2016,HanerSteiger2017,PednaultWisnieff2017}. As a result the amount of memory required to store an $n$-qubit quantum state grows exponentially with $n$. To given some explicit numbers of the memory complexity using the Schr$\ddot{\text{o}}$dinger simulator (we assume that single precision complex numbers are used), it takes around $1$ GB memory for $n=27$, around $0.5$ TB memory for $n=36$, and around $2$ PB memory for $n=48$ which is already beyond the reach of most of the existing supercomputers in the world! In fact till now  the largest classical simulation based on the Schr$\ddot{\text{o}}$dinger simulator has only reached $45$ qubits which uses about $0.5$ PB memory~\cite{HanerSteiger2017}.

For the Schr$\ddot{\text{o}}$dinger simulator, the initialization is straightforward since the full quantum state is stored in memory. One can simply set the amplitude corresponding to the computational basis $\vert 0, \dots, 0\rangle$ to $1$ and the amplitudes for the rest to be $0$. The quantum gate operations on the state vector can be simulated by following their mathematical definitions. For example a single-qubit gate operation acting on the $i$-th qubit, denoted as $Q^{\sigma_i'}_{\sigma_i}$, can be simulated as
\begin{align}\label{eq:sa_onebody}
c_{\sigma_1, \dots, \sigma_i', \dots, \sigma_n} = \sum_{\sigma_i} Q_{\sigma_i}^{\sigma_i'} c_{\sigma_1, \dots, \sigma_i, \dots, \sigma_n},
\end{align}
and a two-qubit gate operation acting on the two qubits $i$ and $j$ ($1 \leq i < j \leq n$), denoted as $Q^{\sigma_{i'}, \sigma_{j'}}_{\sigma_{i}, \sigma_j}$,  can be simulated as
\begin{align}\label{eq:sa_twobody}
c_{\sigma_1, \dots, \sigma_i', \dots, \sigma_j', \dots, \sigma_n} = \sum_{\sigma_i, \sigma_j} Q_{\sigma_i, \sigma_j}^{\sigma_i', \sigma_j'} c_{\sigma_1,\dots, \sigma_i, \dots, \sigma_j, \dots, \sigma_n}.
\end{align}
Eqs.(\ref{eq:sa_onebody},\ref{eq:sa_twobody}) are simply tensor contractions. However, directly implementing them as tensor contractions on classical computers would be extremely slow for large $n$ since tensor contraction would generally involve copies of the tensors (a standard implementation for tensor contraction is to first permute the tensor indices of each tensors and then do matrix multiplication, where the tensors will be copied during the tensor index permutation). Additionally, in actual implementation one usually prefers to implement the gate operations in an inplace fashion to reduce the memory usage, namely one directly uses the result to overwrite the input tensor for the quantum state instead of generating a new one. 

The memory-efficient way to implement Eqs.(\ref{eq:sa_onebody},\ref{eq:sa_twobody}) is to directly follow the definition of tensor contraction, that is, iterating over all the uncontracted indices and performing inplace operations for the contracted indices inside each iteration. 
Taking the two-qubit gate operation in Eq.(\ref{eq:sa_twobody}) as a concrete example, one could first group all the uncontracted indices into three contiguous sets to reduce the number of for loops, which results in a rank-$5$ tensor denoted as $\tilde{c}$:
\begin{align}\label{eq:grouping}
\tilde{c}_{(\sigma_1, \dots, \sigma_{i-1}), \sigma_i, (\sigma_{i+1}, \dots, \sigma_{j-1}), \sigma_j, (\sigma_{j+1}, \dots, \sigma_n)} =c_{\sigma_1, \dots, \sigma_n} ,
\end{align}
where the set of indices insider the parentheses mean that they are treated together as a single index. We note that in real coding nothing needs to be done in Eq.(\ref{eq:grouping}), it is just an equivalent way to view the same state vector.
The tensor $\tilde{c}$ have a shape $2^{i-1} \times 2 \times 2^{j-i-1} \times 2 \times 2^{n-j}$, and $Q^{\sigma_{i'}, \sigma_{j'}}_{\sigma_{i}, \sigma_j}$ can be applied onto $\tilde{c}$ using three for loops as shown in Algorithm.~\ref{alg:sv_two_body}.

\begin{algorithm}[H]
\caption{Implementation of a two-qubit gate operation $Q^{\sigma_{i'}, \sigma_{j'}}_{\sigma_{i}, \sigma_j}$ on an $n$-qubit quantum state stored as a state vector. Column-major storage and zero-based indexing are assumed for all the tensors.} \label{alg:sv_two_body}
\begin{algorithmic}[1]
\For{$s_3 = 0:2^{n-j}-1$}
\For{$s_2=0:2^{j-i-1}-1$}
\For{$s_1 = 0:2^{i-1}-1$}
	\State $\tilde{c}[s_1, \sigma_i', s_2, \sigma_j', s_3] = \sum_{\sigma_i, \sigma_j} Q^{\sigma_{i'}, \sigma_{j'}}_{\sigma_{i}, \sigma_j} \times \tilde{c}[s_1, \sigma_i, s_2, \sigma_j, s_3] $ \label{alg:inner}
\EndFor
\EndFor
\EndFor

\end{algorithmic}
\end{algorithm}

We note that the operation inside the for loops of Algorithm.~\ref{alg:sv_two_body} is simply a $4\times 4$ matrix times a vector of size $4$.
The overall floating number operations in Algorithm.~\ref{alg:sv_two_body} is thus $O(4\times 2^n)$. At the same time one needs to fetch all the elements of the state vector from the main memory into the CPU cache and then send the results back after the computation for at least once, namely the amount of memory access is $O(2\times 2^n)$. As a result the number of memory access compared to the number of floating point arithmetic is roughly $1/2$, while for modern classical computing hardware the memory bandwidth is usually much lower compared to the Flops efficiency. For example the V100 GPU has a memory bandwidth of $1.9$ TB/s but has a single-precision performance of $19.5$ TFlops (the former is only about $10\%$ of the latter). Therefore the efficiency of the Schr$\ddot{\text{o}}$dinger simulator is essentially bounded by the memory bandwidth. Nevertheless, one could slightly increase the compute density by unrolling the inner most for loop such that several vectors of size $4$ are processed simultaneously, that is, several matrix-vector multiplications are grouped together into a single matrix-matrix multiplication to optimize the usage the CPU cache (one would also make sure that the sizes of the matrices exactly fit into the CPU cache size for the best performance). The loop unrolling will also have an another advantage that one could fetch a bunch of numbers which are contiguous in memory simultaneously, which is an operation that is vectorized by most modern computing hardware. Additionally, the two outer for loops in Algorithm.~\ref{alg:sv_two_body} can be parallelized straightforwardly on a shared-memory architecture. The sparsity of gate operations could also be explored to further speed up the calculation, for example, the control-control-Z gate would only have a single non-trivial operation for its inner matrix-vector multiplication, which means that applying such a three-qubit gate would have a complexity which is only $1/8$ of the complexity of a dense single-qubit gate operation.

After all the gate operations have been performed, one obtains the final state $\vert \psi\rangle$ in the form of a state vector. No additional computations need to be done to obtain the amplitudes since they have been directly stored. The expectation value of a local observable $\hat{O}^i$ on the $i$-th qubit, whose corresponding tensor is $O^{\sigma'_i}_{\sigma_i}$, can be computed as
\begin{align}\label{eq:sa_localexp}
\langle \psi \vert \hat{O}^i \vert\psi\rangle =& \sum_{\sigma_1, \dots, \sigma_{i-1}, \sigma_i, \sigma'_i, \sigma_{i+1}, \dots, \sigma_n} c_{\sigma_1, \dots, \sigma_i, \dots, \sigma_n} \nonumber \\
& \times O^{\sigma'_i}_{\sigma_i} c_{\sigma_1, \dots, \sigma'_i, \dots, \sigma_n}^{\ast}.
\end{align}
One way to implement Eq.(\ref{eq:sa_localexp}) is to treat $\hat{O}^i$ as a single-qubit operator and apply it onto $\vert \psi\rangle$ to obtain $\hat{O}^i \vert \psi\rangle$, and then compute the dot product between $\vert \psi\rangle$ and $\hat{O}^i \vert \psi\rangle$. The advantage of this approach is that it could easily be generalized to more general observables such as a general Pauli string written as
\begin{align}
\hat{P} = M^1 \otimes M^2 \otimes \cdots \otimes M^n,
\end{align}
where each $M^j \in \{X, Y, Z, I_2\}$ ($X$, $Y$, $Z$ are Pauli matrices and $I_2$ is the $2\times 2$ identity matrix) is a local matrix, for which one can simply apply $M^j$ sequentially onto $\vert \psi\rangle$ to obtain $\hat{P} \vert \psi\rangle$ and then compute the dot product between $\vert \psi\rangle$ and $\hat{P} \vert \psi\rangle$. The disadvantage is that one has to use an additional copy of the quantum state to store the intermediate state $\hat{O}^i \vert \psi\rangle$ or $\hat{P} \vert \psi\rangle$.
A better approach to take is to directly evaluate Eq.(\ref{eq:sa_localexp}) using several for loops similar to Algorithm.~\ref{alg:sv_two_body}, for which no copy of the quantum state is required, however for a general Pauli string this approach would require $n$ for loops which is likely to be less efficient.

As long as one can compute the expectation of a local observable and can apply gate operations onto the quantum state for some classical simulator, one is able to faithfully simulate the quantum measurement process (the following procedures to simulate quantum measurement are valid for all classical simulators which support those operations). Assuming that one wants to simulate a local quantum measurement on the $i$-th qubit, one can first compute the expectation value of the local operator $\vert 1_i\rangle \langle 1_i\vert$ which is the probability that the $i$-th qubit is in state $1$, then one generates a random number $p$ according to the uniform distribution and outputs $1$ if $p < \langle \psi \vert 1_i\rangle \langle 1_i\vert\psi\rangle$ and outputs $0$ otherwise. The ``collapse'' the quantum state after measurement can be simulated by applying a local operator $\vert 1_i\rangle \langle 1_i\vert$ onto $\vert \psi\rangle$ if $1$ is obtained from the measurement, or applying $\vert 0_i\rangle \langle 0_i\vert$ if $0$ is obtained (the resulting quantum state should also be normalized by the corresponding probability). 



An important advantage of the Schr$\ddot{\text{o}}$dinger simulator is that its computational time scales strictly linearly against the number of gate operations (therefore it would also be beneficial to perform gate fusion at the beginning to absorb smaller gates into larger ones to reduce the total number of gate operations). There already exists several excellent libraries which are based on the Schr$\ddot{\text{o}}$dinger simulator, such as ProjectQ~\cite{projectq}, Yao~\cite{Yao}, Qiskit~\cite{Aleksandrowicz2019qiskit}, Quest~\cite{Quest}, Qulacs~\cite{Qulacs}.

\subsubsection{Schr$\ddot{\text{o}}$dinger-Feynman simulator}
In the Schr$\ddot{\text{o}}$dinger algorithm the whole quantum state is directly stored in memory. There is another extreme, the path-integral (Feynman) algorithm, which does not store any quantum state at all. In the Feynman algorithm, the amplitude of a specific bitstring $\vec{b}$ is computed as 
\begin{align}\label{eq:path_integral}
&\langle \vec{b}\vert \psi\rangle = \langle \vec{b}\vert \Qop^m \dots  \Qop^1 \vert 0^{\otimes n}\rangle \nonumber \\
=& \sum_{\vec{x}_1, \dots, \vec{x}_{m-1}} \langle \vec{b}\vert \Qop^m \vert \vec{x}_{m-1}\rangle\langle \vec{x}_{m-1}\vert \dots  \vert \vec{x}_1\rangle\langle \vec{x}_1\vert \Qop^1 \vert 0^{\otimes n}\rangle,
\end{align}
where each $\vec{x}_j$ denotes a specific computational basis. To evaluate Eq.(\ref{eq:path_integral}), all one needs are terms $\langle x_{j} \vert \Qop^{j} \vert x_{j-1}\rangle $ which are simply the entries of the tensor $Q^j$. Therefore the quantum state is never explicitly stored in the Feynman algorithm.
Each combination $\{\vec{x}_{1}, \dots, \vec{x}_{m-1}\}$ specifies a particular path and the total number of different paths (thus the time complexity) scales exponentially with the number of gate operations $m$, while in the meantime the memory complexity is negligible. The Feynman algorithm is of course impractical for any quantum circuit containing a few hundreds of gate operations due to its scaring time complexity. 

The Schr$\ddot{\text{o}}$dinger-Feynman algorithm aims to take advantages of both the Schr$\ddot{\text{o}}$dinger algorithm and the Feynman algorithm~\cite{MarkovBoixo2018}. Concretely, it divides the whole system into two subsystems and stores two quantum states corresponding to the two subsystems instead of storing the global quantum state. For the Sycamore quantum processor with $53$ qubits, one only needs to store two quantum states of $26$ and $27$ qubits respectively (one could of source increase the number of subsystems if there is not enough memory on the classical computer to store the quantum states of the subsystems). Once the partition is fixed, the gate operations acting on qubits which belong to one subsystem can be simply applied onto the quantum state corresponding to this subsystem. For gate operations which act on qubits from both subsystems, denoted as $\Qop^{g_l}$ ($1\leq l \leq m_c$ with $m_c$ the total number of such cross gates), one could first decompose it into the following tensor product form (the details of the decomposition will be shown later) 
\begin{align}\label{eq:spliting}
\Qop^{g_l} = \sum_{s_l} \hat{U}^{g_l}_{s_l} \otimes \hat{V}^{g_l}_{s_l}, 
\end{align}
where the operators $\hat{U}^{g_l}_{s_l}$ and $\hat{V}^{g_l}_{s_l}$ act on the two subsystems respectively for each $s_l$, then the amplitude in Eq.(\ref{eq:path_integral}) can be computed as
\begin{align}\label{eq:sfa}
\langle \vec{b}\vert \psi\rangle =& \sum_{s_1, \dots, s_{m_c}} \langle \vec{b}\vert \dots  \hat{U}^{g_2}_{s_2} \hat{V}^{g_2}_{s_2} (\Qop^{g_2-1} \dots \Qop^{g_1+1}) \times \nonumber \\ 
&\times \hat{U}^{g_1}_{s_1}  \hat{V}^{g_1}_{s_1} (\Qop^{g_1-1} \dots  \Qop^1) \vert 0^{\otimes n}\rangle.
\end{align}
The operators in the summand of Eq.(\ref{eq:sfa}) act either on one subsystem or the other subsystem, therefore one only needs to store the two quantum states of the two subsystems.
Compared to Eq.(\ref{eq:path_integral}), the total number of paths in Eq.(\ref{eq:sfa}) will only grow exponentially against the total number of cross gates, namely $m_c$, instead of $m$. One could also evaluate the expectation values of observables (local operators or Pauli strings) similar to Eq.(\ref{eq:sfa}), however the number of possible paths will generally be squared.

From Eq.(\ref{eq:sfa}), the different paths specified by $\{s_{1}, \dots, s_{m_c}\}$ can be computed in parallel with no data communication, therefore the Schr$\ddot{\text{o}}$dinger-Feynman algorithm is extremely friendly for large-scale parallelization on distributed architectures. Moreover, one could easily use the Schr$\ddot{\text{o}}$dinger-Feynman algorithm to simulate the simplified quantum circuits by removing a number of cross gates from the original quantum circuit. For these reasons it is used as the first classical benchmarking baseline to characterize \textit{quantum supremacy} for the Sycamore quantum processor~\cite{arute2019quantum}.

\subsubsection{Matrix product state based simulator}

Tensor network states based algorithms belong to an important class of algorithms which could often leverage the exponential memory requirement when simulating quantum algorithms. Generally, in tensor network states based simulators one represents the underlying quantum state as a tensor network which only consists of low-rank tensors, then the gate operations are simulated by updating the local tensors of the tensor network been acted on and quantum measurements could be simulated by contracting some tensor network. 

Tensor network states are originally introduced to solve quantum many-body problems, which are mostly used to approximate the low-energy states of quantum many-body Hamiltonians~\cite{Orus2014}. Depending on the structure of the physical problem, various tensor network ansatz have been used to represent the underlying quantum state, such as matrix product states (MPS)~\cite{Vidal2003,Vidal2004,DaleyVidal2004}, projected entangled pair states (PEPS)~\cite{VerstraeteCirac2004}, tree tensor network (TTN)~\cite{ShiVidal2006}, multiscale entanglement renormalization ansatz (MERA)~\cite{Vidal2007}. In particular, MPS and PEPS are designed for one-dimensional and two-dimensional quantum many-body systems respectively~\cite{Schollwock2011,Orus2014}, and have achieved great success therein for the past thirty years. 

Out of all tensor network states, MPS is a very special and important one, mostly because that the errors in MPS based algorithms can be often well controlled and that it allows very efficient and stable ground state searching~\cite{White1992,White1993} and time evolution algorithms~\cite{Vidal2003,Vidal2004,DaleyVidal2004}. Therefore although MPS is mostly designed to represent the ground states of one-dimensional quantum many-body systems, it could also be used as a general-purpose ansatz similar to the state vector for arbitrary quantum states (in the general case the cost of MPS could also grow exponentially)~\cite{McCaskeyHumble2018}. MPS based simulator has been used to simulate Shor's algorithm up to $60$ qubits~\cite{WangHollenberg2017,DangHollenberg2019}, to simulate (noisy) optical quantum circuits~\cite{HuangGuo2019}, to simulate the Sycamore RQCs~\cite{ZhouWaintal2020}, and to simulate VQE algorithms for quantum chemistry problems with up to $100$ qubits~\cite{ShangLi2022}. Some typical simulations of VQE using the Schr$\ddot{\text{o}}$dinger simulator and the MPS based simulator are listed in TABLE.~\ref{tab:vqe}.

\begin{table}[!htb]
\tabcolsep 1mm \caption{Typical simulations of molecular systems with classical simulators. The number of atoms ($N_a$), number of qubits ($N_q$), the estimated number of CNOT gates ($N_{\rm CNOT}$), and the algorithms (Alg.) used are listed for comparison. SA is a shorthand for the Schr$\ddot{\text{o}}$dinger algorithm. 
}
\begin{center}
\begin{tabular}{c|c|c|c|c|c}\hline \hline
Work & System & $N_a$ & $N_q$ & $N_{\rm CNOT}$ & Alg.  \\
\hline
Microsoft QDK~\cite{bylaska2021quantum}  & H$_2$ & 2 & 4 & 64 & \multirow{10}{*}{SA} \\ 
Cirq~\cite{YalSenGun21}  & CH$_2$O &  4 & 6 & 272  \\ 
Qulacs~\cite{ManKhaYam21}  &He crystal & 1 & 8 & $1.6\times10^3$ \\ 
Qiskit~\cite{XiaKai20}   &N$_2$ & 2 & 16 & $2.7\times10^4$ \\ 
Yao.jl~\cite{LiLv2022}   & C$_{18}$ & 18 & 16 & $5.4\times10^4$ \\ 
VQEChem~\cite{LiuWanLi20}   & H chain & 2 & 16 & $5.4\times10^4$ \\ 
 QCQC~\cite{FanLiuLi21}  & Si crystal & 2 & 16 & $1.1\times10^5$ \\ 
 Tequlia~\cite{KotSchTam21}  & BH & 2 & 22 & $1.6\times10^5$ \\  
 HiQ~\cite{CaoHuZha21}  & C$_2$H$_4$ & 6 & 28 & $8.6\times10^5$ \\ 
 iQCC-VQE~\cite{RyaIZmGen21}  & Ir(ppy)$_3$ & 61 & 56 & $\sim 60$ \\ 
\hline
Differentiable MPS~\cite{XuGuo2022} &  CO$_2$      & 3 &  30 & $3.7\times10^4$      \\
MPS-VQE~\cite{ShangLi2022} &  H$_2$      & 2 &  92 & $1.4\times10^5$ & \multirow{2}{*}{MPS}  \\
MPS-VQE~\cite{ShangLi2022} &  C$_2$H$_6$      & 8 &  32 & $4.4\times10^5$      \\
\hline \hline 
\end{tabular}
\end{center}
\label{tab:vqe}
\end{table}

To simulate quantum circuits based on the MPS representation of the quantum state, one could directly use an MPS based time evolution algorithm such as time evolving block decimation (TEBD)~\cite{Vidal2004}. In the following we will introduce a variant of TEBD to enhance the stability of the algorithm. 
MPS rewrites the rank-$n$ coefficient tensor $c_{\sigma_1, \dots, \sigma_n}$ in Eq.(\ref{eq:phi}) as the product of a chain of rank-$3$ tensors as
\begin{align}\label{eq:mps}
c_{\sigma_1, \sigma_2, \dots, \sigma_n} = \sum_{a_0,\dots,a_{n}}  
B^{\sigma_1}_{a_0, a_1} B^{\sigma_2}_{a_1,a_2} \ldots  B^{\sigma_{n}}_{a_{n-1},a_n},
\end{align}
where $\sigma_j \in \{0,1\}$ is the ``physical'' index and $a_j$ the ``auxiliary'' index. The indices $a_0$ and $a_n$ at the boundaries are trivial indices added for notational convenience. The size of an MPS can be conveniently characterized by the largest size of the auxiliary indices
\begin{align}
D = \max_{1\leq j \leq n-1}\left(\dim (a_j)\right),
\end{align}
which is usually referred to as the \textit{bond dimension} of MPS. In principle, Eq.(\ref{eq:mps}) is able to represent any quantum states exactly if $D$ increases exponentially. A quantum state is said to be efficiently representable as an MPS if $D$ is nearly a constant as $n$ grows.
Generally, for MPS based algorithms the memory complexity scales as $O(nD^2)$ and the computational complexity scales as $O(nD^3)$.

Slightly different from the TEBD algorithm, we assume that the MPS is prepared in a right-canonical form, where each site tensor satisfies the right-canonical condition
\begin{align}\label{eq:rightcanonicalcondition}
\sum_{\sigma_j, a_j} (B^{\sigma_j}_{a_{j-1}', a_j})^{\ast} B^{\sigma_j}_{a_{j-1}, a_j} = \delta_{a_{j-1}, a_{j-1}'}.
\end{align}
We also assume that the bipartition singular vectors, denoted as $\lambda_{a_j}$,  which are the Schmidt numbers when splitting the system into two subsystems from qubits $1$ to $j$ and from qubits $j+1$ to $n$, are also stored. 
The initial state $\vert 0^{\otimes n}\rangle$ can be easily written as an MPS with $D=1$, for which each site tensor satisfies
\begin{align}
B^{\sigma_j=0}_{a_{j-1}=0, a_j=0} = 1, B^{\sigma_j=1}_{a_{j-1}=0, a_j=0} = 0,
\end{align}
(In fact any separable state can be written as an MPS with $D=1$ similarly). The corresponding singular vectors are simply initialized as $\lambda_{a_j=0}=1$. It is straightforward to verify that the site tensors initialized in this way satisfy Eq.(\ref{eq:rightcanonicalcondition}) and that the singular vectors are the correct Schmidt numbers.
The gate operations are then applied sequentially onto the MPS. 
A single-qubit gate operation acting on the $j$-th qubit, denoted as $Q^{\sigma_j'}_{\sigma_j}$, changes the $j$-th site tensor of the MPS as
\begin{align}
\tilde{B}^{\sigma_j'}_{a_{j-1},a_{j}}  = \sum_{\sigma_j} Q^{\sigma_j'}_{\sigma_j} B^{\sigma_j}_{a_{j-1},a_{j}}.
\end{align}
Since $Q^{\sigma_j'}_{\sigma_j}$ is unitary, it is straightforward to verify that $\tilde{B}^{\sigma_j'}_{a_{j-1},a_{j}}$ will satisfy Eq.(\ref{eq:rightcanonicalcondition}) as long as $B^{\sigma_j}_{a_{j-1},a_{j}}$ does.
For a nearest-neighbour two-qubit gate $Q^{\sigma_j',\sigma_{j+1}'}_{\sigma_j,\sigma_{j+1}}$ acting on two qubits $j$ and $j+1$ (the $j$-th bond), it will change the both the $j$ and $j+1$-th site tensors of the MPS. To simulate a two-qubit gate operation while preserving the right canonical property of the underlying MPS, we use the technique first introduced in Ref.~\cite{Hastings2009}, which is shown as follows. First we contract the two site tensors $B^{\sigma_j}_{a_{j-1},a_j}$ and $B^{\sigma_{j+1}}_{a_j,a_{j+1}}$ with $Q_{\sigma_j,\sigma_{j+1}}^{\sigma'_j,\sigma'_{j+1}}$ to get a two-site tensor
\begin{align}
C_{a_{j-1},a_{j+1}}^{\sigma_j',\sigma'_{j+1}}=\sum_{a_j, \sigma_j,\sigma_{j+1}} Q_{\sigma_j,\sigma_{j+1}}^{\sigma'_j,\sigma'_{j+1}} B^{\sigma_j}_{a_{j-1},a_j} B^{\sigma_{j+1}}_{a_j,a_{j+1}}.
\end{align}
Then we contract $C_{a_{j-1},a_{j+1}}^{\sigma'_j,\sigma'_{j+1}}$ with the singular vector $\lambda_{a_{j-1}}$ at the $j-1$-th bond to get a new two-site tensor as
\begin{align}
\tilde{C}_{a_{j-1},a_{j+1}}^{\sigma'_j,\sigma'_{j+1}}= \lambda_{a_{j-1}} C_{a_{j-1},a_{j+1}}^{\sigma'_j,\sigma'_{j+1}}.
\end{align}
Now we perform singular value decomposition (SVD) on the tensor $\tilde{C}_{a_{j-1},a_{j+1}}^{\sigma'_j,\sigma'_{j+1}}$ and get
\begin{equation}\label{eq:mpssvd}
\SVD(\tilde{C}_{a_{j-1},a_{j+1}}^{\sigma'_j,\sigma'_{j+1}})=\sum_{a_j} U^{\sigma'_j}_{a_{j-1},a_j} \tilde{\lambda}_{a_j} V^{\sigma'_{j+1}}_{a_j,a_{j+1}},
\end{equation}
during which we will also truncate the small singular values below a certain threshold $\epsilon$ or simply reserve the $D$ largest singular values if the singular values larger than $\epsilon$ is more than $D$. Here we note that this truncation step is the only step that would introduce errors in the algorithm, and such errors could be made arbitrarily small in principle if one sets $\epsilon$ to be very small and $D$ to be very large (the actual size of $D$ required is of course problem-dependent). After the SVD, the updated site tensors $\tilde{B}^{\sigma'_j}_{a_{j-1},a_j}$ and $\tilde{B}^{\sigma'_{j+1}}_{a_j,a_{j+1}}$ can be obtained as
\begin{align}
\tilde{B}^{\sigma'_j}_{a_{j-1},a_j} &= \sum_{\sigma'_{j+1},a_{j+1}} C_{a_{j-1},a_{j+1}}^{\sigma'_j,\sigma'_{j+1}} \left(V^{\sigma'_{j+1}}_{a_j,a_{j+1}}\right)^{\ast}; \\
\tilde{B}^{\sigma'_{j+1}}_{a_j,a_{j+1}} &= V^{\sigma'_{j+1}}_{a_j,a_{j+1}}.
\end{align}
To verify that $\tilde{B}^{\sigma'_j}_{a_{j-1},a_j}$ and $\tilde{B}^{\sigma'_{j+1}}_{a_j,a_{j+1}}$ are indeed the solutions, we can see that $\sum_{a_j} \tilde{B}^{\sigma'_j}_{a_{j-1},a_j} \tilde{B}^{\sigma'_{j+1}}_{a_j,a_{j+1}} = C_{a_{j-1},a_{j+1}}^{\sigma'_j,\sigma'_{j+1}}$ up to truncation errors. The reason that one does not directly do SVD on $C_{a_{j-1},a_{j+1}}^{\sigma'_j,\sigma'_{j+1}}$ to get the updated site tensors is to preserve the right canonical properties of the solutions, in fact it could be verified that $\tilde{B}^{\sigma'_j}_{a_{j-1},a_j}$ and $\tilde{B}^{\sigma'_{j+1}}_{a_j,a_{j+1}}$ both satisfy Eq.(\ref{eq:rightcanonicalcondition}) ($\tilde{B}^{\sigma'_{j+1}}_{a_j,a_{j+1}}$ is right canonical by construction from Eq.(\ref{eq:mpssvd})).
The new singular vector $\tilde{\lambda}_{\alpha_j}$ contains the correct Schmidt numbers on bond $j$ of the updated MPS and is used to replace the old $\lambda_{\alpha_j}$. Similar to the Schr$\ddot{\text{o}}$dinger simulator, one can also absorb the single-qubit gate operations into two-qubit gate operations before hand, but the speedup will not be as significant since the complexity of a single-qubit gate operations is negligible compared to a two-qubit gate for MPS based simulator. A three-qubit gate which acts on three contiguous qubits could also be simulated similarly, where the three site tensors as well as the two singular vectors in between will be updated. 
A non-nearest-neighbour gate operation could be simulated by decomposing it into a series of nearest-neighbour gate operations using the SWAP gate. 

With a right-canonical MPS, there is a very efficient way to compute the expectation of a local operator or a Pauli string. For example, the expectation value of a single-qubit observable $\hat{O}^j$ on the $j$-th qubit can be simply computed as
\begin{align}\label{eq:expec1}
\langle \psi\vert \hat{O}^j\vert \psi\rangle = \sum_{a_{j-1},a_j,\sigma_j,\sigma'_j} \lambda^2_{a_{j-1}}O_{\sigma_j}^{\sigma'_j} B^{\sigma'_j}_{a_{j-1},a_j}(B^{\sigma_j}_{a_{j-1},a_j})^{\ast},
\end{align}
and the expectation value of a generic two-qubit observable $\hat{O}^{i,j}$ on the $i$ and $j$-th qubits ($i < j$) can be computed as
\begin{align}\label{eq:expec2}
\langle \psi\vert \hat{O}^{i,j}\vert \psi\rangle = \sum_{a_{j:i-1},\sigma_{j:i},\sigma'_{j:i}} &\lambda^2_{a_{i-1}}O^{\sigma_i', \sigma'_j}_{\sigma_i, \sigma_j} B^{\sigma'_i}_{a_{i-1},a_i}(B^{\sigma_i}_{a_{i-1},a_i})^{\ast} \times \nonumber \\ 
&\dots \times B^{\sigma'_j}_{a_{j-1},a_j}(B^{\sigma_j}_{a_{j-1},a_j})^{\ast},
\end{align}
where we have used $x_{j:i} = \{x_i, x_{i+1}, \dots, x_j\}$ as an abbreviation for a list of indices.
The expectation value of a general $n$-qubit Pauli string could be computed similarly, with a complexity that scales as $O(nD^3)$. 

After one obtains the quantum state $\vert \psi\rangle$ in MPS form, the amplitude of a given bitstring $\vec{b}$ can be computed as
\begin{align}
\langle \vec{b}\vert \psi\rangle = \sum_{a_0, \dots, a_n} B^{\sigma_1 = b_1}_{a_0, a_1} B^{\sigma_2 = b_2}_{a_1, a_2} \dots B^{\sigma_n = b_n}_{a_{n-1}, a_n},
\end{align}
for which only $O(n)$ matrix-vector multiplications need to be performed. Given the ability to efficiently compute the amplitude, one could also simulate the sampling problem by further using a sampling algorithm. However, sampling an MPS can be done exactly and more efficiently~\cite{HanZhang2018}. Given an MPS in the right-canonical form as in Eq.(\ref{eq:mps}), the reduced density matrix for qubits from $1$ to $j$ can be simply computed as
\begin{align}\label{eq:reducedm}
\rho^{\sigma_1', \dots, \sigma_j'}_{\sigma_1, \dots, \sigma_j} = &\sum_{a_{j-1:0}, a_{j-1:0}', a_j} (B^{\sigma_1'}_{a_0', a_1'})^{\ast} B^{\sigma_1}_{a_0, a_1} \times \cdots \nonumber \\ 
&\times (B^{\sigma_{j-1}'}_{a_{j-2}', a_{j-1}'})^{\ast} B^{\sigma_{j-1}}_{a_{j-2}, a_{j-1}} (B^{\sigma_j'}_{a_{j-1}', a_j})^{\ast} B^{\sigma_j}_{a_{j-1}, a_j}.
\end{align}
To generate a single sample (bitstring), one could first compute the reduced density matrix of the first qubit, $\rho^{\sigma_1'}_{\sigma_1}$, using Eq.(\ref{eq:reducedm}), and then one could perform a local sampling on $\rho^{\sigma_1'}_{\sigma_1}$ to get the sampling output for the first qubit, denoted as $\nu_1$. After that, the reduced density matrix of the second qubit conditioned on the sampling outcome of the first qubit, denoted as $\rho^{\sigma_2'}_{\sigma_2}|_{\nu_1}$, can be computed as
\begin{align}
\rho^{\sigma_2'}_{\sigma_2}|_{\nu_1} = \frac{1}{p(\nu_1)} \sum_{a_{1:0}, a_{1:0}', a_2} (B^{\sigma_1' = \nu_1}_{a_0', a_1'})^{\ast} B^{\sigma_1 = \nu_1}_{a_0, a_1} (B^{\sigma_2'}_{a_1', a_2})^{\ast} B^{\sigma_2}_{a_1, a_2} ,
\end{align}
where $p(\nu_1) = \rho^{\sigma_1' = \nu_1}_{\sigma_1 = \nu_1} $ is the probability to get $\nu_1$ when measuring the first qubit. By a local sampling on $\rhoop^{\sigma_2'}_{\sigma_2}|_{\nu_1}$ on could obtain the sampling outcome of the second qubit, denoted as $\nu_2$. Repeating this process from left to right, one can obtain one sample $\vec{\nu} = \{\nu_1, \dots, \nu_n\}$. The complexities of computing one amplitude and generating one sample both scale as $O(nD^2)$.

\subsubsection{Projected entangled pair states based simulator}
PEPS is a natural extension of MPS to two-dimensional systems~\cite{VerstraeteCirac2004}. Currently, the mainstream quantum processors all have a two-dimensional geometry, therefore PEPS could be a natural tensor network ansatz to represent quantum states generated on those quantum processors. 
A PEPS on a square lattice can be written as     
\begin{align}
\vert\psi\rangle = \sum_{\sigma_1, \sigma_2, \dots, \sigma_n}\mathcal{F}(A_1^{\sigma_1}A_2^{\sigma_2}\dots A_n^{\sigma_n}) \vert \sigma_1, \sigma_2, \dots, \sigma_n\rangle,
\end{align}
where each $A_j^{\sigma_j}$ is a short hand for the rank-$5$ site tensor $A_{l,r,u,d}^{\sigma_j}$ with a physical index $\sigma_j$ and $4$ auxiliary indices $l,r,u,d$ representing the connecting of this tensor with the ones on the left, right, upper and lower respectively. $\mathcal{F}(\dots)$ means to contract the pairs of connected auxiliary indices. Similar to MPS, the size of a PEPS can also be characterized by its \textit{bond dimension} defined as the largest size of its auxiliary indices, which we will still denote as $D$. The memory complexity to store a PEPS scales as $O(nD^4)$.


The initial state $\vert 0^{\otimes n}\rangle$ can be easily constructed as a PEPS with $D=1$, with site tensors satisfying
\begin{align}
A^{\sigma_j=0}_{l=0,r=0,u=0,d=0} = 1, A^{\sigma_j=1}_{l=0,r=0,u=0,d=0} = 0.
\end{align}
To simulate the gate operations on a PEPS, one needs to properly update the site tensors of the PEPS. However for a PEPS the updating is much less straightforward than for an MPS. In the context of quantum many-body physics, many sophisticated methods have been proposed to approximately update the site tensors of a PEPS after a gate operation has been applied on the PEPS, such that the bond dimension of the updated PEPS is bounded by a given bond dimension $D$~\cite{JiangXiang2008,CorbozVidal2010,JordanCirac2008,LubaschBanuls2014b}. Here we will only present the method used in Ref.~\cite{guo2019general} to simulate RQCs, which allows $D$ to grow after each gate operation and is essentially exact.


For single-qubit gate operation $Q_{\sigma_j}^{\sigma_j'}$ acting on the $j$-th qubit, it only updates the $j$-th site tensor of the PEPS as
\begin{align}
\tilde{A}^{\sigma_j}_{l,r,u,d} = \sum_{\sigma_j'} Q_{\sigma_j}^{\sigma_j'} A^{\sigma_j'}_{l,r,u,d}.
\end{align}
For a two-qubit gate operation $Q_{\sigma_i, \sigma_j}^{\sigma_i', \sigma_j'}$, one first splits it into the product of two rank-$3$ tensors using SVD as in Eq.(\ref{eq:spliting}). Concretely, this can be done by first decomposing it as
\begin{align}\label{eq:svd}
\SVD(Q_{\sigma_i, \sigma_j}^{\sigma_i', \sigma_j'}) = \sum_s \tilde{U}^{\sigma'_i}_{\sigma_i, s} \lambda_s \tilde{V}^{\sigma'_j}_{s, \sigma_j},
\end{align}
then Eq.(\ref{eq:spliting}) can be satisfied by choosing the tensors $U$ and $V$ as
\begin{align}
U^{\sigma'_i}_{\sigma_i, s} &= \sum_s \tilde{U}^{\sigma'_i}_{\sigma_i, s} \sqrt{\lambda_s} ; \\
V^{\sigma'_j}_{s, \sigma_j} &= \sum_s \sqrt{\lambda_s} \tilde{V}^{\sigma'_j}_{s, \sigma_j}.
\end{align}
Assuming that the qubits $i$ and $j$ are a pair of nearest-neighbour qubits in the horizontal direction, then the site tensors $A^{\sigma_i}_i$ and $A^{\sigma_j}_j$ can be updated as
\begin{align}
\tilde{A}^{\sigma_i'}_{l,r',u,d} &= \sum_{\sigma_i} U^{\sigma'_i}_{\sigma_i, s} A^{\sigma_i}_{l,r,u,d}; \\
\tilde{A}^{\sigma_j'}_{l',r,u,d} &= \sum_{\sigma_j} V^{\sigma'_j}_{s, \sigma_j} A^{\sigma_j}_{l,r,u,d},
\end{align}
where we have used $r' = (r, s)$ and $l' = (s, l)$. We note that the size of the right auxiliary index of the site tensor $A^{\sigma_i}_i$ (and the left auxiliary index of the site tensor $A^{\sigma_j}_j$) will be enlarged by $\chi$ times after updating where $\chi$ is the number of nonzero Schmidt number in Eq.(\ref{eq:svd}).
A two-qubit gate operation on a vertical pair of qubits can be done similarly by updating the auxiliary indices in the vertical directions ($d$ and $u$) instead.

After one obtains the final quantum state $\vert \psi\rangle$ as a PEPS, the amplitude of a specific bitstring $\vec{b}$ can be computed as
\begin{align}\label{eq:pepsamp}
\langle \vec{b}\vert \psi\rangle = \mathcal{F}(A_1^{\sigma_1 = b_1} A_2^{\sigma_1 = b_2} \cdots A_n^{\sigma_n = b_n}),
\end{align}
which amounts to contracting a two-dimensional tensor network with no output indices. For an arbitrary PEPS, the contraction of Eq.(\ref{eq:pepsamp}) is a hard problem. Nevertheless in the context of quantum many-body physics there has been quite a few approximate methods to contract a two-dimensional tensor network similar to Eq.(\ref{eq:pepsamp}), which are fairly efficient and accurate~\cite{LevinNave2007,VerstraeteCirac2008,JordanCirac2008,OrusVidal2009,XieXiang2009,ZhaoXiang2010,XieXiang2012}. However the two dimensional tensor networks formed in quantum many-body problems usually has a much clear physical origin, such as resulting from computing a local observable on a PEPS which represents the ground state of a local Hamiltonian. When applied to computing amplitudes of RQCs, the physical context is not as clear, and the approximate methods to contract Eq.(\ref{eq:pepsamp}) could easily result in significant errors. Therefore in Ref.~\cite{guo2019general} Eq.(\ref{eq:pepsamp}) is contracted exactly.
For a rectangular lattice with size $n_h \times n_v$, it is estimated that the time complexity $\mathcal{C}^t$ of directly contracting Eq.(\ref{eq:pepsamp}) is 
\begin{align}
\mathcal{C}^t = (n_h -2)(n_v - 2) D^{\min(n_h, n_v) + 3}.
\end{align}
One could also compute local observables similar to Eq.(\ref{eq:pepsamp}), however the resulting two-dimensional tensor network will have a bond dimension $D^2$ instead, for which one may have to resort to those approximate methods. The PEPS simulator could also be easily adapted for other two-dimensional geometries~\cite{guo2021verifying}.
Additionally, the algorithms used in Ref.~\cite{VillalongaMandra2018,VillalongaMandra2019} are essentially equivalent to the PEPS algorithm presented here, although they are not presented using the PEPS language. 

Currently, PEPS based simulator has only been applied to simulate the RQC sampling problem. For RQC on a square lattice, the largest simulation scale till now is reached by using the PEPS based simulator, where one amplitude of a $10\times 10$ RQC with depth $40$ is computed within $60$ hours~\cite{liu2021closing}.

Besides the RQC sampling problem, PEPS based simulator is also promising for simulating other quantum algorithms which are designed for two-dimensional architectures, such as VQE which uses a physically motivated ansatz. In such situations one may also explore those approximate PEPS updating and contraction algorithms to significantly speedup the calculations. 


\subsubsection{Tensor network contraction based simulator}

Instead of representing the quantum state as a tensor network state and then applying the gate operations sequentially onto it, the quantum circuit itself, as in Eq.(\ref{eq:circuitevolve}), can be naturally viewed as a large tensor network, in which the initial state $\vert 0^{\otimes n}\rangle$ is simply the tensor product of $n$ vectors $\vert 0\rangle = [1, 0]$, each single-qubit gate operation is a rank-$2$ tensor and each two-qubit gate operation is a rank-$4$ tensor~\cite{MarkovShi2008}. There are $n$ uncontracted (output) indices in the end of the tensor network which correspond to the final quantum state. Contracting this tensor network one would directly obtain the final quantum state $\vert \psi\rangle$ as a state vector, which is of source only feasible for very small problems. Depending on the specific task to do, the tensor network could be simplified. For example for computing the amplitude of a bitstring, one would project each output index into a particular qubit state of the bitstring, resulting in a new tensor network with no output indices. For computing a local observables, one needs to contract a new tensor network where the observable is sandwiched between the two tensor networks formed by $\vert \psi\rangle$ and its conjugate $\langle \psi\vert$. In either case, the solution of the problem could be obtained by contracting a large tensor network with no or very few output indices.

Once the problem has been equivalently transformed into a tensor network contraction (TNC) problem, the key for the performance is to identify a near optimal tensor network contraction order (TNCO). 
Here we note that there is an important difference between the TNC algorithm and the tensor network states based algorithms for simulating quantum circuits, although in both cases the target problem will eventually be converted into the problem of contracting a tensor network. Taking the task of computing one amplitude as an example, for a quantum circuit which only contains single-qubit and two-qubit gate operations, the total number of tensors of the tensor network formed in the TNC algorithm will be approximately equal to the number of two-qubit gates, while for MPS or PEPS algorithms the number of tensors in the final tensor network is strictly equal to the total number of qubits (but in the latter case each tensor could be very large). In fact, the PEPS algorithm can also be viewed as a specialized instance of the TNC algorithm by choosing a particular TNCO: one first contracts all the tensors in the time direction, which results in a two-dimensional tensor network in the spatial directions (assuming that the quantum processor has a two-dimensional geometry), and then contracts the resulting two-dimensional tensor network. For PEPS based simulator since the resulting two-dimensional tensor network has a much smaller number of tensors and also has a very regular structure, one could easily identify an optimal TNCO. Therefore when the tensor network is contracted exactly, the TNC based simulator has the maximum flexibility to choose an optimal TNCO which can result in lower computational complexity compared to the PEPS based simulator in principle.

The TNC algorithm is powered by the highly efficient heuristic methods to search for near optimal TNCOs developed in recent years~\cite{GrayKourtis2020,HuangChen2021,PanZhang2021}. When applied to RQCs on Sycamore(-like) quantum processors, the TNC algorithm gives much lower computational complexity than PEPS. The central idea of those heuristic methods is to use existing graph partitioning methods to first divide the large tensor network with hundreds of tensors (corresponding to a graph with hundreds of nodes) into many smaller tensor networks such that one could easily compute the TNCO for each smaller tensor network and then assembly then together into a large TNCO for the original tensor network~\cite{GrayKourtis2020}. For very large tensor networks, slicing is also an indispensable technique which ``cuts'' a number of edges in the original tensor network such that the original tensor network is transformed into the sum of many smaller tensor networks with fewer edges. The slicing technique could significantly reduce the memory usage since only simpler tensor networks are being contracted, and if used properly, it could only result in moderate computational overhead compared to directly contracting the original tensor network~\cite{HuangChen2021}. By understanding the mechanism for the computational overhead induced by slicing, a lifetime based heuristic method is proposed to systematically find a near optimal slicing scheme~\cite{ChenYang2022}.

For the task of computing one amplitude or a correlated batch of amplitudes for the hardest Sycamore RQC (with a circuit depth $20$), various works have all given a computational complexity of the order $O(10^{18})$. While such a complexity scale can be accomplished on a exascale supercomputer within seconds in principle, the fastest record till now takes around $150$ seconds to compute a batch of correlated amplitudes~\cite{ChenYang2022}. For computing a number of uncorrelated amplitudes (or generating uncorrelated samples), a recent work develops a multiple-amplitude TNC algorithm which caches intermediate computations to reduce the overall complexity, and based on this method millions of uncorrelated amplitudes for the Sycamore RQCs with depths up to $16$ are successfully computed~\cite{KalachevYung2021}. A sparse-state method is also proposed to compute a large number of amplitudes with a limited fidelity, which is used to generate one million uncorrelated samples for the hardest Sycamore RQC using $512$ GPUs for $15$ hours, with a higher fidelity than in the quantum supremacy experiment~\cite{PanZhang2021b}. For RQC on a square lattice the largest reported simulation scale using the TNC algorithm reaches $9\times 9$ qubits with a circuit depth $40$~\cite{ChenShi2018}. Typical large-scale classical simulations of RQCs have also been summarized in TABLE.~\ref{tab:rqc}.

\begin{table*}[!htb]
\centering
\caption{
Typical large-scale classical simulations of random quantum circuits, where the results for two different circuit geometries, rectangular (Rect.) lattice and Sycamore are listed. For rectangular lattice the circuit size is shown by $height\times width$. The column \#Amplitudes show that number of amplitudes been computed. QTIA is a shorthand for a quantum teleportation inspired algorithm used in Ref.~\cite{ChenPan2020}.
}
\label{tab:rqc}
\begin{tabular}{lclcccc}
\hline \hline
Hardware & RQC (Two-qubit gate) & Circuit Size (depth) & Time & Fidelity & \#Amplitudes & Algorithm\\
\hline
$8,192$ nodes of Cori II~\cite{HanerSteiger2017} & Rect. (CZ) &  $5\times 9$  (1+25+1)  &   10 minutes & 100\% & all & SA \\
$4,096$ nodes of Blue Gene/Q~\cite{PednaultWisnieff2017} & Rect. (CZ)  & $7\times 7$   (1+27+1)  &   2 days  & 100\% & 1 & SA \\
$32,768$ nodes of Sunway TaihuLight~\cite{LiYang2019} & Rect. (CZ)  &   $7\times 7$   (1+39+1)  &   4.2 hours & 100\% & all & TNC \\
$4,600$ Summit~\cite{VillalongaMandra2019} & Rect. (CZ)  &   $7\times 7$   (1+40+1)  &   2.4 hours & 0.5\% & $1.01\times 10^6$ & PEPS \\
$4,096$ nodes of Tianhe-2A~\cite{guo2019general} & Rect. (CZ) & $7\times 7$ (1+40+1) & 31 minutes & 100\% &1 & PEPS \\
$131,072$ nodes of Alibaba~\cite{ChenShi2018} & Rect. (CZ) & $9\times 9$ (40) & 13 hours & 100\% &1 & TNC \\
$16,384$ nodes of Sunway TaihuLight~\cite{ChenPan2020} & Rect. (CZ) & $8\times 125$ (1+40+1) & 131.6 minutes & 100\% &1 & QTIA \\
$60$ GPUs~\cite{PanZhang2021} & Sycamore (FSIM) & $53$ (20) & 5 days & 100\% &1 & TNC \\
$512$ GPUs~\cite{PanZhang2021b} & Sycamore (FSIM) & $53$ (20) & 15 hours & 0.37\% & $10^6$  & TNC \\
$107,520$ nodes of New Sunway~\cite{liu2021closing} & Sycamore (FSIM) & $53$ (20) & 304 seconds & 100\% &1 & TNC \\
$107,520$ nodes of New Sunway~\cite{liu2021closing} & Rect. (CZ) & $10\times 10$ (1+40+1) & 60 hours & 100\% &1 & PEPS \\
\hline\hline
\end{tabular}
\end{table*}


\subsection{Simulating noisy quantum circuits}
In certain situations it is also important to simulate noisy quantum circuits, for example, to understand the behaviors of quantum algorithms under certain level of noises, to test quantum noise models for quantum computers or to test quantum error mitigations schemes. 

In a noisy quantum circuit, the underlying quantum state should be represented as a mixed state (density operator), and the quantum gate operations should be generalized to quantum channels which are super operators acting on the density operator. To simulate a noisy quantum circuit classically, two very different but in principle equivalent approaches can be used.
In the first approach, one simply inserts random Pauli errors into the original noiseless quantum circuit with certain probabilities~\cite{Nielsen2002QuantumComputation}. By averaging the results over a large number of error realizations, one could reproduce the result of a noisy quantum circuit. Since for each noise realization one only needs to deal with a quantum circuit which only contains unitary quantum gate operations, the classical simulators for noiseless quantum circuits can be directly used. 

In the second approach, one directly simulates the noisy quantum circuit by representing the quantum state as a density operator, then one applies the quantum channels onto the density operator and performs quantum measurements based on the density operator in the end. Compared to the first approach, the second approach is free of the errors resulting from a finite number of noise realizations. However, the memory cost of the second approach could be significantly larger since the size of a density operator is the square of a pure state in general. Although a noisy quantum circuit seems very different from a noiseless one mathematically, one can equivalently map it into a problem which is very similar to the noiseless case, after which one could directly make use of all the classical algorithms used for simulating noiseless quantum circuits. This mapping is demonstrated as follows, which is described mostly based on the state vector representation of the quantum state but is completely general for all the classical simulators we have mentioned.

First we map the density operator into an enlarged ``pure state''. Mathematically, a density operator $\rhoop$ for an $n$-qubit quantum system can be denoted as
\begin{align}
\rhoop = \sum_{\sigma_{n:1}, \sigma'_{n:1}} \rho_{\sigma_1, \dots, \sigma_n}^{\sigma'_1, \dots, \sigma'_n} \vert \sigma_1, \dots, \sigma_n\rangle\langle \sigma'_1, \dots, \sigma'_n\vert,
\end{align}
where $\rho_{\sigma_1, \dots, \sigma_n}^{\sigma'_1, \dots, \sigma'_n}$ is the rank-$2n$ coefficient tensor (density matrix). One can equivalently view the density operator $\rhoop$ as a pure state with $2n$ qubits as
\begin{align}
\vert \rhoop\rangle = \sum_{\sigma_{n:1}, \sigma'_{n:1}} \tilde{\rho}_{\sigma_1, \dots, \sigma_n, \sigma'_1, \dots, \sigma'_n} \vert \sigma_1, \dots, \sigma_n, \sigma'_1, \dots, \sigma'_n\rangle,
\end{align}
where $\tilde{\rho}_{\sigma_1, \dots, \sigma_n, \sigma'_1, \dots, \sigma'_n}$ is similar to $c_{\sigma_1, \dots, \sigma_n}$ in the noiseless case. Again in the mapping of the coefficient tensor from $\rho$ to $\tilde{\rho}$, nothing needs to be done actually, it is just a different way to view the same tensor. We will also refer to $\tilde{\rho}$ as the \textit{squashed state vector}.

The quantum channel is defined as a super operator acting on the density operator $\rhoop$. In the squashed state vector representation, the quantum channel will be a normal quantum operator acting on $\vert \rhoop\rangle$ instead. Taking the case of a single-qubit quantum channel for example, which is denoted as $\mathcal{K}$ and acts on the $j$-th qubit, we assume that it is written in the standard Sudarshan-Kraus-Choi form as~\cite{SudarshanRau1961,JordanSudarshan1961,Choi1975}
\begin{align}
\mathcal{K}(\rho) &= \sum_{s, \sigma_j, \sigma'_j} K^s_{\tau'_j, \sigma'_j} \rho_{\sigma_1, \dots, \sigma_n}^{\sigma'_1, \dots, \sigma'_n} (K^s_{\tau_j, \sigma_j})^{\ast}.
\end{align}
In the squashed state vector representation, the quantum channel $\mathcal{K}$ will be mapped into a normal operator acting on $\tilde{\rho}$, denoted as $\tilde{\mathcal{K}}$, which can be writen as
\begin{align}\label{eq:squashqc}
\tilde{\mathcal{K}}(\tilde{\rho}) 
&= \sum_{s, \sigma_j, \sigma'_j} K^s_{\tau'_j, \sigma'_j}  (K^s_{\tau_j, \sigma_j})^{\ast} \tilde{\rho}_{\sigma_1, \dots, \sigma_n, \sigma'_1, \dots, \sigma'_n} \nonumber \\
& = M^{\tau_j, \tau'_j}_{\sigma_j, \sigma'_j} \tilde{\rho}_{\sigma_1, \dots, \sigma_n, \sigma'_1, \dots, \sigma'_n},
\end{align}
where in the second line we have defined 
\begin{align}
M^{\tau_j, \tau'_j}_{\sigma_j, \sigma'_j} = \sum_{s} K^s_{\tau'_j, \sigma'_j}  (K^s_{\tau_j, \sigma_j})^{\ast} .
\end{align}
We can see from Eq.(\ref{eq:squashqc}) that the effect of a single-qubit quantum channel acting on $\rhoop$ is exactly equivalent to a two-qubit ``gate operation'' acting on $\vert\rhoop\rangle$. Similarly any the $l$-qubit quantum channel acting on $\rhoop$ can be mapped to a $2l$-qubit gate operation on $\vert\rhoop\rangle$. Thus all the quantum channels can be applied onto the squashed state vector using the same algorithm which applies unitary quantum gate operations onto a state vector.

For a density operator, the amplitude of a given bitstring $\vec{b}$ is not well defined, but one can compute the probability of $\vec{b}$ instead, which is
\begin{align}
\langle \vec{b} \vert \rhoop \vert \vec{b}\rangle &= \rho^{\sigma'_1=b_1, \dots, \sigma'_n = b_n}_{\sigma_1=b_1, \dots, \sigma_n = b_n} \nonumber \\ 
&= \tilde{\rho}_{\sigma_1=b_1, \dots, \sigma_n = b_n, \sigma'_1=b_1, \dots, \sigma'_n = b_n}.
\end{align}
Therefore the probability $\langle \vec{b} \vert \rhoop \vert \vec{b}\rangle$ can be computed by computing the ``amplitude'' of the squashed state vector $\vert \rhoop\rangle$ on an enlarged bitstring $\{b_1, \dots, b_n, b_1, \dots, b_n\}$. A local observable $\hat{O}^j$ acting on the $j$-th qubit is by definition
\begin{align}
\trace(\hat{O}^j \rhoop) = \sum_{\sigma_{j-1:1}, \sigma_{n:j+1}, \sigma_j, \sigma'_j} O^{\sigma'_j}_{\sigma_j} \rho^{\sigma_1, \dots, \sigma'_j, \dots, \sigma_n}_{\sigma_1, \dots, \sigma_j, \dots, \sigma_n},
\end{align}
which can be computed by applying a two-qubit gate operation on $\vert \rhoop\rangle$, and then view $\vert \rhoop\rangle$ as a density operator again and take the trace of it (instead of taking the dot product between two state vectors as in the noiseless case).

Therefore we can see that the second approach to simulate a noisy quantum circuit could be equivalently mapped into simulating a noiseless quantum circuit, with some minor differences that the effective number of qubits is doubled and that one may need to slightly adapt the algorithms used to simulate quantum measurements.
There is another subtle difference that in case of noisy quantum circuits the ``gate operations'' on the squashed state vector are no longer unitary, which would affect the classical simulators which explicitly make use of this property.
For the classical simulators that we have introduced, only the MPS based simulator has explicitly used the unitary property to preserve the right canonical form of MPS. In the noisy case, the gate operation algorithms based on MPS can still be used for $\vert \rhoop\rangle$ in principle if the truncation error is set to be very small. However it could be less accurate and less stable compared to the unitary case since the singular values being truncated no longer correspond to the correct Schmidt numbers. The non-unitarity of noisy quantum circuits could also affect the stability of the PEPS based simulator if the gate operation and the tensor network contraction are not performed exactly. The density operator based approach for noisy quantum circuits has been implemented in Qiskit~\cite{Aleksandrowicz2019qiskit}, Quest~\cite{Quest}, Qulacs~\cite{Qulacs}, TensorCircuit~\cite{TensorCircuit}, which mostly uses the Schr$\ddot{\text{o}}$dinger algorithm.


\subsection{Computing gradients for parametric quantum circuits}
In the VQE algorithm, the quantum circuit often contains a number of tunable parameters, and the solution is approached by iteratively updating those parameters. To accelerate convergence, a gradient-based optimization algorithm is usually preferred, especially when there is a large number of parameters.   
To simulate VQE on a classical computer, it would also be helpful to efficiently compute the gradients on classical computers. Compared to quantum computers, computing gradients on classical computers is much more flexible. In the following we will show several different approaches to compute the gradients of parametric quantum circuits, which could be useful in different situations.

We assume that the quantum circuit which is parameterized by a list of parameters $\vec{\theta} = \{\theta_1, \theta_2, \dots, \theta_m\}$, denoted as $\circuit(\vec{\theta})$. We also assume that $\circuit(\vec{\theta})$ can be written as
\begin{align}\label{eq:pqc}
\circuit(\vec{\theta}) = \Qop(\theta_m) \cdots  \Qop(\theta_2) \Qop(\theta_1)
\end{align}
where each parametric gate $\Qop(\theta_j)$ contains a single parameter $\theta_j$ (the case that there exists several parameters in one gate could be analyzed similarly). A quite generic task to do using parametric quantum circuits is to minimize some loss function in the form
\begin{align}\label{eq:loss}
\loss(\vec{\theta}) = \langle 0^{\otimes n} \vert \circuit^{\dagger}(\vec{\theta}) \Hop \circuit(\vec{\theta}) \vert 0^{\otimes n} \rangle,
\end{align}
where $\Hop$ is a Hermitian quantum operator (which does not have to be local).

The most straightforward approach to compute the gradient of the loss function in Eq.(\ref{eq:loss}) is the finite difference method, in which the gradient with respect to one parameter $\theta_j$ is evaluated as
\begin{align}\label{eq:finitediff}
\frac{\partial \loss(\vec{\theta})}{\partial \theta_j} \approx \frac{\loss( \dots, \theta_j + \delta, \dots) - \loss(\vec{\theta})}{\delta},
\end{align}
with $\delta$ a small positive number. Denoting the complexity of evaluating Eq.(\ref{eq:loss}) as $\cpx$, we can see that the complexity of evaluating the gradients with respect to all the parameters using Eq.(\ref{eq:finitediff}) is $O(m\cpx)$ with $m$ the total number of parameters.
Eq.(\ref{eq:finitediff}) evaluates the first-order derivative against $\delta$, which can be slightly modified to be second-order:
\begin{align}\label{eq:finitediff2}
\frac{\partial \loss(\vec{\theta})}{\partial \theta_j} \approx \frac{\loss(\dots, \theta_j + \delta, \dots) - \loss(\dots, \theta_j - \delta, \dots) }{2\delta}.
\end{align}
The complexity of evaluating Eq.(\ref{eq:finitediff2}) is $O(2m\cpx)$. One could also use a higher-order finite-difference method for computing the gradients, which would in general be more accurate by also incur higher computational cost.

For gradient-based algorithms one often wants to obtain gradients which are \textit{numerically exact}. For those parametric quantum circuits whose parameters are all encoded in the single-qubit rotational gates $R_x$, $R_y$, $R_z$ defined as
\begin{align}
R_x(\theta) &= \left[\begin{array}{cc} \cos(\frac{\theta}{2}) & -\im \sin(\frac{\theta}{2}) \\ -\im \sin(\frac{\theta}{2}) & \cos(\frac{\theta}{2}) \end{array} \right]; \\
R_y(\theta) &= \left[\begin{array}{cc} \cos(\frac{\theta}{2}) & -\sin(\frac{\theta}{2}) \\ \sin(\frac{\theta}{2}) & \cos(\frac{\theta}{2}) \end{array} \right]; \\
R_z(\theta) &= \left[\begin{array}{cc} e^{-\im \frac{\theta}{2}} & 0 \\ 0 & e^{\im \frac{\theta}{2}} \end{array} \right],
\end{align}
the exact gradients can be computed using the parameter shift rule as~\cite{mitarai2018quantum}
\begin{align}\label{eq:psr}
\frac{\partial \loss(\vec{\theta})}{\partial \theta_j} = \loss(\dots, \theta_j + \frac{\pi}{2} , \dots) - \loss(\dots, \theta_j - \frac{\pi}{2}, \dots) .
\end{align}
We can see that the complexity of evaluating Eq.(\ref{eq:psr}) is the same to Eq.(\ref{eq:finitediff2}). For more general parametric quantum gate operations Eq.(\ref{eq:psr}) may not hold, nevertheless more sophisticated methods have been proposed evaluate the gradients exactly~\cite{SchuldKilloran2019}.
Importantly, Eqs.(\ref{eq:finitediff},\ref{eq:finitediff2},\ref{eq:psr}) can all be evaluated on quantum computers, since they only require the ability for \textit{forward evaluation} of the loss function in Eq.(\ref{eq:loss}).

Similar to computing the expectation values, for classical simulators there often exists shortcuts to evaluate the gradients. In fact the key for the success of classical deep learning is the \textit{automatic differentiation} algorithm which allows to compute the gradients very efficiently as evaluating the loss function itself~\cite{autodiff}. Since for classical simulators the loss function only runs on a classical computer, in principle one can also benefit from the automatic differentiation algorithm by generalizing it to complex numbers~\cite{GuoPoletti2021}. However, if one directly applies the automatic differentiation algorithm in the classical simulator, then the intermediate outputs after applying each $\Qop(\theta_j)$ onto the quantum state have to be cached so as to enable efficient back propagation, namely one has to store at least $m$ copies of the quantum state. This would be impossible (or highly inefficient) even for moderate values of $m$. Nevertheless, by exploring the reversibility of the quantum circuit, one could use a memory efficient back propagation algorithm for parametric quantum circuits where only two copies of the quantum state have to be used at least. The algorithm is shown in the following based on the Schr$\ddot{\text{o}}$dinger simulator.

Taking the partial derivative of Eq.(\ref{eq:loss}) with respect to one of the parameters, $\theta_j$, we get
\begin{align}\label{eq:grad1}
\frac{\partial \loss(\vec{\theta})}{\partial \theta_j} = \langle \psi(\vec{\theta}) \vert \Hop \circuit_{m:j+1} \frac{d \Qop(\theta_j)}{d\theta_j}\circuit_{j-1:1}\vert 0^{\otimes n}\rangle + \hc,
\end{align}
where we have used $\vert \psi(\vec{\theta})\rangle = \circuit(\vec{\theta}) \vert 0^{\otimes n}\rangle $ and $\circuit_{b:a} = \Qop(\theta_b)\Qop(\theta_{b-1})\cdots \Qop(\theta_a)$. Now we define
\begin{align}
\vert \Phi_j\rangle &= \circuit_{j:1} \vert 0^{\otimes n}\rangle; \\
\vert \Psi_j\rangle &= \circuit_{m:j+1}^{\dagger} \Hop \vert \psi(\vec{\theta})\rangle, 
\end{align}
then Eq.(\ref{eq:grad1}) can be rewritten as
\begin{align}\label{eq:grad2}
\frac{\partial \loss(\vec{\theta})}{\partial \theta_j} = \langle \Psi_{j}\vert \frac{d \Qop(\theta_j)}{d\theta_j} \vert \Phi_{j-1}\rangle + \hc.
\end{align}
Now the memory efficient back propagation algorithm to compute all the partial derivatives runs as follows. First we run a forward evaluation of Eq.(\ref{eq:loss}), during which the two intermediate states $\vert \Phi_m\rangle = \vert \psi(\vec{\theta})\rangle $ and $\vert \Psi_m\rangle = \Hop \vert \psi(\vec{\theta})\rangle $ are both saved. Then we do a backward evolution of the two states $\vert \Phi_m\rangle$ and $\vert \Psi_m\rangle$ by applying each of the gate operations in $\circuit^{\dagger}(\vec{\theta})$ onto them. The details of the backward evolution algorithm is shown in Algorithm.~\ref{alg:ad}.

\begin{algorithm}[H]
\caption{Memory-efficient back propagation algorithm for computing the gradient of a parametric quantum circuit, where $\vert \Phi_m\rangle$ and $\vert \Psi_m\rangle$ are produced during the forward evaluation.} \label{alg:ad}
\begin{algorithmic}[1]
\State $grads = zeros(m)$
\For{$j = m-1:-1:0$}
	\State $\vert \Phi_j \rangle = \Qop(\theta_{j+1})^{\dagger} \vert \Phi_{j+1} \rangle $
	\State $grads[j] = 2\real( \langle \Psi_{j+1} \vert \frac{d \Qop(\theta_j)}{d\theta_j} \vert \Phi_j \rangle  )$
	\State $ \vert \Psi_j\rangle = \Qop(\theta_{j+1})^{\dagger} \vert \Psi_{j+1}\rangle   $
\EndFor
\State Return $grads$

\end{algorithmic}
\end{algorithm}

We can see that in Algorithm.~\ref{alg:ad} no additional memory needs to be allocated in principle, since the gate operations can be done in an inplace fashion, and the ``expectation value'' on two different states as in Eq.(\ref{eq:grad2}) can also be implemented without allocating new memory. Therefore only two copies of the quantum states are required. We can also see that the time complexity of Algorithm.~\ref{alg:ad} is approximately two times the complexity of the forward evaluation, namely $O(2\cpx)$, since one needs to (backward) evolve the two states $\vert \Phi_m\rangle$ and $\vert \Psi_m\rangle$ .

To this end we stress that the $O(2\cpx)$ scaling for Algorithm.~\ref{alg:ad} is only an ideal estimation, in which we have implicitly assumed that the complexity of the state $\vert \Psi_j\rangle$ is equal to $\vert \Phi_m\rangle$ ($\vert \Psi_j\rangle$ is not a proper quantum state since $\Hop$ may not be unitary in general), which is true for the Schr$\ddot{\text{o}}$dinger simulator but may not be true for other classical simulators. Taking the MPS based simulator for example, the state $\vert \Psi_m\rangle$ requires to apply $\Hop$ onto $\vert \psi(\vec{\theta})\rangle$, however, since $\Hop$ could be a complex summation of a large number of Pauli strings, the $\vert \Psi_m\rangle$ would have a much larger bond dimension than $\vert \psi(\vec{\theta})\rangle$ if this operation is performed accurately and the gate operations on $\vert \Psi_m\rangle$ has to be simulated with higher complexity. The integration of the MPS based simulator into the automatic differentiation framework has been proposed and implemented in Ref.~\cite{XuGuo2022}, referred to as the \textit{differentiable MPS}.
It is also possible to generalize Algorithm.~\ref{alg:ad} to compute the gradients of noisy parametric quantum circuits (but the details of the algorithm have to be significantly modified), as long as the quantum channels are reversible. 

\section{CONCLUSION AND OUTLOOK}

This review comprehensively summarizes near-term quantum computing techniques, including variational quantum algorithms, quantum error mitigation, quantum circuit compilation, benchmarking protocols, and classical simulation, from basic concepts to current progress. To develop a practical near-term quantum computing system, a high level of interaction and collaboration between quantum hardware and these near-term quantum computing techniques is required. Over the last years, crucial theoretical and experimental advancements have been made. To realize the leap from the quantum computational advantage for the sampling task without too much practical use to the application-oriented quantum computational advantage, however, greater efforts are required.

First, more profound research are required to fully grasp the potential of NISQ devices and what is the right goal for the NISQ era. Fortunately, we have seen a shift in the focus of research from an initial blind pursuit to serious consideration of these issues~\cite{schuld2022quantum}. We need to find the  ``killer apps” for NISQ, evaluate the resources they consume, and determine whether they can provide us with speed, accuracy or other advantages (especially whether it will be de quantized~\cite{tang2019quantum,ding2021quantum,arrazola2019quantum}), to fully unleash the power of the near-term quantum computing systems.

Second, to continually enhance the capabilities quantum computing hardware, and make it grow massively, a large number of cutting-edge experimental techniques should be conquered. For example, advanced processes, materials and designs are required for the fabrication of quantum computing processors; Dilution refrigerators with larger space and higher cooling power; Cryogenic electronic control techniques; and so on. We believe that the influx of more companies will be of great help to the benign development of the quantum computing industry. Besides, the deep integration of classical and quantum computing is also a preferable way to further enhance the computational power, such as the circuit-cutting method~\cite{peng_simulating_2019,ying2022experimental}.

The future quantum computers may shape our daily lives and science and technology in ways that we cannot currently foresee. Near-term quantum computing techniques are crucial to the entire development stage of quantum computing, serving as an enabler from proof-of-principle demonstrations to engineering scaling. In addition, the available NISQ era cloud-based platforms provide great convenience for exploring and developing practical quantum computing algorithms and methods~\cite{huang2017homomorphic,huang2018demonstrationessentiality}. We expect more young, bright, energetic people entering this field to accelerate the pace of realizing the full promise of quantum computing.

\begin{acknowledgments}
H.-L. H. acknowledges support from the Youth Talent Lifting Project (Grant No. 2020-JCJQ-QT-030), National Natural Science Foundation of China (Grants No. 11905294, 12274464), China Postdoctoral Science Foundation, and the Open Research Fund from State Key Laboratory of High Performance Computing of China (Grant No. 201901-01). C. G. acknowledges support from National Natural Science Foundation of China under Grants No.~11805279, No.~12074117, No.~61833010 and No.~12061131011. G. T. acknowledges support from the Strategic Priority Research Program of Chinese Academy of Sciences Grant No. XDB28000000, and the National Natural Science Foundation of China Grants No. 61832003,  61872334, 61801459. S.-J. W. acknowledge  the National Natural Science Foundation of China under Grants No. 12005015. G.-L. L. acknowledges support from the National Natural Science Foundation of China under Grants No. 11974205, and No. 11774197. The National Key Research and  Development Program of China (2017YFA0303700); The Key Research and  Development Program of Guangdong province (2018B030325002)
\end{acknowledgments}

\bibliographystyle{apsrev4-1}

\bibliography{references}

\end{document}